\pdfoutput=1

\documentclass[preprint,10pt,authoryear]{elsarticle}
\usepackage[utf8]{inputenc}
\usepackage[T1]{fontenc}
\usepackage{lmodern}
\usepackage[left=25mm,right=25mm,top=125.27pt,textheight=550pt]{geometry}
\usepackage{amsmath,amssymb}
\usepackage{textcomp}
\usepackage{graphicx}
\usepackage{longtable,booktabs,array,calc}
\usepackage{microtype}
\usepackage{caption}
\usepackage[hidelinks,unicode]{hyperref}

\allowdisplaybreaks
\providecommand{\doi}[1]{\href{https://doi.org/#1}{\nolinkurl{#1}}}
\usepackage{iftex}
\ifPDFTeX
\DeclareUnicodeCharacter{03B1}{\ensuremath{\alpha}}
\DeclareUnicodeCharacter{03B2}{\ensuremath{\beta}}
\DeclareUnicodeCharacter{03B3}{\ensuremath{\gamma}}
\DeclareUnicodeCharacter{03B4}{\ensuremath{\delta}}
\DeclareUnicodeCharacter{03B5}{\ensuremath{\varepsilon}}
\DeclareUnicodeCharacter{2207}{\ensuremath{\nabla}}
\DeclareUnicodeCharacter{2212}{\ensuremath{-}}
\DeclareUnicodeCharacter{2192}{\ensuremath{\rightarrow}}
\DeclareUnicodeCharacter{207B}{\ensuremath{{}^{-}}}
\DeclareUnicodeCharacter{FB02}{fl}
\DeclareUnicodeCharacter{2010}{-}
\fi

\makeatletter
\newcommand{\FloatBarrier}{\par\ifx\@deferlist\@empty
  \ifx\@toplist\@empty\ifx\@botlist\@empty\else\clearpage\fi
  \else\clearpage\fi\else\clearpage\fi}
\def\ps@pprintTitle{\let\@oddhead\@empty\let\@evenhead\@empty
  \def\@oddfoot{\hfil\thepage\hfil}\let\@evenfoot\@oddfoot}
\makeatother

\newsavebox{\reactionbox}
\newcommand{\reactionline}[1]{\sbox{\reactionbox}{\(#1\)}%
  \ifdim\wd\reactionbox>\linewidth\resizebox{\linewidth}{!}{\usebox{\reactionbox}}%
  \else\usebox{\reactionbox}\fi}
\hypersetup{pdftitle={marineGeochemFoam: A multipurpose framework for simulating reactive transport and biogeochemical cycling in marine sediments and at seawater-groundwater interfaces},pdfauthor={Navid Ahmadi; Evgenii Kortunov; Karim Norouzi Moghanjoghi; Joonas J. Virtasalo; Muhammad Muniruzzaman}}

\begin{document}
\begin{frontmatter}
\title{\emph{marineGeochemFoam}: A multipurpose framework for simulating reactive transport and biogeochemical cycling in marine sediments and at seawater-groundwater interfaces}
\author[aff1,aff2]{Navid Ahmadi\fnref{equal}}
\author[aff3]{Evgenii Kortunov\fnref{equal}}
\author[aff4]{Karim Norouzi Moghanjoghi}
\author[aff5]{Joonas J. Virtasalo}
\author[aff6,aff7]{Muhammad Muniruzzaman\corref{cor1}}
\ead{m.muniruzzaman@uni-bonn.de}
\cortext[cor1]{Corresponding author}
\fntext[equal]{These authors contributed equally to this work.}
\address[aff1]{Global Wetland Center, Department of Geosciences and Natural Resource Management, University of Copenhagen, Copenhagen, Denmark}
\address[aff2]{DHI A/S, Hørsholm, Denmark}
\address[aff3]{Energy and Construction Solutions, Geological Survey of Finland (GTK), Espoo, Finland}
\address[aff4]{Geosciences Department/MARUM-Center for Marine Environmental Sciences, University of Bremen, Bremen, Germany}
\address[aff5]{Marine Geology, Geological Survey of Finland (GTK), Espoo, Finland}
\address[aff6]{Institute of Geosciences, University of Bonn, Bonn, Germany}
\address[aff7]{Water and Mining Solutions, Geological Survey of Finland (GTK), Espoo, Finland}

\begin{abstract}
We present \emph{marineGeochemFoam,} an open-source reactive transport modeling framework for simulating early diagenesis, sediment burial, submarine groundwater discharge, and biogeochemical cycling in marine sediments and groundwater-seawater transition zones. The code is implemented in the widely used numerical toolbox OpenFOAM\textsuperscript{®} and extends the capabilities of the existing diagenetic models by supporting, among other features, i) coupled reactive transport of dissolved solute and solid species, (ii) a flexible native reaction package, supporting arbitrary reaction networks with \emph{n}-th order and Monod-type kinetics, and (iii) coupling with an external geochemical code, PHREEEQC, providing access to a wide range of thermodynamically based geochemical processes. We benchmarked different features of the code across various examples with increasing complexity. Successively, the code was applied to two case studies reproducing field observations from the Hanko groundwater discharge site, Finland, and the Santa Barbara Basin, California. Beyond diagenetic applications, the code also serves as a general-purpose reactive transport simulator.
\end{abstract}

\begin{keyword}
diagenetic modeling \sep sediment burial \sep submarine groundwater discharge \sep reactive transport modeling \sep OpenFOAM \sep PHREEQC \sep marine sediment \sep biogeochemical cycling \sep seawater-groundwater interface
\end{keyword}
\end{frontmatter}

\FloatBarrier
\section{Introduction}\label{sec:1}

Marine sediments play a central role in ocean biogeochemistry by regulating the transformation, recycling, and long-term burial of organic matter, nutrients, and redox-sensitive elements \citep{Boudreau1997,Henrichs1992,Hulse2018,Paraska2014}. The organic matter deposited at the seafloor undergoes a range of physical, chemical, and microbially mediated transformations collectively known as early diagenesis. The degradation of organic matter is driven by a sequence of redox reactions with successive utilization of available electron acceptors, including oxygen, nitrate, manganese and iron oxides, sulfate, and finally methanogenesis, thereby controlling porewater composition, mineral dissolution and precipitation, and the redistribution of carbon, nutrients, sulfur, iron, manganese, and other elements within the sediment column as well as across sediment-water interface \citep{Boudreau1997,Soetaert1996,WangCappellen1996}. Furthermore, continuous deposition of sediments progressively transport particulate solids downward, leading to a competition between remineralization and burial and ultimately determining the fraction of deposited material that is recycled to the ocean or retained in the sedimentary record \citep{Boudreau1997,Henrichs1992}.

In coastal environments, the flow of terrestrial groundwater and recirculated seawater from coastal aquifers into the ocean, referred to as submarine groundwater discharge (SGD), can exert additional controls, delivering large amounts of nutrients and carbon to marine ecosystems that can exceed global riverine inputs \citep{Burnett2003,Cho2018,Church1996,Moore2010,Zhu2025}, thereby substantially impacting coastal ecosystems, primary productivity and water quality on regional and local scales \citep{DiegoFeliu2022,Lecher2016,Rodellas2015,Santos2021,Szymczycha2012}. The fluid that seeps out of the sediment during SGD can create local depressions on the seafloor known as pockmarks \citep{Hoffmann2020,Jensen2002,Virtasalo2019}, which act as physical and biogeochemical hotspots within the mixing zone at the seawater--groundwater interface where strong gradients arise \citep{Alperin1999,Santschi1990}. Thus, early diagenesis, sediment burial, and SGD exert an important control on benthic--pelagic exchange and on marine elemental budgets over shorter to longer timescales \citep{Hulse2018}. Therefore, understanding such environments requires simultaneous consideration of sedimentary reactions, groundwater flow, solute/solid transport, and their spatial variability \citep{Soetaert2012}.

Simulation of early diagenetic processes has evolved from a relatively simple steady-state description of individual geochemical cycles toward more complex reactive transport models (RTMs) that couple multiple transport and biogeochemical processes across sedimentary environments \citep[e.g.,][]{Boudreau1996a,Boudreau1996b,Boudreau2004,Bruggeman2014,Couture2010,Donis2017,Munhoven2021,Purkamo2022,Rooze2016,Sulpis2022,WangCappellen1996}. The most comprehensive models rely on a mass-balance framework in which dissolved solutes and particulate solids are transported by molecular diffusion, burial, bioturbation and bioirrigation while being coupled through networks of primary and secondary biogeochemical reactions \citep{Boudreau1996a,Du2023,Meysman2003}. In such models, SGD is represented by an imposed vertical porewater flux, describing transport of groundwater solute constituents \citep{Purkamo2022}. Despite these advances, most existing models remain tailored to specific reaction networks and parameterization of transport equations that are implemented directly within the source code. This limits their flexibility to extend reaction networks and to incorporate different definitions of rate laws and transport parameters, often requiring users to modify existing model implementations or develop new codes from scratch. Moreover, traditional diagenetic models do not explicitly incorporate aqueous speciation reactions, pH evolution and thermodynamic databases, which are essential in studying mineral reactions as well as dynamics of trace elements and isotopes during diagenesis \citep{Du2023}. While the general-purpose reactive-transport codes can resolve multicomponent transport and thermodynamic calculations, they are typically not formulated around sediment-specific processes such as solid burial, bioturbation, and benthic transport \citep{Steefel2015}. Another limitation is that the relevant capabilities remain fragmented across existing modeling approaches, with individual models incorporating only subsets of processes such as sediment-specific transport, separate treatment of dissolved and solid phases, burial, bioturbation, and flexible reaction networks \citep{Shen2018}. Particularly, the explicit integration of SGD processes into diagenetic models remains uncommon. Finally, as the landscape of scientific computing is rapidly evolving with more computationally efficient and accessible codes, it is important to take advantage of current developments and build open-source, extensible and modular biogeochemical models capable of simultaneously resolving these interacting processes.

In this study, we present \emph{marineGeochemFoam}, a multipurpose open-source modeling framework implemented in the popular computational fluid dynamics platform OpenFOAM\textsuperscript{®} for simulating coupled advective-dispersive transport of solutes, transport of solids via advection and bioturbation, and biogeochemical reactions in marine sediments. \emph{marineGeochemFoam} builds upon the \emph{porousMedia4Foam} framework \citep{Soulaine2021}, extending its capabilities for multiscale flow and reactive transport in porous media to marine biogeochemical applications. We leverage the flexibility of OpenFOAM\textsuperscript{®} including the feasibility of multi-dimensional simulations, unstructured grids and parallel computing, which enables efficient simulation of complex coupled processes \citep{LiWallace2025,Li2025}. The proposed code provides a flexible, generic reaction package capable of simulating a wide range of kinetic reactions with various rate laws, specifically relevant for marine systems. It also offers coupling with the popular geochemical speciation code PHREEQC that facilitates access to a wide range of mineral and speciation reactions, pH and redox calculations, and other geochemical processes (e.g., surface complexation, ion-exchange) by taking advantage of the PhreeqcRM module \citep{ParkhurstWissmeier2015}. In addition, \emph{marineGeochemFoam} framework provides flexibility in choosing between different diffusion and dispersion parameterizations. In the following sections, we first present an overview of the code, followed by a set of benchmark examples in which key features of the proposed model are tested against analytical solutions, data, and other numerical simulators. Once validated, we present several application examples in which the model was used to simulate the measured evolution of dissolved and solid species within different marine settings subject to different degrees of groundwater influence.

\FloatBarrier
\section{Modeling Approach}\label{sec:2}

We developed a tailored reactive transport package for marine biogeochemical applications, enabling us to mechanistically represent transient diagenetic, burial, and SGD processes in sediments. However, the code is flexible and can be used for reactive transport simulations across a wide range of applications. The governing equations and the key features of \emph{marineGeochemFoam} are presented in the following sections.

\subsection{Multi-species reactive transport in marine sediments}\label{sec:2.1}

The reactive transport of solutes and solids was described by two coupled sets of mass balance equations:

\begin{equation}\label{eq:1}
\frac{\partial\left( \varepsilon c_{i}^{w} \right)}{\partial t} + \nabla \cdot \left( {\mathbf{u}c}_{i}^{w} \right) - \nabla \cdot \left( \varepsilon\left( \mathbf{D}_{h,i} + \mathbf{D}_{b} \right)\nabla c_{i}^{w} \right) = R_{i}^{w}
\end{equation}

\begin{equation}\label{eq:2}
\frac{\partial\left( (1 - \varepsilon)c_{j}^{s} \right)}{\partial t} + \nabla \cdot \left( {\mathbf{v}c}_{j}^{s} \right) - \nabla \cdot \left( {(1 - \varepsilon)\mathbf{D}}_{b}\nabla c_{j}^{s} \right) = R_{j}^{s}
\end{equation}

where \(\varepsilon\) is the porosity, \(c_{i}^{w}\) is the concentration of solute \emph{i}, \(c_{j}^{s}\) is the concentration of solid \emph{j}, \(\mathbf{u}\) and \(\mathbf{v}\) are the Darcy flux of solutes and solids, respectively, \(\mathbf{D}_{h,i}\) is the hydrodynamic dispersion tensor of solute \emph{i}, \(\mathbf{D}_{b}\) is the tensor form of bioturbation coefficient, and \(R_{i}^{w}\) and \(R_{j}^{s}\) are the source/sink terms for solute \emph{i} and solid \emph{j}, respectively.

Due to sediment compaction and deposition, the porosity is typically found to be spatially variable in marine sediments and is commonly described by an exponential relationship:

\begin{equation}\label{eq:3}
\varepsilon\left( \mathbf{x} \right) = \varepsilon_{\infty} + {(\varepsilon_{0} - \varepsilon}_{\infty})\exp\left( - \frac{\left( \mathbf{x} - \mathbf{x}_{0,\varepsilon} \right) \cdot \mathbf{i}_{\varepsilon}}{\zeta} \right)
\end{equation}

where, \(\varepsilon_{0}\) denotes the porosity at the sediment-water interface, \(\varepsilon_{\infty}\) represents the asymptotic porosity at greater depth within the sediment, \(\zeta\) is the porosity attenuation factor that referred to as the \emph{e}-folding distance, typically obtained by fitting against measured porosity profiles, \textbf{x}\textsubscript{0\emph{,ε}} is the coordinate of the sediment-water interface, and \textbf{i}\emph{\textsubscript{ε}} is a unit vector in the direction along which porosity changes (for instance, −∇\emph{z} for the vertical downward direction). However, the code is flexible enough to define other arbitrary functions or customized porosity distributions.

The tensor for hydrodynamic dispersion and bioturbation can be expressed as:

\begin{equation}\label{eq:4}
\mathbf{D}_{h,i} = \frac{\mathbf{u}\otimes\mathbf{u}}{\left\| \mathbf{u} \right\|^{2}}\left( D_{\ell/t,i} - D_{t,i} \right) + \mathbf{I}D_{t,i}
\end{equation}

\begin{equation}\label{eq:5}
\mathbf{D}_{b} = D_{b}\left( \mathbf{x} \right)\left( \mathbf{i}_{D_{b}}\otimes\mathbf{i}_{D_{b}} \right)
\end{equation}

where \textbf{I} is an identity matrix, \(\mathbf{i}_{D_{b}}\) is a user-defined unit vector specifying the direction along which bioturbation decays as a function of distance (for instance, −∇\emph{z} for the direction along depth), and \(D_{\ell,i},\ \ D_{t,i}\) and \(D_{b}\) are the longitudinal and transverse dispersion coefficients as well as bioturbation coefficient, respectively:

\begin{equation}\label{eq:6}
D_{\ell/t,i} = \frac{D_{aq,i}}{\tau} + \ D_{m_{\ell /t},i}
\end{equation}

\begin{equation}\label{eq:7}
D_{b}\left( \mathbf{x} \right) = D_{b0}\exp\left( - \frac{\left( \mathbf{x} - \mathbf{x}_{0,D_{b}} \right) \cdot \mathbf{i}_{D_{b}}}{\xi} \right)
\end{equation}

in which \(D_{aq,i}\) is the aqueous diffusion coefficient of species \emph{i}, \(\tau\) is the tortuosity, \(D_{m_{\ell/t,i}}\) is the mechanical dispersion coefficient in the longitudinal or transverse direction. Bioturbation is typically parameterized with a depth-dependent exponential relation, where \(D_{b0}\) is the bioturbation coefficient at the sediment-water interface, \textbf{x}\textsubscript{0\emph{,Db}} is the coordinate of this interface or reference location and \(\xi\) is the mixed layer depth. The proposed code allows users to select between different parameterizations of tortuosity and mechanical dispersion coefficients that are summarized in Table~\ref{tab:1}. Note that mechanical dispersion is typically neglected in early diagenetic models because porewater velocity in sediments is low, but our modeling framework incorporates the dispersion term for completeness, allowing users to select an appropriate dispersion model where relevant.

\begin{table}[htb]
\caption{Overview of the tortuosity and mechanical dispersion models incorporated in marineGeochemFoam. Note that the code allows the mixed selection of the dispersion models for the longitudinal or transverse directions.}\label{tab:1}
\begingroup\linespread{1}\fontsize{9}{10.8}\selectfont
\renewcommand{\arraystretch}{1.1}\setlength{\tabcolsep}{3pt}
\begin{tabular}{@{}>{\raggedright\arraybackslash}p{\dimexpr.48\linewidth-\tabcolsep\relax}>{\raggedright\arraybackslash}p{\dimexpr.52\linewidth-\tabcolsep\relax}@{}}
\toprule
\multicolumn{2}{@{}l}{\textbf{Tortuosity models}}\\\midrule\textbf{Model / Expression} & \textbf{Parameter}\\\midrule
Archie's Law \citep{Archie2003}, \(\tau = \ \varepsilon^{n}\) & \(n\): empirical coefficient,\par \(\varepsilon\): porosity\\[3pt]
Boudreau \citep{Boudreau1996b}, \(\tau = 1 - 2\ln(\varepsilon)\) & \(\varepsilon\): porosity\\[3pt]
\hline
\multicolumn{2}{@{}l}{\textbf{Dispersion models}}\\[3pt]
\hline
\textbf{Model / Expression} & \textbf{Parameter}\\[3pt]
\hline
Scheidegger \citep{Scheidegger1961},\par \[D_{m_{\ell/t},i} = \ \alpha_{\ell/t}\frac{\left\| \mathbf{u} \right\|}{\varepsilon}\] & \(\alpha_{\ell}\): longitudinal dispersivity,\par \(\alpha_{t}\): transverse dispersivity\\[3pt]
Bear and Bachmat \citep{BearBachmat1967},\par \[D_{m_{\ell /t},i} = \ D_{aq,\ \ i}\left\lbrack \frac{{Pe}^{2}}{Pe + 2 + 4\delta_{\ell /t}} \right\rbrack^{\beta_{\ell /t}}\] & \emph{Pe}: Peclet number \(= \frac{\left\| \mathbf{u} \right\| d_{g}}{\varepsilon D_{aq,i}}\), \(d_{g}\): grain diameter,\par \(\beta_{\ell /t}\): empirical exponent for incomplete mixing, \(\delta_{\mathcal{l/}t}\): ratio between a pore-channel length and its hydraulic radius\\[3pt]
Delgado \citep{Delgado2006},\par \[D_{m_{\ell /t},i} = x_{\ell /t}d_{g}\frac{\left\| \mathbf{u} \right\|}{\varepsilon}\] & \(d_{g}\): grain diameter, \(x_{\ell /t}\): empirical coefficient in the longitudinal or transverse direction\\[3pt]
\bottomrule\end{tabular}\endgroup\end{table}

The reactive transport of solutes and solids is tightly coupled through a network of biogeochemical reactions, the rate of which is specified in the sink and source terms of Eq.~\eqref{eq:1} and Eq.~\eqref{eq:2}. In \emph{marineGeochemFoam}, we implemented two reaction packages for simulating arbitrary reaction networks: 1) a \emph{native} reaction package (\emph{multiKinetics}) incorporating a variety of kinetic rate laws, and ii) a PHREEQC coupling package (\emph{phreeqcRM}) providing access to broader range of geochemical reactions and thermodynamic databases. The \emph{multiKinetics} package offers a flexible framework that allows users to define arbitrary reactions with user-specified stoichiometry and incorporates two main types of rate laws: a generic \emph{n}-th-order kinetic formulation and a generalized Monod-type expression. Accordingly, the \emph{n}-th order rate expression for reaction \emph{m,} involving \emph{N\textsubscript{D}} dissolved and \emph{N\textsubscript{S}} solid species can be written as:

\begin{equation}\label{eq:8}
r_{m} = k_{m}\left( \prod_{i = 1}^{N_{D,m}}{{(c}_{i}^{w})}^{\alpha_{i,m}}\  \right)\left( \prod_{j = 1}^{N_{S,m}}{{(c}_{j}^{s})}^{\beta_{j,m}}\  \right)
\end{equation}

where \(k_{m}\) is the rate constant for reaction \emph{m}, and \(\alpha_{i,m}\) and \(\beta_{j,m}\) are empirical exponents representing the reaction order associated with dissolved species \emph{i} and solid species \emph{j}, respectively. Similarly, the generalized Monod-type kinetic expression for reaction \emph{m} containing \emph{N\textsubscript{A}} electron acceptors, and \emph{N\textsubscript{L}} inhibitory species can be written as follows:

\begin{equation}\label{eq:9}
r_{m} = k_{\max,m}\left( \prod_{i = 1}^{N_{A,m}}\frac{c_{i\ }^{w,s}}{c_{i}^{w,s} + k_{c_{i}^{w,s},m}}\  \right)\left( \prod_{j = 1}^{N_{L,m}}\frac{k_{{L,c}_{j}^{w,s},m}}{c_{j}^{w,s} + k_{{L,c}_{j}^{w,s},m}}\  \right)
\end{equation}

where \(k_{max,m}\) is the maximum rate for reaction \emph{m}, \(c_{i}^{w,s}\)and \(c_{j}^{w,s}\)denote the concentration of electron acceptors and inhibitory species, respectively, and \(k_{c_{i}^{w,s},m}\) and \(k_{{L,c}_{j}^{w,s},m}\) are the corresponding half-saturation and inhibition constants. \emph{multiKinetics} allows users to combine these rate expressions in different ways to construct mixed kinetic formulations. A relevant example in marine systems is the organic matter degradation, which commonly uses expressions involving multiple Monod and inhibition terms for different electron acceptors but a first-order dependence on organic matter concentration. Such mixed rate laws can be represented in \emph{multiKinetics} by combining Eq.~\eqref{eq:8} and (9) to include \emph{N\textsubscript{S}} solid species that follow an \emph{n}-th order rate law:

\begin{equation}\label{eq:10}
r_{m} = k_{\max,m}\left( \prod_{k = 1}^{N_{S,m}}{{(c}_{k}^{s})}^{\beta_{k,m}}\  \right)\left( \prod_{i = 1}^{N_{A,m}}\frac{c_{i\ }^{w,s}}{c_{i}^{w,s} + k_{c_{i}^{w,s},m}}\  \right)\left( \prod_{j = 1}^{N_{L,m}}\frac{k_{L,c_{j}^{w,s},m}}{c_{j}^{w,s} + k_{{L,c}_{j}^{w,s},m}}\  \right)
\end{equation}

where \(c_{k}^{s}\) denotes the concentration of solid species (e.g., organic matter) and \(\beta_{k,m}\) is the kinetic reaction order associated with solid species \emph{k} in reaction \emph{m}.

For a reactive system consisting of \emph{M} coupled reactions, the net production/consumption rate for species \emph{i} or \emph{j}, as defined in the reactive transport equations (Eq.~\eqref{eq:1} and Eq.~\eqref{eq:2}) through the source/sink terms, is described as:

\begin{equation}\label{eq:11}
R_{i}^{w} = \ \sum_{m = 1}^{M}{{\varepsilon\Psi_{i,m}\nu}_{i,\ m}r_{m}}
\end{equation}

\begin{equation}\label{eq:12}
R_{j}^{s} = \ \sum_{m = 1}^{M}{{(1 - \varepsilon)\Psi_{j,m}\nu}_{j,\ m}r_{m}}
\end{equation}

where \(\nu_{i,\ m}\) and \(\nu_{j,\ m}\) are the stoichiometric coefficients of solute species \emph{i} and solid species \emph{j} in reaction \emph{m}, respectively. Note that negative values of the stoichiometric coefficients are used for reactants and positive values for products. \(\Psi_{i,m}\) and \(\Psi_{j,m}\) denote unit-conversion factors to ensure dimensional consistency with the left-hand side of the transport equation:

\begin{equation}\label{eq:13}
\Psi_{i,m} = \ \left\{ \begin{array}{r}
\begin{matrix}
\frac{(1 - \varepsilon)}{\varepsilon}, & when\ r_{m}\ has\ a\ unit\ of\ \lbrack mol/m_{solid}^{3}\rbrack
\end{matrix} \\
\begin{matrix}
1, & when\ r_{m\ }has\ a\ unit\ of\ \lbrack mol/m_{water}^{3}\rbrack
\end{matrix}
\end{array} \right.\ 
\end{equation}

\begin{equation}\label{eq:14}
\Psi_{j,m} = \ \left\{ \begin{array}{r}
\begin{matrix}
1, & when\ r_{m}\ has\ a\ unit\ of\ \lbrack mol/m_{solid}^{3}\rbrack
\end{matrix} \\
\begin{matrix}
\frac{\varepsilon}{(1 - \varepsilon)}, & when\ r_{m\ }has\ a\ unit\ of\ \lbrack mol/m_{water}^{3}\rbrack
\end{matrix}
\end{array} \right.\ 
\end{equation}

This formulation offers great flexibility as it can be applied to any user-specified reaction network with selectable rate laws rather than being limited to a set of hardcoded ordinary differential equations (ODEs) for a particular reaction network.

In addition to the native reaction package, the \emph{phreeqcRM} package of \emph{marineGeochemFoam} provides coupling with the widely used geochemical code PHREEQC, hence providing access to its extensive geochemical capabilities, including various equilibrium and kinetic reactions such as aqueous speciation, mineral dissolution and precipitation, ion exchange, surface complexation, redox reactions, and user-defined rate laws. This package is particularly advantageous for systems in which resolving aqueous speciation, calculating pH, and accounting for thermodynamic databases are critical.

\subsection{Numerical implementation}\label{sec:2.2}

Reactive transport equations of solutes and solids (Eq.~\eqref{eq:1}-(2)) were solved using a cell-centered finite volume formulation on arbitrary polyhedral meshes. This was done by implementing \emph{marineGeochemFoam} within the open-source C++ finite volume library OpenFOAM, which offers a flexible platform for solving partial differential equations (PDEs) and supports a wide variety of mesh structures, discretization methods, and numerical schemes for differential operators. Specifically, we used the \emph{porousMedia4Foam} package \citep{Soulaine2021} as a starting point, as it provides a generic implementation of advection-dispersion equation for solutes and a coupling to PHREEQC via PhreeqcRM module \citep{ParkhurstWissmeier2015}. \emph{marineGeochemFoam} extended these features by incorporating the transport of solid species, reactive source and sink terms with flexible rate-law definitions, comprehensive dispersion and tortuosity parameterizations, reaction-controlled adaptive time stepping, and an updated coupling to PHREEQC that accommodates solid transport. The general code structure of \emph{porousMedia4Foam} was preserved, while libraries and packages that were redundant for marine system simulations were simplified or removed. Thus, \emph{marineGeochemFoam} inherits the core features of OpenFOAM and \emph{porousMedia4Foam}, including automatic matrix assembly for finite-volume discretization, parallel computing capabilities, established input-file structures, and support for a variety of mesh types. The discretized algebraic system of equations was solved sequentially for each species. In the native reaction package (\emph{multiKinetics}), the nonlinear reaction terms were solved with two different linearization approaches, implemented in two separate libraries: \emph{multiKineticsPicard}, and \emph{multiKineticsLinearized}. In contrast, the PHREEQC coupling package (\emph{phreeqcRM}) relies on an operator-splitting approach, where reactions are solved externally by PhreeqcRM.

We carried out the full code development and numerical implementation, whereas AI tools were used only for limited technical assistance with specific tasks. Particularly, ChatGPT (GPT-5.4) was used to support the derivation and assembly of Jacobian in the \emph{multiKineticsLinearized} package as well as debugging of this library. Furthermore, Codex (GPT-5.6 Sol) was used to assist with calibration, data fitting, and parameter exploration exercises in the Santa Barbara Basin application example (see section 3 and 4). We made all scientific decisions, final implementation and modeling choices.

\subsubsection{multiKineticsPicard: Explicit nonlinear treatment of reactive source/sink term}\label{sec:2.2.1}

\begin{figure}[h]
\centering
\includegraphics[width=\linewidth,height=.70\textheight,keepaspectratio]{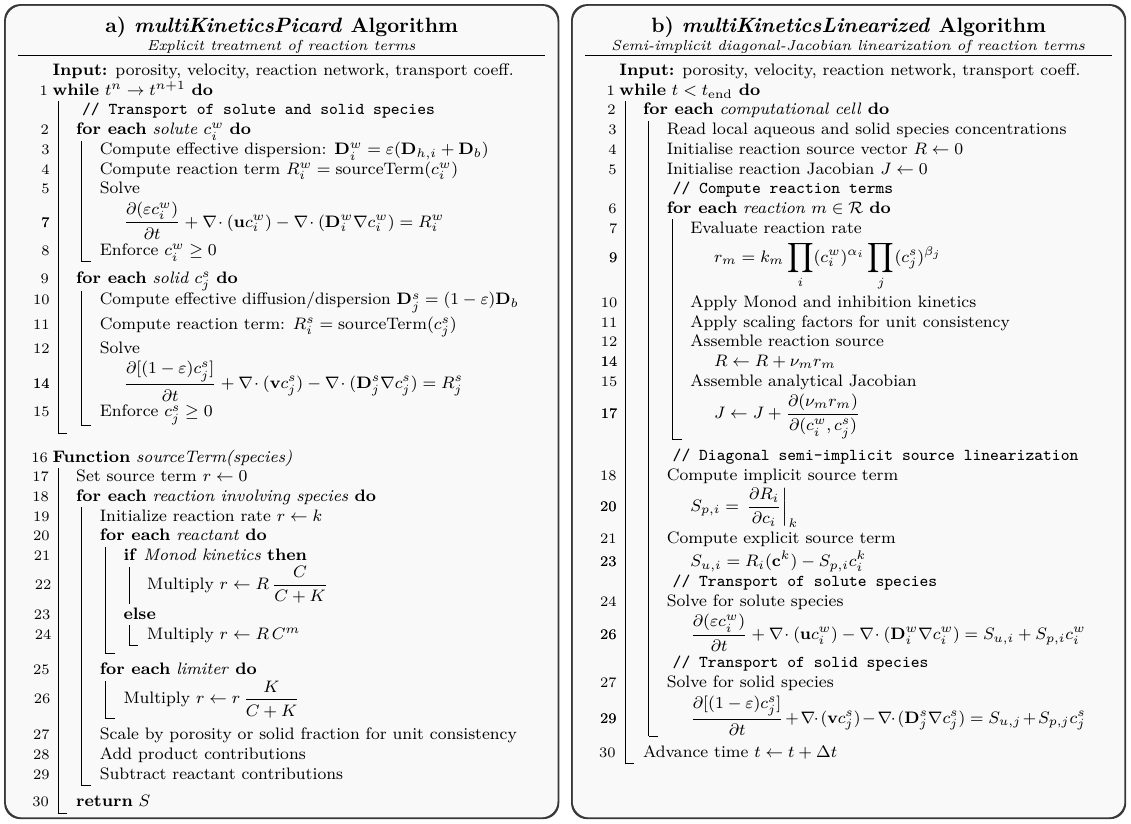}
\caption{Overview of the two solution procedures implemented in the native reaction package (\emph{multiKinetics}).}\label{fig:1}
\end{figure}

The first approach employs a Picard-based loop, where nonlinear reaction terms were evaluated explicitly using concentrations from the previous iteration or time step (Figure~\ref{fig:1}a). The reactive source/sink term was computed independently for each species and inserted directly into the transport equation as an explicit volumetric forcing term. This approach yields a simple operator-splitting style formulation and avoids the construction of nonlinear Jacobian matrices. However, as reaction terms are treated explicitly, this method may suffer from stability restrictions for strongly nonlinear or stiff reaction systems. In particular, highly reactive systems may require small time steps to avoid oscillations or negative concentrations. To maintain physical admissibility, species concentrations were constrained to remain non-negative after each transport step.

\subsubsection{multiKineticsLinearized: Semi-implicit diagonal-Jacobian linearization of reaction terms}\label{sec:2.2.2}

The second approach introduces a quasi-Newton treatment of reaction terms with a semi-implicit linearization of the nonlinear reaction terms through construction of local diagonal-Jacobian contributions \citep{Patankar1981}. Instead of treating reaction rates explicitly, the reactive source terms are linearized around the current solution estimate, \emph{k}+1 (Figure~\ref{fig:1}b). For each species, the reaction term is expanded as:

\begin{equation}\label{eq:15}
R_{i}\left( \mathbf{C}^{k + 1} \right) \approx R_{i}\left( \mathbf{C}^{k} \right) + \sum_{j}^{}\frac{\partial R_{i}}{\partial C_{j}}\left( C_{j}^{k + 1} - C_{j}^{k} \right),
\end{equation}

where \(\mathbf{C}\) denotes the vector of all dissolved and solid species concentrations. The linearized source term can be recast into the standard finite volume form:

\begin{equation}\label{eq:16}
R_{i}^{k + 1} = S_{u,i} + S_{p,i}C_{i}^{k + 1},
\end{equation}

where

\begin{equation}\label{eq:17}
S_{p,i} = \left. \ \frac{\partial R_{i}}{\partial C_{i}} \right|_{\mathbf{C}=\mathbf{C}^{k}}
\end{equation}

and

\begin{equation}\label{eq:18}
S_{u,i} = R_{i}^{k} - S_{p,i}C_{i}^{k}.
\end{equation}

The diagonal Jacobian contribution \(S_{p,i}\) is incorporated implicitly into the matrix diagonal, while the remaining source contribution \(S_{u,i}\) is retained on the right-hand side. The resulting discretized finite volume equations for solute and solid species thus become:

\begin{equation}\label{eq:19}
\frac{\varepsilon c_{i}^{w,\ n + 1} - \varepsilon c_{i}^{w,n}}{\Delta t} + \nabla \cdot \left( \mathbf{u}c_{i}^{w,n + 1} \right) - \nabla \cdot \left( \varepsilon\left( \mathbf{D}_{h,i} + \mathbf{D}_{b} \right)\nabla c_{i}^{w,n + 1} \right) = S_{u,i} + S_{p,i}c_{i}^{w,n + 1}.
\end{equation}

\begin{equation}\label{eq:20}
\frac{(1 - \varepsilon)c_{j}^{s,n + 1} - (1 - \varepsilon)c_{i}^{s,n}}{\Delta t} + \nabla \cdot \left( \mathbf{v}c_{j}^{s,n + 1} \right) - \nabla \cdot \left( {(1 - \varepsilon)\mathbf{D}}_{b}\nabla c_{j}^{s,n + 1} \right) = S_{u,j} + S_{p,j}c_{j}^{s,n + 1}
\end{equation}

with \emph{n} and \emph{n}+1 being the previous and current time step. The Jacobian entries were derived analytically from the nonlinear kinetic expressions. For a general rate law,

\[R = k\prod_{i}^{}f_{i}\left( C_{i} \right),\]

the logarithmic derivative form was used,

\[\frac{\partial R}{\partial C_{j}} = R\frac{\partial\ln R}{\partial C_{j}},\]

which allows efficient evaluation of nonlinear sensitivities associated with power-law, Monod, and inhibitor terms. In this formulation, only the diagonal Jacobian terms were included implicitly in the transport matrix, whereas the off-diagonal couplings between species were retained explicitly. This yields a semi-implicit scheme that substantially improves numerical stability while avoiding the complexity and computational cost of assembling and solving a fully coupled global nonlinear Jacobian system \citep{Eberhardt1992}. Compared to the first method, the semi-implicit Jacobian formulation provides improved robustness for stiff geochemical systems and permits significantly larger time steps while maintaining stability and positivity of the solution.

\subsubsection{phreeqcRM: Coupling to PHREEQC for geochemical reactions}\label{sec:2.2.3}

This library relies on a coupling between OpenFOAM and the widely used geochemical solver PHREEQC through the PhreeqcRM module, proving a flexible and versatile framework for accessing the full range of PHREEQC reaction capabilities. This was achieved using a sequential non-iterative operator-splitting approach \citep[e.g.,][]{MuniruzzamanRolle2016}, in which the transport step is first solved within OpenFOAM and, successively, the concentrations of solutes and solids are passed to PhreeqcRM to solve geochemical reactions. The newly updated concentrations from PhreeqcRM are passed back to the transport solver in a subsequent time step. We took advantage of the general coupling routine implemented in \emph{porousMedia4Foam}, which provides two coupling strategies: (i) a simple sequential approach, in which transport is solved over an entire time step and the resulting concentrations are subsequently passed to PhreeqcRM, and (ii) an approach based on Strang's algorithm. However, we substantially extended and adapted this library for application to marine systems, where the transport of solid species introduces additional complexities that require further treatment. One inherent limitation of this library is that small time steps are required to reduce numerical errors associated with operator splitting. Nevertheless, this can be beneficial when resolving detailed geochemical processes is important.

\FloatBarrier
\section{Benchmark and application examples}\label{sec:3}

The capabilities and key features of the proposed modeling framework were systematically tested in a series of benchmark examples with increasing complexity. The code was then applied to simulate previously published biogeochemical datasets collected from different sediments to demonstrate its applicability to real-world settings. An overview of the benchmark and application problems with the key features addressed in each problem is presented in Table~\ref{tab:2}.

\begin{table}[p]
\caption{Overview of the benchmark and application examples considered to verify the implementation of different features of marineGeochemFoam and demonstrate its applicability to real-world settings}\label{tab:2}
\begingroup\linespread{1}\fontsize{8.5}{10}\selectfont
\renewcommand{\arraystretch}{1.04}\setlength{\tabcolsep}{3pt}
\begin{tabular}{@{}>{\centering\arraybackslash}p{\dimexpr.14\linewidth-2\tabcolsep\relax}>{\raggedright\arraybackslash}p{\dimexpr.29\linewidth-\tabcolsep\relax}>{\raggedright\arraybackslash}p{\dimexpr.57\linewidth-\tabcolsep\relax}@{}}
\toprule
\textbf{Benchmark} & \textbf{Problem} & \textbf{Description of key features}\\\midrule
1 & Conservative transport of solute and solid species in porous media & - Advective-dispersive transport of two solute species in 2-D and 3-D homogenous porous media (Benchmark 1A)\par - Transport of two solid species with advection and bioturbation in a 1-D domain with spatially variable porosity (Benchmark 1B)\par - Testing the implementation of the transport module\\[3pt]
\midrule
2 & Kinetically-controlled reactions presented with different rate laws in 0-D systems & - Linear decay, quadratic, bimolecular and dual-Monod type reaction rates\par - Multi-Monod reaction rates with inhibition terms for simulation of organic matter degradation with successive utilization of available electron acceptors\par - Benchmarking the implementation of the reaction package\\[3pt]
\midrule
3 & Reactive transport of multiple solute and solid species (early diagenesis) & - 1-D transport of multiple solute species through advection/dispersion and bioturbation mechanisms, coupled with reactions involving immobile solid phases (Benchmark 3A)\par - 1-D coupled transport of multiple solute and solid species in the opposite direction, coupled with early diagenetic reaction network involving both solutes and solids (Benchmark 3B)\par - Testing the implementation of the coupled solute/solid reactive transport capability with native reaction package (\emph{multiKinetics})\\[3pt]
\midrule
4 & Multicomponent solute transport (speciation, mineral reactions and surface reactions) & - 1-D transport of ion species and cation exchange reactions (Benchmark 4A)\par - 1-D solute transport with dissolution/precipitation of minerals and pH evolution (Benchmark 4B)\par - 1-D solute transport and sulfide oxidation in a marine sediment (Benchmark 4C)\par - Testing the implementation of the reactive transport capability with PhreeqcRM coupling package\\[3pt]
\midrule
\textbf{Application} & \textbf{Problem} & \textbf{Description of key features}\\[3pt]
\midrule
1 & Biogeochemical cycling during submarine groundwater discharge in the Hanko site, Finland & - Coupled reactive transport of multiple solute and solid species in marine sediments with and without the influence of submarine groundwater discharge\par - A complex reaction network including organic matter decomposition and reoxidation of reaction products\par - Testing the flexibility of the native reaction package (\emph{multiKinetics}) for simulating real-world systems with extensive reaction networks over long timescales.\\[3pt]
\midrule
2 & Marine sediment diagenesis in the Santa Barbara Basin, California & - Coupled reactive transport of multiple solute and solid species in a marine sediment\par - A complex reaction network including diagenetic processes, mineral reactions, speciation of major and trace elements, and isotope fractionation\par - Testing the capability of the \emph{phreeqcRM} package for reproducing field observations by incorporating thermodynamic databases\\[3pt]
\bottomrule\end{tabular}\endgroup\end{table}
\FloatBarrier

\subsection{Benchmark examples}\label{sec:3.1}

In the first benchmark exercise (Benchmark 1), we tested the conservative transport capability of the model in two separate examples. Benchmark 1A focuses on 2-D and 3-D transport simulation of two solute tracers (i.e., fluorescein and oxygen) in homogenous porous media (i.e., Eq.~\eqref{eq:1} with \(R_{i}^{w} = 0\)). The simulation results were compared against an experimental dataset from \citet{Ye2015} and an analytical solution. In this study, a series of flow-through experiments was performed in quasi 2-D and 3-D chambers operated under constant seepage velocities between 0.5 and 8 m/day. After establishing steady-state flow conditions, a solution containing sodium fluorescein and oxygen was injected through the central inlet port of the setups, while a tracer-free solution was injected from the surrounding inlet ports, generating a localized tracer plume that evolved through advection and dispersion within the saturated porous medium. When the plume reached steady state conditions, the spatial distribution of fluorescein and oxygen concentration was measured at the outlet ports. The benchmark is particularly suitable for testing the implementation of the conservative solute transport equation in a multi-dimensional system, as well as for validating the implementation of the non-linear, compound-specific parameterization of the dispersion coefficient (see Table~\ref{tab:1}) that was introduced by \citet{BearBachmat1967} and experimentally validated by several studies \citep[e.g.,][]{Chiogna2010,Rolle2012,Ye2015}. The details of the simulations including model input parameters are provided in Table S1, whereas experimental details can be found in \citet{Ye2015}.

Benchmark 1B focuses on transport of two solid species (A and B) in a 1-D domain representing a sediment profile with spatially variable porosity. The problem was designed to validate the implementation of conservative solid transport equation (i.e., Eq.~\eqref{eq:2} with \(R_{j}^{s} = 0\)). Initially, the sediment domain was completely occupied by species A with concentration of 1 mol/m\textsuperscript{3}. A constant concentration of 0 mol/m\textsuperscript{3} for species A and 1 mol/m\textsuperscript{3} for species B was imposed on the left boundary, leading to the progressive displacement of A by B throughout the domain. The development of the A and B front was simulated and compared against another PDE solver, COMSOL Multiphysics. The model input parameters for this exercise are provided in Table S1.

Benchmark 2 comprises a set of subproblems dedicated to validating the implementation of the reaction package (see Table S2 for details). First, simple first-order (Benchmark 2A) and second-order (Benchmark 2B) reaction kinetics were tested against analytical solutions to validate the numerical implementation of linear decay, quadratic, and bimolecular reactions. Successively, the implementation of Monod-type microbial reaction kinetics was validated against an ODE solver (Benchmark 2C). Finally, coupled reaction networks representing the primary redox processes in early diagenesis were simulated to assess the robustness of \emph{marineGeochemFoam} for handling multi-species reactive systems with nonlinear and interdependent reaction pathways (Benchmark 2D).

Benchmark 3 presents two examples of increasing complexity for testing the implementation of coupled multi-species reactive transport features. Benchmark 3A is dedicated to testing the reactive transport of solute species while solids are present in the domain and participate in the reaction network but are not transported (results are given in the Supplementary Material). Benchmark 3B, however, considers migration of solid species in the opposite direction of solutes, representing submarine groundwater discharge and burial of solids and allowing us to test the full complexity. The reaction network considered for this example was the primary redox processes in early diagenesis (Table S2). The simulation results of these two examples were compared against the outcomes of COMSOL Multiphysics, and the details regarding model input parameters, initial and boundary conditions are given in Table S3.

Eventually, we tested the PHREEQC coupling approach as implemented in the \emph{phreeqcRM} library in Benchmark 4 through three examples. Benchmark 4A represents example 11 in the PHREEQC version 3 manual and focuses on the simulation of transport and ion exchange reactions \citep{ParkhurstAppelo2013}. The problem has been widely used in previous studies validating PhreeqcRM coupling and involves a column containing a cation exchanger initially in equilibrium with sodium, potassium and nitrate. A calcium chloride solution was then injected into the column, promoting competitive cation exchange reactions in the column and thus different arrival times of solutes at the column outlet. The solute breakthrough curves at the outlet were simulated by \emph{marineGeochemFoam} and were compared with the PHREEQC results. Further details of this numerical example are given in Table S4, and the results are shown in the Supplementary Material (Figure S4).

Benchmark 4B is dedicated to the simulation of transport and mineral dissolution/precipitation reactions. This example was originally presented by \citet{EngesgaardKipp1992} and involves a 1-D porous medium domain in which calcite is initially present and is in equilibrium with a water composition. Successively, the system is flushed with a different composition and lower pH, leading to calcite dissolution and dolomite precipitation in the domain. The simulated spatial profiles of minerals, aqueous species and pH by \emph{marineGeochemFoam} were compared with those of PHREEQC. The speciation reactions used to simulate this problem are provided in \emph{phreeqc.dat} database and the model input parameters are given in Table S4.

Benchmark 4C focuses on the simulation of sulfide oxidation in a dense mat of sulfide-oxidizing bacteria (\emph{Beggiato}). This example is based on the reactive transport experiment performed by \citet{JorgensenRevsbech1983}, in which the vertical gradients of oxygen, sulfide, and pH were measured in the black mud of a sediment core. We simulated this experiment by including diffusion of dissolved species and representing the oxidation reaction using a depth-dependent kinetic rate in the form of a Gaussian distribution as \({R = k}_{ox}\exp\left( - \alpha{(x - x_{0})}^{2} \right)\). In this kinetic model, \(k_{ox}\) is the rate coefficient, \(x\) is the depth, \(x_{0}\) is the location of the \emph{Beggiatoa} mat and \(\alpha\) controls the sharpness of the reaction zone around the mat. Additionally, the model accounts for the association/dissociation of carbonic and boric acids and hydrogen sulfide present in the porewater (defined in \emph{phreeqc.dat} database). The simulation results were compared with the measurements under steady-state conditions established after almost one hour. The model input parameters for this example are given in Table S4.

\subsection{Application examples}\label{sec:3.2}

After benchmarking different features of the code, it was applied to simulate two selected case studies: i) biogeochemical cycling during submarine groundwater discharge different pockmarks in the Hanko site, Finland, and ii) marine sediment diagenesis in the Santa Barbara Basin (SBB), one of the California borderland basins.

In the first application example (Hanko SGD site, Finland), we used \emph{marineGeochemFoam} to simulate the measured pore water and sediment profiles in two of the pockmarks (D and E) presented in \citet{Purkamo2022}. Pockmark D was formed in fine sand and showed the strongest groundwater discharge; hence, it was simulated considering advection, diffusion and bioturbation for solutes and advection and bioturbation for solids. Conversely, pockmark E was inactive and covered by an organic-rich mud layer, showing no observable groundwater velocity, and it was thus simulated by neglecting solute advection. In both cases, porosity was spatially variable, with a gradually decreasing trend along the depth. The reaction network comprises a set of 34 complex biogeochemical reactions, including primary and secondary reactions occurring during early diagenesis (see Table~\ref{tab:3}). The primary redox reactions describe organic matter degradation using a sequence of terminal electron acceptors (O\textsubscript{2}, NO\textsubscript{3}⁻, MnO\textsubscript{2}, Fe(OH)\textsubscript{3} and SO\textsubscript{4}⁻\textsuperscript{2}), followed by methanogenesis once these oxidants are depleted. Additionally, the model also considers multiple reactivity pools for organic matter and metal oxides corresponding to highly reactive (\emph{α}), less reactive (\emph{β}), and nonreactive phases (\emph{γ}) \citep{Jorgensen1978,WestrichBerner1984}. These biogeochemical processes were simulated using Monod-type expressions with inhibition terms to represent the preferential use of electron acceptors (Table S8).

\begin{table}[p]
\caption{Overview of the biogeochemical reaction network used in the simulation of pockmark E and in D in the Hanko site, Finland.}\label{tab:3}
\begingroup
\linespread{1}\fontsize{9.5}{11}\selectfont
\renewcommand{\arraystretch}{1.0}\setlength{\tabcolsep}{3pt}

\begin{tabular}{@{}>{\raggedright\arraybackslash}m{.065\linewidth}>{\raggedright\arraybackslash}m{\dimexpr.935\linewidth-2\tabcolsep\relax}@{}}
\toprule \textbf{ID} & \textbf{Reaction}\\\midrule
\multicolumn{2}{@{}l}{\textbf{Primary redox reactions}}\\*[2pt]
\hypertarget{pockmark-reaction:1}{R1} & \reactionline{\text{O}\text{M}^{\alpha,\beta} + a\ \text{O}_{2} \rightarrow a\ \text{C}\text{O}_{2} + b\ \text{N}\text{H}_{4}^{+} + c\text{ H}_{3}\text{P}\text{O}_{4} + a\ \text{H}_{2}\text{O}}\\[1pt]
\hypertarget{pockmark-reaction:2}{R2} & \reactionline{\text{O}\text{M}^{\alpha,\beta} + \frac{4a}{5}\text{N}\text{O}_{3}^{-} + \frac{4a}{5}\text{ H}^{+} \rightarrow a\ \text{C}\text{O}_{2} + b\ \text{N}\text{H}_{4}^{+} + c\ \text{H}_{3}\text{P}\text{O}_{4} + \frac{2a}{5}\text{ N}_{2} + \frac{7a}{5}\text{ H}_{2}\text{O}}\\[1pt]
\hypertarget{pockmark-reaction:3}{R3} & \reactionline{\text{O}\text{M}^{\alpha,\beta} + 2a\ \ \text{Mn}\text{O}_{2}^{\alpha} + 4a\text{ H}^{+} \rightarrow a\ \text{C}\text{O}_{2} + b\ \text{N}\text{H}_{4}^{+} + c\text{ H}_{3}\text{P}\text{O}_{4} + 2a\text{ Mn}^{2 +} + 3a\text{ H}_{2}\text{O}}\\[1pt]
\hypertarget{pockmark-reaction:4}{R4} & \reactionline{\text{O}\text{M}^{\alpha,\beta} + 4a\ \text{Fe}(\text{OH})_{3}^{\alpha} + 4a\chi^{\alpha}\ \text{Fe}_{ox}\text{P} + 12a\text{ H}^{+} \rightarrow a\ \text{C}\text{O}_{2} + b\ \text{N}\text{H}_{4}^{+} + (c + 4a\chi^{\alpha})\ \text{H}_{3}\text{P}\text{O}_{4} + 4a\text{ Fe}^{2 +} + 13a\ \text{H}_{2}\text{O}}\\[1pt]
\hypertarget{pockmark-reaction:5}{R5} & \reactionline{\text{O}\text{M}^{\alpha,\beta} + \frac{a}{2}\text{ S}\text{O}_{4}^{2 -} + a\text{ H}^{+} \rightarrow a\ \text{C}\text{O}_{2} + b\ \text{N}\text{H}_{4}^{+} + c\ \text{H}_{3}\text{P}\text{O}_{4} + \frac{a}{2}\text{ H}_{2}\text{S} + a\ \text{H}_{2}\text{O}}\\[1pt]
\hypertarget{pockmark-reaction:6}{R6} & \reactionline{\text{O}\text{M}^{\alpha,\beta} \rightarrow \frac{a}{2}\text{ C}\text{O}_{2} + b\ \text{N}\text{H}_{4}^{+} + c\text{ H}_{3}\text{P}\text{O}_{4} + \frac{a}{2}\text{ C}\text{H}_{4}}\\[1pt]
\hypertarget{pockmark-reaction:7}{R7} & \reactionline{\text{C}\text{O}_{2} + 4\ \text{H}_{2} \rightarrow \text{C}\text{H}_{4} + 2\text{ H}_{2}\text{O}}\\[1pt]
\multicolumn{2}{@{}l}{\textbf{Secondary redox and other reactions}}\\*[2pt]
\hypertarget{pockmark-reaction:8}{R8} & \reactionline{2\ \text{O}_{2} + \text{N}\text{H}_{4}^{+} + 2\ \text{HC}\text{O}_{3}^{-} \rightarrow \text{N}\text{O}_{3}^{-} + 2\ \text{C}\text{O}_{2} + 3\text{ H}_{2}\text{O}}\\[1pt]
\hypertarget{pockmark-reaction:9}{R9} & \reactionline{\text{O}_{2} + 2\ \text{Mn}^{2 +} + 4\ \text{HC}\text{O}_{3}^{-} \rightarrow 2\ \text{Mn}\text{O}_{2}^{\alpha} + 4\ \text{C}\text{O}_{2} + 2\text{ H}_{2}\text{O}}\\[1pt]
\hypertarget{pockmark-reaction:10}{R10} & \reactionline{\text{O}_{2} + 4\ \text{Fe}^{2 +} + 8\ \text{HC}\text{O}_{3}^{-} + 2\text{ H}_{2}\text{O} + 4\chi^{\alpha}\text{ H}_{2}\text{P}\text{O}_{4}^{-} \rightarrow 4\ \text{Fe}(\text{OH})_{3}^{\alpha} + 4\chi^{\alpha}\ \text{Fe}_{ox}\text{P} + 8\ \text{C}\text{O}_{2}}\\[1pt]
\hypertarget{pockmark-reaction:11}{R11} & \reactionline{2\text{ O}_{2} + \text{FeS} \rightarrow \text{S}\text{O}_{4}^{2 -} + \text{Fe}^{2 +}}\\[1pt]
\hypertarget{pockmark-reaction:12}{R12} & \reactionline{7\ \text{O}_{2} + 2\ \text{Fe}\text{S}_{2} + 2\text{ H}_{2}\text{O} \rightarrow 4\ \text{S}\text{O}_{4}^{2 -} + 2\ \text{Fe}^{2 +} + 4\ \text{H}^{+}}\\[1pt]
\hypertarget{pockmark-reaction:13}{R13} & \reactionline{2\ \text{O}_{2} + \text{H}_{2}\text{S} + 2\ \text{HC}\text{O}_{3}^{-} \rightarrow \text{S}\text{O}_{4}^{2 -} + 2\ \text{C}\text{O}_{2} + 2\ \text{H}_{2}\text{O}}\\[1pt]
\hypertarget{pockmark-reaction:14}{R14} & \reactionline{2\ \text{O}_{2} + \text{C}\text{H}_{4} \rightarrow \text{C}\text{O}_{2} + 2\text{ H}_{2}\text{O}}\\[1pt]
\hypertarget{pockmark-reaction:15}{R15} & \reactionline{\text{Mn}\text{O}_{2}^{\alpha} + 2\ \text{Fe}^{2 +} + 2\chi^{\alpha}\text{ H}_{2}\text{P}\text{O}_{4}^{-} + 2\ \text{H}_{2}\text{O} + 2\ \text{HC}\text{O}_{3}^{-} \rightarrow 2\ \text{Fe}(\text{OH})_{3}^{\alpha} + 2\chi^{\alpha}\text{ Fe}_{ox}\text{P} + \text{Mn}^{2 +} + 2\ \text{C}\text{O}_{2}}\\[1pt]
\hypertarget{pockmark-reaction:16}{R16} & \reactionline{\text{Mn}\text{O}_{2}^{\beta} + 2\text{ Fe}^{2 +} + 2\chi^{\alpha}\ \text{H}_{2}\text{P}\text{O}_{4}^{-} + 2\ \text{H}_{2}\text{O} + 2\ \text{HC}\text{O}_{3}^{-} \rightarrow 2\ \text{Fe}(\text{OH})_{3}^{\alpha} + 2\chi^{\alpha}\ \text{Fe}_{ox}\text{P} + \text{Mn}^{2 +} + 2\ \text{C}\text{O}_{2}}\\[1pt]
\hypertarget{pockmark-reaction:17}{R17} & \reactionline{\text{Mn}\text{O}_{2}^{\alpha} + \text{H}_{2}\text{S} + 2\ \text{C}\text{O}_{2} \rightarrow \text{Mn}^{2 +} + \text{S}_{0} + 2\ \text{HC}\text{O}_{3}^{-}}\\[1pt]
\hypertarget{pockmark-reaction:18}{R18} & \reactionline{\text{Mn}\text{O}_{2}^{\beta} + \text{H}_{2}\text{S} + 2\ \text{C}\text{O}_{2} \rightarrow \text{Mn}^{2 +} + \text{S}_{0} + 2\ \text{HC}\text{O}_{3}^{-}}\\[1pt]
\hypertarget{pockmark-reaction:19}{R19} & \reactionline{2\ \text{Fe}(\text{OH})_{3}^{\alpha} + 2\chi^{\alpha}\text{ Fe}_{ox}\text{P} + \text{H}_{2}\text{S} + 4\ \text{C}\text{O}_{2} \rightarrow 2\text{ Fe}^{2 +} + 2\chi^{\alpha}\text{ H}_{2}\text{P}\text{O}_{4}^{-} + \text{S}_{0} + 4\ \text{HC}\text{O}_{3}^{-} + 2\text{ H}_{2}\text{O}}\\[1pt]
\hypertarget{pockmark-reaction:20}{R20} & \reactionline{2\ \text{Fe}(\text{OH})_{3}^{\beta} + 2\chi^{\beta}\text{ Fe}_{ox}\text{P} + \text{H}_{2}\text{S} + 4\ \text{C}\text{O}_{2} \rightarrow 2\text{ Fe}^{2 +} + 2\chi^{\beta}\ \text{H}_{2}\text{P}\text{O}_{4}^{-} + \text{S}_{0} + 4\ \text{HC}\text{O}_{3}^{-} + 2\text{ H}_{2}\text{O}}\\[1pt]
\hypertarget{pockmark-reaction:21}{R21} & \reactionline{\text{Fe}^{2 +} + \text{H}_{2}\text{S} \rightarrow \text{FeS} + 2\ \text{H}^{+}}\\[1pt]
\hypertarget{pockmark-reaction:22}{R22} & \reactionline{\text{FeS} + \text{H}_{2}\text{S} \rightarrow \text{Fe}\text{S}_{2} + \text{H}_{2}}\\[1pt]
\hypertarget{pockmark-reaction:23}{R23} & \reactionline{4\text{ S}_{0} + 4\ \text{H}_{2}\text{O} \rightarrow 3\text{ H}_{2}\text{S} + \text{S}\text{O}_{4}^{2 -} + 2\text{ H}^{+}}\\[1pt]
\hypertarget{pockmark-reaction:24}{R24} & \reactionline{\text{FeS} + \text{S}_{0} \rightarrow \text{Fe}\text{S}_{2}}\\[1pt]
\hypertarget{pockmark-reaction:25}{R25} & \reactionline{\text{S}\text{O}_{4}^{2 -} + \text{C}\text{H}_{4} + \text{C}\text{O}_{2} \rightarrow 2\ \text{HC}\text{O}_{3}^{-} + \text{H}_{2}\text{S}}\\[1pt]
\hypertarget{pockmark-reaction:26}{R26} & \reactionline{\text{C}\text{H}_{4} + 8\ \text{Fe}(\text{OH})_{3}^{\alpha,\beta} + 8\chi^{\alpha,\beta}\ \text{Fe}_{ox}\text{P} + 15\text{ H}^{+} \rightarrow \text{HC}\text{O}_{3}^{-} + 8\text{ Fe}^{2 +} + 8{\ \chi}^{\alpha,\beta}\text{H}_{2}\text{P}\text{O}_{4}^{-} + 21\text{ H}_{2}\text{O}}\\[1pt]
\hypertarget{pockmark-reaction:27}{R27} & \reactionline{\text{Fe}(\text{OH})_{3}^{\alpha} + (\chi^{\alpha} - \chi^{\beta})\ \text{Fe}_{ox}\text{P} \rightarrow \text{Fe}(\text{OH})_{3}^{\beta} + (\chi^{\alpha} - \chi^{\beta})\text{ H}_{2}\text{P}\text{O}_{4}^{-}}\\[1pt]
\hypertarget{pockmark-reaction:28}{R28} & \reactionline{\text{Fe}(\text{OH})_{3}^{\beta} + (\chi^{\beta} - \chi^{\gamma})\ \text{Fe}_{ox}\text{P} \rightarrow \text{Fe}(\text{OH})_{3}^{\gamma} + (\chi^{\beta} - \chi^{\gamma})\ \text{H}_{2}\text{P}\text{O}_{4}^{-}}\\[1pt]
\hypertarget{pockmark-reaction:29}{R29} & \reactionline{\text{Mn}\text{O}_{2}^{\alpha} \rightarrow \text{Mn}\text{O}_{2}^{\beta}}\\[1pt]
\hypertarget{pockmark-reaction:30}{R30} & \reactionline{\text{Mn}\text{O}_{2}^{\beta} \rightarrow \text{Mn}\text{O}_{2}^{\gamma}}\\[1pt]
\hypertarget{pockmark-reaction:31}{R31} & \reactionline{3\ \text{Fe}^{2 +} + 2\ \text{HP}\text{O}_{4}^{2 -} \rightarrow \text{Fe}_{3}(\text{P}\text{O}_{4})_{2} + 2\ \text{H}^{+}}\\[1pt]
\hypertarget{pockmark-reaction:32}{R32} & \reactionline{\text{Fe}^{2 +} + \text{C}\text{O}_{3}^{2 -} \rightarrow \text{FeC}\text{O}_{3}}\\[1pt]
\hypertarget{pockmark-reaction:33}{R33} & \reactionline{\text{FeC}\text{O}_{3} + \text{H}_{2}\text{S} \rightarrow \text{FeS} + \text{HC}\text{O}_{3}^{-} + \text{H}^{+}}\\[1pt]
\hypertarget{pockmark-reaction:34}{R34} & \reactionline{\text{Fe}_{3}(\text{P}\text{O}_{4})_{2} + 3\ \text{H}_{2}\text{S} \rightarrow 3\ \text{FeS} + 2\ \text{HP}\text{O}_{4}^{2 -} + 4\ \text{H}^{+}}\\[1pt]
\bottomrule\end{tabular}
\par\vspace{3pt}{\fontsize{8.5}{10}\selectfont\noindent\emph{Notes.} \(\alpha\), \(\beta\) and \(\gamma\) identify highly reactive, less reactive and nonreactive pools. For C:N:P \(= 106:16:1\), \(a = 1\), \(b = \frac{16}{106}\) and \(c = \frac{1}{106}\). The pool-specific P:Fe ratios (\(\chi^{\alpha\ },\ \chi^{\beta\ }\)and \(\chi^{\gamma\ }\)) are 0.1, 0.055 and 0.03, respectively.\par}
\endgroup
\end{table}

The secondary biogeochemical reactions included mineral precipitation/dissolution, reoxidation of reduced metabolites, methane oxidation, and other homogeneous and heterogeneous reactions. Similar to \citet{Purkamo2022}, these reactions were represented by either bimolecular or first-order kinetic rates (Table S8). The simulations were performed with the native reaction package (\emph{multiKinetics}), for a period of 50 years, and a summary of the input parameters including boundary conditions used in the simulations is given in the Supplementary Material (Section S1.2).

In the second application example, the model was used to simulate the sediment early- diagenetic processes in the SBB organic-rich mud. The model accounts for diffusive transport of dissolved species as well as advection of solutes and solids due to burial in a 1-D domain with spatially variable porosity. Bioturbation was included in the model using an exponential profile that decays from the sediment-bottom water interface. The biogeochemical reaction network was inspired from \citet{Du2023} and consisted of (i) organic matter degradation with a sequence of terminal electron acceptors, (ii) oxidation of products from organic matter degradation, (iii) adsorption of freshly produced mineralization products onto the sediment matrix, and (iv) speciation of major and trace elements (see Table~\ref{tab:4} for the reaction network).

The organic matter reactivity was represented by considering a single pool with first-order degradation rate, in which the rate coefficient is time-dependent and gradually decreases with the sediment age (\(k_{OM} = \upsilon/(a_{0} + \tau_{age})\), where \(\tau_{age}\) is the sediment age, and \(\upsilon\) and \(a_{0}\) are constants). For the trace element, we specifically focused on Molybdenum (Mo) cycling and considered both dissolved and solid-phase Mo. The model incorporates the speciation of dissolved Mo into molybdate and thiomolybdate species and represents the transfer of the dissolved Mo to the solid phase by a first-order kinetic rate. We also included the two Mo isotopes (\textsuperscript{98}Mo and \textsuperscript{95}Mo) in the model and represented equilibrium isotope fractionation during thiomolybdate formation. To this end, the fractionation factors are incorporated into the four equilibrium constants associated with thiomolybdate formation (more details in Table S15). The objective was to examine how sedimentary redox conditions and sulfide availability control Mo speciation and its isotopic composition in the sediment. The simulation was performed using the PHREEQC coupling package (\emph{phreeqcRM}). The standard PHREEQC thermodynamic database (\emph{phreeqc.dat}), including speciation reactions for common aqueous components, was used as the basis and customized to include the specific reaction network considered in the model. The simulation was performed for 200 years, and the resulting pseudo steady-state profiles were then compared with the measured spatial profiles of solute and solid constituents provided by \citet{Reimers1996} and \citet{Zheng2000}. The model input parameters including rate coefficients and boundary conditions are provided in the Supplementary Material (Section S1.3).

\begin{table}[p]
\caption{Overview of the biogeochemical reaction network used in the simulation of the Santa Barba Basin, California.}\label{tab:4}
\begingroup
\linespread{1}\fontsize{9.5}{11}\selectfont
\renewcommand{\arraystretch}{1.0}\setlength{\tabcolsep}{3pt}

\begin{tabular}{@{}>{\raggedright\arraybackslash}m{.065\linewidth}>{\raggedright\arraybackslash}m{\dimexpr.935\linewidth-2\tabcolsep\relax}@{}}
\toprule \textbf{ID} & \textbf{Reaction}\\\midrule
\multicolumn{2}{@{}l}{\textbf{Primary redox reactions}}\\*[2pt]
\hypertarget{sbb-reaction:1}{R1} & \reactionline{\text{OM} + a\ \text{O}_{2} + (b - c)\text{H}^{+} \rightarrow a\ \text{C}\text{O}_{2} + b\ \text{N}\text{H}_{4}^{+} + c\ \text{H}_{2}\text{P}\text{O}_{4}^{-}+ a\text{ H}_{2}\text{O}}\\[1pt]
\hypertarget{sbb-reaction:2}{R2} & \reactionline{\text{OM} + 2a\ \text{N}\text{O}_{3}^{-} + (b - c)\text{H}^{+} \rightarrow a\ \text{C}\text{O}_{2} + b\ \text{N}\text{H}_{4}^{+} + c\ \text{H}_{2}\text{P}\text{O}_{4}^{-} + 2a\ \text{N}\text{O}_{2}^{-}+ a\text{ H}_{2}\text{O}}\\[1pt]
\hypertarget{sbb-reaction:3}{R3} & \reactionline{\text{OM} + \frac{4a}{3}\text{N}\text{O}_{2}^{-} + (\frac{4a}{3} + b - c)\text{H}^{+} \rightarrow a\ \text{C}\text{O}_{2} + b\ \text{N}\text{H}_{4}^{+} + c\ \text{H}_{2}\text{P}\text{O}_{4}^{-} + \frac{2a}{3}\text{N}_{2}+\frac{5a}{3}\text{ H}_{2}\text{O}}\\[1pt]
\hypertarget{sbb-reaction:4}{R4} & \reactionline{\text{OM} + 2a\ \text{Mn}\text{O}_{2} + (4a + b - c)\text{H}^{+} \rightarrow a\ \text{C}\text{O}_{2} + b\ \text{N}\text{H}_{4}^{+} + c\ \text{H}_{2}\text{P}\text{O}_{4}^{-} + 2a\ \text{Mn}^{2 +}+ 3\text{a H}_{2}\text{O}}\\[1pt]
\hypertarget{sbb-reaction:5}{R5} & \reactionline{\text{OM} + 4a\ \text{FeOOH} + (8a + b - c)\ \text{H}^{+} \rightarrow a\ \text{C}\text{O}_{2} + b\ \text{N}\text{H}_{4}^{+} + c\ \text{H}_{2}\text{P}\text{O}_{4}^{-} + 4a\ \text{Fe}^{2 +} + \ 3{\ H}_{2}O}\\[1pt]
\hypertarget{sbb-reaction:6}{R6} & \reactionline{\text{OM} + \frac{a}{2}\text{ S}\text{O}_{4}^{2 -} + (a + b - c)\ \text{H}^{+} \rightarrow a\ \text{C}\text{O}_{2} + b\ \text{N}\text{H}_{4}^{+} + c\ \text{H}_{2}\text{P}\text{O}_{4}^{-} + \frac{a}{2}\text{ H}_{2}\text{S}+ 7a\text{ H}_{2}\text{O}}\\[1pt]
\hypertarget{sbb-reaction:7}{R7} & \reactionline{\text{OM} + (b - c)\ \text{H}^{+} \rightarrow \frac{a}{2}\text{ C}\text{O}_{2} + \frac{a}{2}\text{ C}\text{H}_{4} + b\ \text{N}\text{H}_{4}^{+} + c\ \text{H}_{2}\text{P}\text{O}_{4}^{-}}\\[1pt]
\multicolumn{2}{@{}l}{\textbf{Secondary reactions}}\\*[2pt]
\hypertarget{sbb-reaction:8}{R8} & \reactionline{\text{N}\text{O}_{2}^{-} + 0.5\ \text{O}_{2} \rightarrow \text{N}\text{O}_{3}^{-}}\\[1pt]
\hypertarget{sbb-reaction:9}{R9} & \reactionline{\text{N}\text{H}_{4}^{+} + 1.5\ \text{O}_{2} \rightarrow \text{N}\text{O}_{2}^{-} + 2\ \text{H}^{+}}\\[1pt]
\hypertarget{sbb-reaction:10}{R10} & \reactionline{\text{Mn}^{2 +} + 0.5\ \text{O}_{2} + \ \text{H}_{2}\text{O} \rightarrow \text{Mn}\text{O}_{2} + 2\ \text{H}^{+}}\\[1pt]
\hypertarget{sbb-reaction:11}{R11} & \reactionline{\text{Fe}^{2 +} + 0.25\ \text{O}_{2} + \ \text{1.5 H}_{2}\text{O} \rightarrow \text{FeOOH} + 2\text{ H}^{+}}\\[1pt]
\hypertarget{sbb-reaction:12}{R12} & \reactionline{\text{H}_{2}\text{S} + 2\ \text{O}_{2} \rightarrow \text{S}\text{O}_{4}^{2 -} + 2\ \text{H}^{+}}\\[1pt]
\hypertarget{sbb-reaction:13}{R13} & \reactionline{\text{FeS} + 2.25\text{ O}_{2} + \ 1.5\ \text{H}_{2}\text{O} \rightarrow \text{FeOOH} + \text{S}\text{O}_{4}^{2 -} + 2\text{ H}^{+}}\\[1pt]
\hypertarget{sbb-reaction:14}{R14} & \reactionline{\text{C}\text{H}_{4} + 2\text{O}_{2} \rightarrow \text{C}\text{O}_{2} + \ \text{2 H}_{2}\text{O}}\\[1pt]
\hypertarget{sbb-reaction:15}{R15} & \reactionline{\text{N}\text{O}_{2}^{-} + \text{N}\text{H}_{4}^{+} \rightarrow \text{N}_{2} + \ \text{2 H}_{2}\text{O}}\\[1pt]
\hypertarget{sbb-reaction:16}{R16} & \reactionline{0.4\ \text{N}\text{O}_{3}^{-} + \text{Mn}^{2 +} + \ \text{0.8 H}_{2}\text{O} \rightarrow \text{Mn}\text{O}_{2} + 0.2\text{ N}_{2} + 1.6\text{ H}^{+}}\\[1pt]
\hypertarget{sbb-reaction:17}{R17} & \reactionline{0.2\ \text{N}\text{O}_{3}^{-} + \text{Fe}^{2 +} + \ \text{1.4 H}_{2}\text{O} \rightarrow \text{FeOOH} + 0.1\ \text{N}_{2} + 1.8\ \text{H}^{+}}\\[1pt]
\hypertarget{sbb-reaction:18}{R18} & \reactionline{\text{S}\text{O}_{4}^{2 -} + \text{C}\text{H}_{4} + 2\text{ H}^{+} \rightarrow \text{C}\text{O}_{2} + \text{H}_{2}\text{S} + \ \text{2 H}_{2}\text{O}}\\[1pt]
\hypertarget{sbb-reaction:19}{R19} & \reactionline{0.5\ \text{Mn}\text{O}_{2} + \text{Fe}^{2 +} + \ \text{H}_{2}\text{O} \rightarrow 0.5\ \text{Mn}^{2 +} + \text{FeOOH} + \text{H}^{+}}\\[1pt]
\hypertarget{sbb-reaction:20}{R20} & \reactionline{\text{Mn}\text{O}_{2} + \text{H}_{2}\text{S} + 2\text{ H}^{+} \rightarrow \text{Mn}^{2 +} + \text{S}^{0} + \ \text{2 H}_{2}\text{O}}\\[1pt]
\hypertarget{sbb-reaction:21}{R21} & \reactionline{2\ \text{FeOOH} + \text{H}_{2}\text{S} + 4\ \text{H}^{+} \rightarrow 2\text{ Fe}^{2 +} + \text{S}^{0} + \ \text{4 H}_{2}\text{O}}\\[1pt]
\hypertarget{sbb-reaction:22}{R22} & \reactionline{\text{FeS} + \text{H}_{2}\text{S} \rightarrow \text{Fe}\text{S}_{2} + \text{H}_{2}}\\[1pt]
\hypertarget{sbb-reaction:23}{R23} & \reactionline{\text{FeS} + \text{H}^{+} \rightarrow \text{Fe}^{2 +} + \text{H}\text{S}^{-}}\\[1pt]
\hypertarget{sbb-reaction:24}{R24} & \reactionline{\text{Fe}^{2 +} + \text{H}\text{S}^{-} \rightarrow \text{FeS} + \text{H}^{+}}\\[1pt]
\multicolumn{2}{@{}l}{\textbf{Mineral and Mo reactions}}\\*[2pt]
\hypertarget{sbb-reaction:25}{R25} & \reactionline{\text{CaC}\text{O}_{3} \rightleftharpoons \text{Ca}^{2 +} + \text{C}\text{O}_{3}^{2 -}}\\[1pt]
\hypertarget{sbb-reaction:26}{R26} & \reactionline{\text{MnC}\text{O}_{3} \rightleftharpoons \text{Mn}^{2 +} + \text{C}\text{O}_{3}^{2 -}}\\[1pt]
\hypertarget{sbb-reaction:27}{R27} & \reactionline{\text{FeC}\text{O}_{3} \rightleftharpoons \text{Fe}^{2 +} + \text{C}\text{O}_{3}^{2 -}}\\[1pt]
\hypertarget{sbb-reaction:28}{R28} & \reactionline{\text{BSi} \rightarrow \text{H}_{4}\text{Si}\text{O}_{4}}\\[1pt]
\hypertarget{sbb-reaction:29}{R29} & \reactionline{\text{H}_{4}\text{Si}\text{O}_{4} \rightarrow \text{ASi}}\\[1pt]
\hypertarget{sbb-reaction:30}{R30} & \reactionline{9.54\text{Ca}^{2 +} + 0.33\text{Na}^{+} + 0.13\text{Mg}^{2 +} + 4.8\text{P}\text{O}_{4}^{3 -} + 1.2\text{C}\text{O}_{3}^{2 -} + 2.48\text{F}^{-} + 0.39\text{O}\text{H}^{-} \rightarrow \text{CFA}}\\[1pt]
\hypertarget{sbb-reaction:31}{R31} & \reactionline{^{i}\text{Mo}_{\text{thio}} \rightarrow^{i}\text{Mo}_{\text{auth}}\ (i = 95,98)}\\[1pt]
\multicolumn{2}{@{}l}{\textbf{Surface adsorption reactions}}\\*[2pt]
\hypertarget{sbb-reaction:32}{R32} & \reactionline{X_{N}\text{OH} + \text{N}\text{H}_{4}^{+} \rightleftharpoons X_{N}\text{OHN}\text{H}_{4}^{+}}\\[1pt]
\hypertarget{sbb-reaction:33}{R33} & \reactionline{X_{M}\text{OH} + \text{Mn}^{2 +} \rightleftharpoons X_{M}\text{OHM}\text{n}^{2 +}}\\[1pt]
\hypertarget{sbb-reaction:34}{R34} & \reactionline{X_{F}\text{OH} + \text{Fe}^{2 +} \rightleftharpoons X_{F}\text{OHF}\text{e}^{2 +}}\\[1pt]
\bottomrule\end{tabular}
\par\vspace{3pt}{\fontsize{8.5}{10}\selectfont\noindent\emph{Notes.} \(a = 1\), \(b = 3/20\) and \(c = 3/200\) for the single OM pool. BSi and ASi denote biogenic and authigenic silica. CFA is \(\text{Ca}_{9.54}\text{Na}_{0.33}\text{Mg}_{0.13}\text{P}_{4.8}\text{C}_{1.2}\text{F}_{2.48}\text{H}_{0.39}\text{O}_{23.19}\). In R31, \(\text{Mo}_{thio}\) denotes the four thiomolybdates.\par}
\endgroup
\end{table}

\FloatBarrier
\section{Results and discussion}\label{sec:4}

\subsection{Conservative transport of solute and solid species (Benchmark 1)}\label{sec:4.1}

\begin{figure}[htbp]
\centering
\includegraphics[width=\linewidth,height=.70\textheight,keepaspectratio]{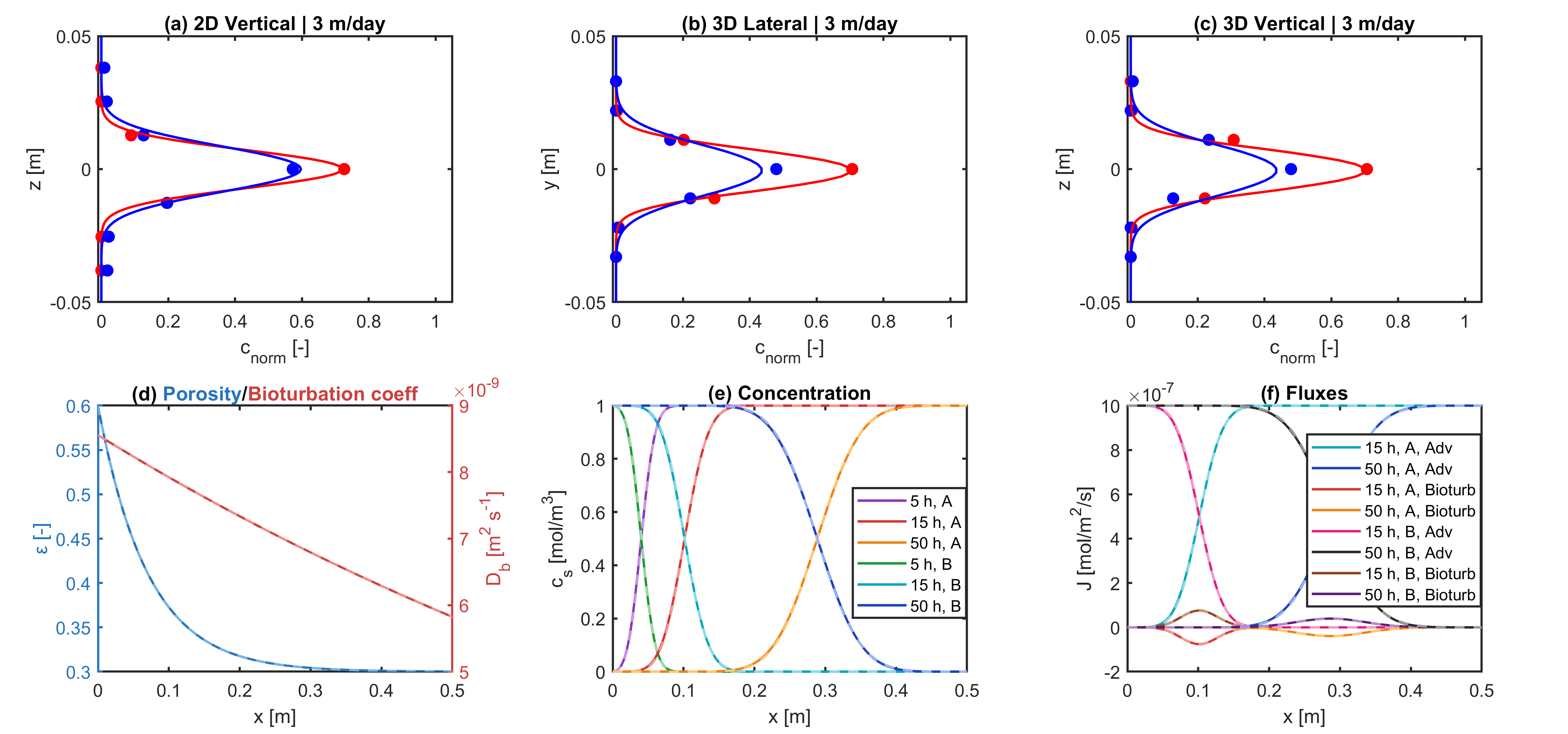}
\caption{Measured (markers) versus simulated (solid lines) spatial profiles of fluorescein and oxygen at the outlet of the quasi 2-D setup (a) and the 3-D system along the horizontal (b) and vertical (c) directions for the seepage velocity of 3 m/d (Benchmark 1A). Simulated spatial profiles of porosity and bioturbation coefficient (d), concentration of solid species A and B at t=5, 15 and 50 h (e) and advective/bioturbation fluxes of species A and B at t=15 and 50 h (f) in a 1-D porous medium domain (Benchmark 1B). In panels d-f, the simulations were performed by marineGeochemFoam (solid lines) and compared with COMSOL Multiphysics (dashed lines).}\label{fig:2}
\end{figure}

In Benchmark 1A, the advective-dispersive transport of two tracers was simulated in quasi 2-D and 3-D porous media. The simulations were performed with different velocities (3, 5 and 8 m/d) and Figure~\ref{fig:2}a-c shows a comparison between the simulated fluorescein and oxygen profiles and the experimental data for the case of 3 m/d. The results at other flow velocities as well as the comparison against an analytical solution for all cases are shown in the Supplementary Material (Figure S1 and S2).

In both experimental setups (2-D and 3-D), the measured outlet concentrations of the two tracers show quasi-Gaussian distributions with distinct peak values (i.e., consistently higher peaks for fluorescein than oxygen). The observed difference in peak concentrations is more evident in the 3-D porous media systems and reflects the compound-specific transport behavior of the tracers, arising from their different transverse dispersion and molecular diffusion coefficients (see Table S1). Such enhanced compound-specific behavior in 3-D is triggered by system dimensionality because tracers can spread in both horizontal and vertical directions, leading to larger differences between the spatial profiles in the 3-D compared to the 2-D systems.

The observed transport behavior for oxygen and fluorescein is captured very well by \emph{marineGeochemFoam} by using the non-linear and compound-specific transverse dispersion parameterization \citep{BearBachmat1967,Chiogna2010} (Table~\ref{tab:1}). Additionally, there is a perfect agreement between the results of \emph{marineGeochemFoam} and the analytical solution (Figure S2), verifying the correct implementation of solute transport as well as dispersion formulations in the code.

Benchmark 1B verifies the implementation of the solid transport equation (Eq.~\eqref{eq:2}) considering conservative transport of two solid species in a 1-D domain (0.5 m length) with variable porosity (Figure~\ref{fig:2}d). Solid species A is initially present in the domain with a concentration of 1 mol/m\textsuperscript{3}, whereas the initial concentration of B is zero. Species B is injected into the system with a fixed concentration of 1 mol/m\textsuperscript{3} from the left boundary, but species A is supplied with a concentration of 0 mol/m\textsuperscript{3} at the same boundary. This results in the gradual advancement of the front of species B and displacement of species A within the domain through advection and bioturbation. The code computes the evolution of concentrations and fluxes in the domain (Figure~\ref{fig:2}e and d). The outcomes of this simulation were compared against COMSOL Multiphysics, and there is an excellent agreement, suggesting that \emph{marineGeochemFoam} correctly (i) simulates conservative transport of multiple solid species via advective and bioturbation mechanisms, and (ii) captures the influence of spatially variable porosity, transport coefficients, and flux components on transport behavior.

\subsection{Reactive systems with multiple solute and solid species (Benchmark 2)}\label{sec:4.2}

\begin{figure}[htbp]
\centering
\includegraphics[width=\linewidth,height=.70\textheight,keepaspectratio]{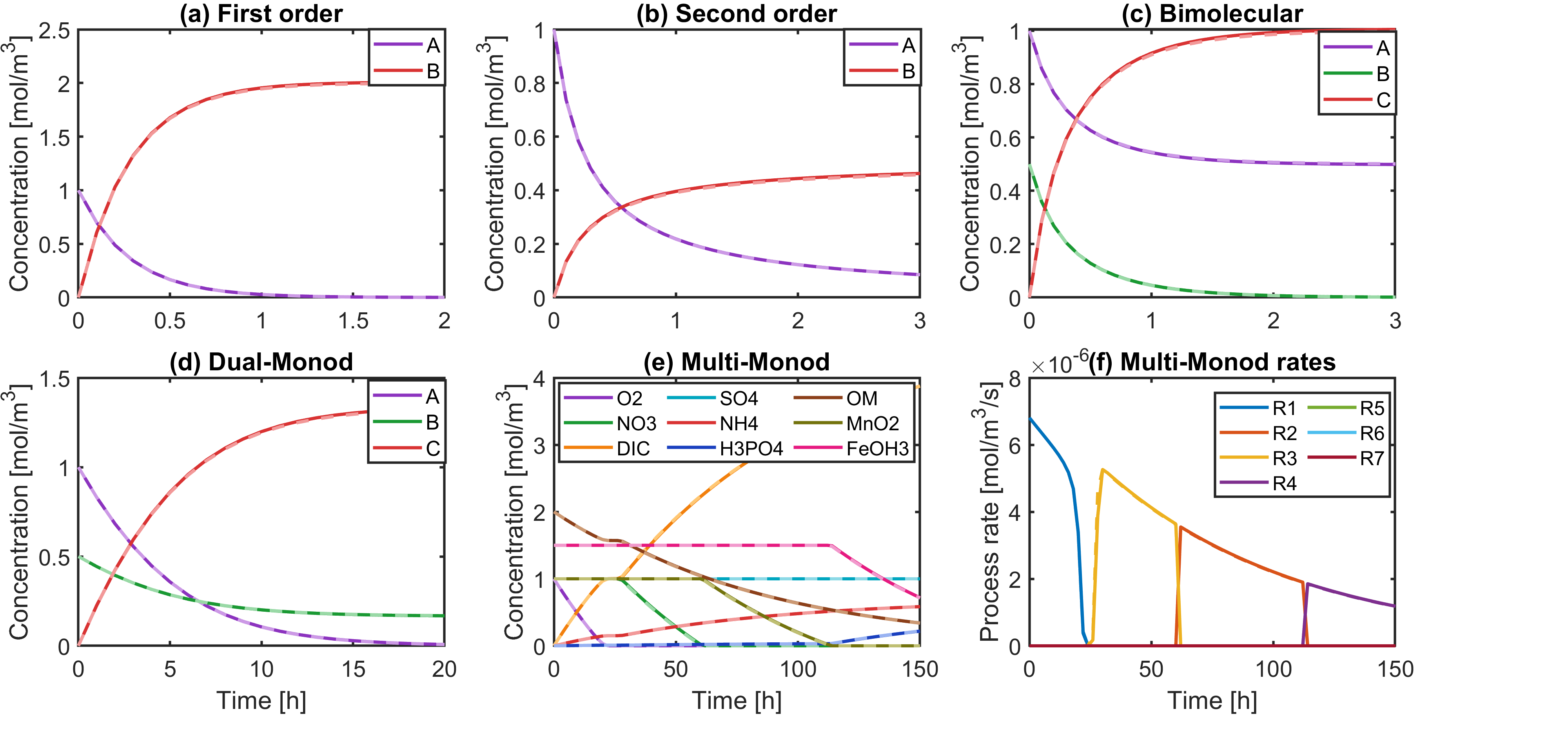}
\caption{Simulated temporal evolution of species' concentration involved in first-order kinetic reaction (a), quadratic second-order kinetic reaction (b), bimolecular kinetic reaction (c), a Monod type of reaction (d), coupled network of reactions representing early diagenetic processes (e), and reaction rates of species involved in early diagenetic processes (f). The solid lines show the marineGeochemFoam results, whereas dashed lines show the results from analytical solutions (first and second order reactions) or MATLAB ode solver ode45 (dual- and multi-Monod type of reactions).}\label{fig:3}
\end{figure}

Benchmark 2 focuses on the validation of the native reaction package, \emph{multiKinetics} by considering a series of batch systems. Specifically, four representative reactions involving various rate laws were considered: (a) a first-order reaction (A → 2B); (b) two second-order reactions including a quadratic second order reaction (2A → B), and a bimolecular reaction (A + B → 2C); (c) a Monod-type reaction (1.5A + 0.5B → 2C); and (d) a network of sequential redox reactions with multi-Monod rate laws. The reactions in the last example are representative of organic matter degradation by terminal electron acceptors in early diagenesis (R1-R7 in Table S2). The results obtained from \emph{marineGeochemFoam} are compared against analytical solutions (first and second order) and outcomes from an ODE solver (dual- and multi-Monod kinetics). In all cases, the \emph{marineGeochemFoam} results perfectly match the analytical solutions or ODE solver outcomes, demonstrating the accuracy and robustness of the reaction package implementation across a range of kinetic formulations and coupled biogeochemical processes (Figure~\ref{fig:3}).

\subsection{Multi-species reactive transport (Benchmarks 3 and 4)}\label{sec:4.3}

\begin{figure}[htbp]
\centering
\includegraphics[width=\linewidth,height=.70\textheight,keepaspectratio]{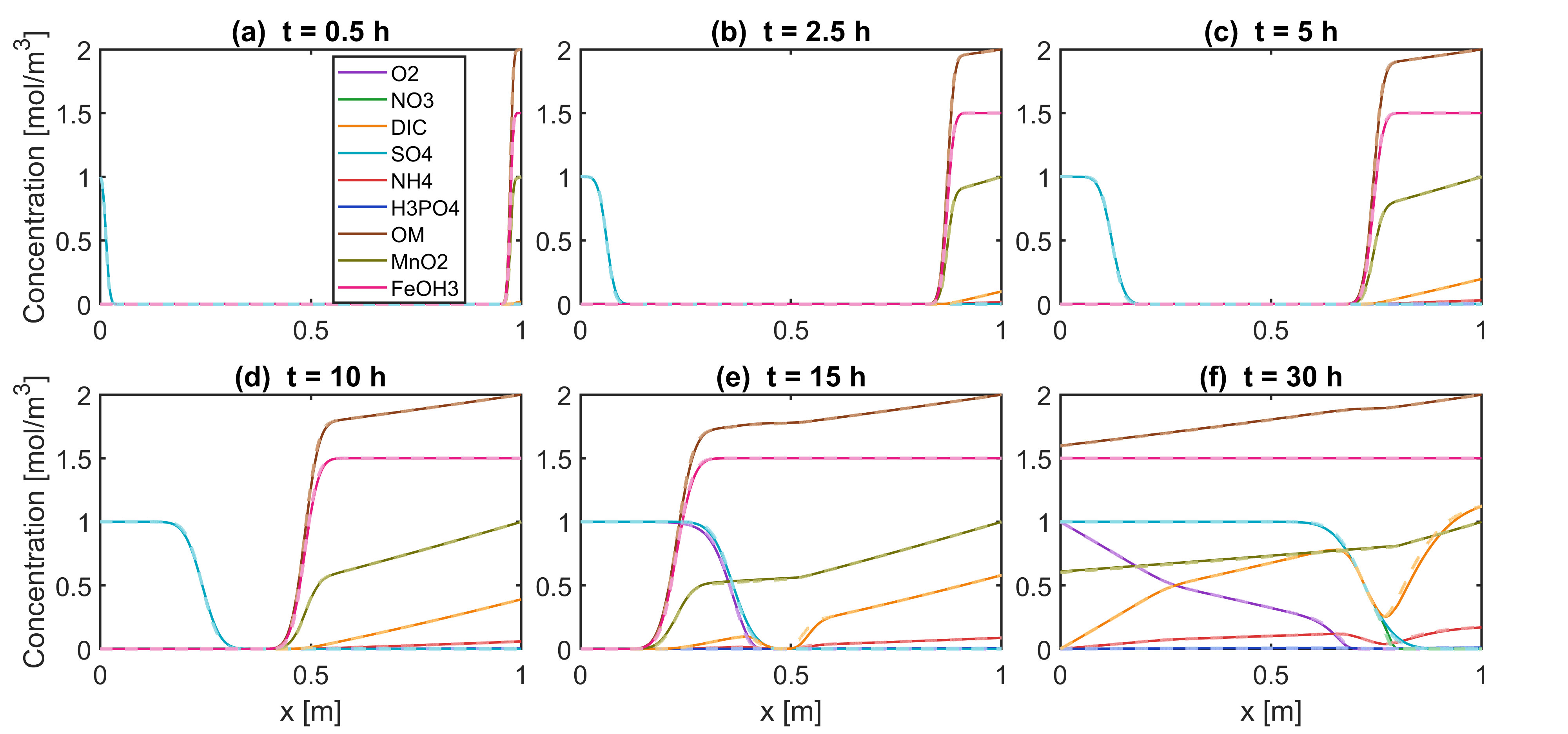}
\caption{Simulated solute and solid species profiles at t=0.5, 2.5, 5, 10, 15 and 30 hours (a-f) using marineGeochemFoam (solid lines) and COMSOL Multiphysics (dashed lines). Solutes and solids enter from the left and right boundary, respectively. Note that the NO\textsubscript{3} and SO\textsubscript{4} profiles mostly overlap in this example due to their similar reactive transport behavior.}\label{fig:4}
\end{figure}

Benchmark 3 and 4 are dedicated to verifying the implementation of coupled reactive transport in \emph{multiKinetics} and \emph{phreeqcRM} package, respectively. The reactive system considered for Benchmark 3 is the primary redox reactions of early diagenesis. In Benchmark 3A, transport of solutes (O\textsubscript{2}, NO\textsubscript{3}, SO\textsubscript{4}, NH\textsubscript{4}, DIC and H\textsubscript{3}PO\textsubscript{4}) was considered, but solids (OM, MnO\textsubscript{2} and Fe(OH)\textsubscript{3} were assumed to be immobile. Three solute species (O\textsubscript{2}, NO\textsubscript{3} and SO\textsubscript{4}) enter the domain from the left boundary, gradually advance and participate in the redox reactions with the initially present solids, resulting in the production of NH\textsubscript{4}, DIC and H\textsubscript{3}PO\textsubscript{4}. The outcome of this simulation is provided in the Supplementary Material (Figure S3). Benchmark 3B extends this example by including coupled reactive transport of both solute and solid species. Initially, the concentration of the solutes and solids is zero. The solutes enter from the left boundary, while solid species are supplied from the right boundary. As these species propagate in the opposite direction, different reaction zones develop in the domain, which highlights a complex interplay between transport and reactive processes (Figure~\ref{fig:4}).

OM migrates into the domain from the right boundary via bioturbation and advection and is oxidized through multiple pathways depending on the locally available electron acceptors. On the right side of the domain, OM oxidation is initially driven by the reduction of solid species like MnO\textsubscript{2}. However, when the advancing solutes reach the solids, a second reaction zone appears that is primarily associated with the oxygen consumption and aerobic oxidation of OM (Figure~\ref{fig:4}e and f). Therefore, redox reactions occur simultaneously in different regions of the domain through distinct pathways and generate different concentration gradients and evolving reaction fronts. The development of such distinct reactive zones also influences the spatial distribution of reaction products like DIC, which is mainly controlled by the O\textsubscript{2} reactive front on the left side of the domain and by the MnO\textsubscript{2} reactive front on the right side. The excellent agreement between the numerical solutions obtained with the two simulators confirms the correct implementation of the full model complexity, including coupled reactive transport of solid and solute species in sediments.

\begin{figure}[htbp]
\centering
\includegraphics[width=\linewidth,height=.70\textheight,keepaspectratio]{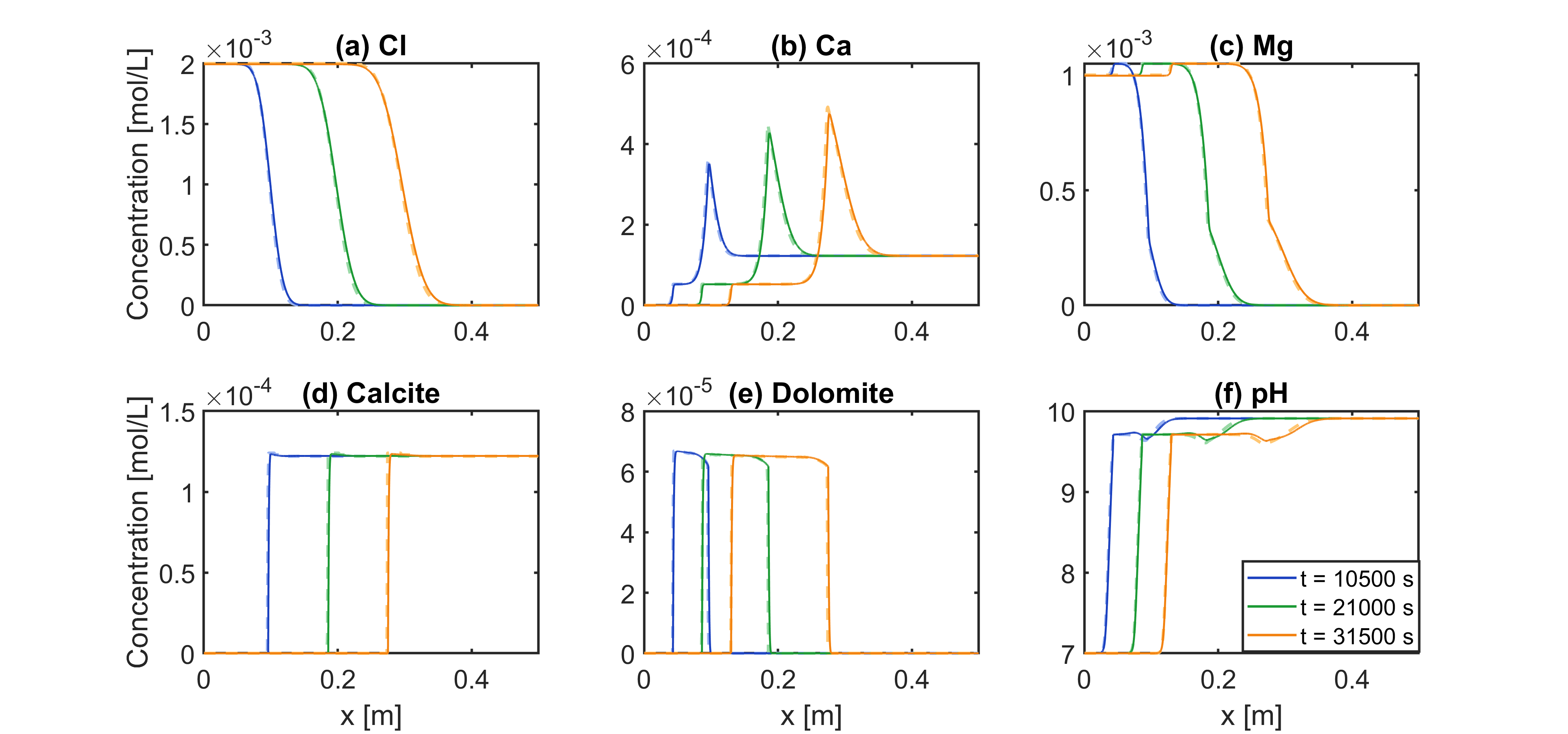}
\caption{Simulated spatial profiles of aqueous and mineral concentrations using marineGeochemFoam (solid lines) and PHREEQC (dashed lines) at t=10500, 21000 and 31500 seconds. In this example, marineGeochemFoam uses the PhreeqcRM coupling package.}\label{fig:5}
\end{figure}

Benchmark 4 evaluates the PHREECQ coupling implementation within the \emph{phreeqcRM} package in three examples. The results from two examples (Benchmark 4B and 4C) are shown below, whereas Benchmark 4A results are presented in the Supplementary Material (Figure S4). Benchmark 4B provides a rigorous test of the coupling framework by evaluating advective-dispersive transport, aqueous speciation, and mineral dissolution/precipitation reactions in a 1-D column. The system is initially in equilibrium with calcite and is flushed by a magnesium chloride solution that is undersaturated with respect to calcite. Consequently, calcite dissolves and dolomite precipitates, leading to a complex sequence of mineral dissolution/dissolution fronts. Figure~\ref{fig:5} shows the evolution of simulated spatial profiles of aqueous species and minerals.

As the injected MgCl\textsubscript{2} solution enters the column, a calcium- and carbonate-depleted front with circumneutral pH progressively displaces the initial alkaline pore water, which had relatively high concentration of Ca\textsuperscript{2+} and Mg\textsuperscript{2+} to maintain equilibrium with calcite. Consequently, calcite becomes undersaturated, promoting its dissolution and Ca\textsuperscript{2+} release. The incoming Mg\textsuperscript{2+} and the released Ca\textsuperscript{2+} subsequently create a supersaturation for dolomite, leading to its precipitation at the mixing zone. The dolomite precipitation front does not overlap with the calcite dissolution front and is not sustained in the upstream part of the domain, because dolomite formation primarily requires the co-existence of Ca\textsuperscript{2+} and Mg\textsuperscript{2+} at sufficiently high concentrations. The perfect agreement between the results obtained from PHREEQC and \emph{marineGeochemFoam} verifies the correct implementation of PHREEQC coupling.

\begin{figure}[htbp]
\centering
\includegraphics[width=\linewidth,height=.70\textheight,keepaspectratio]{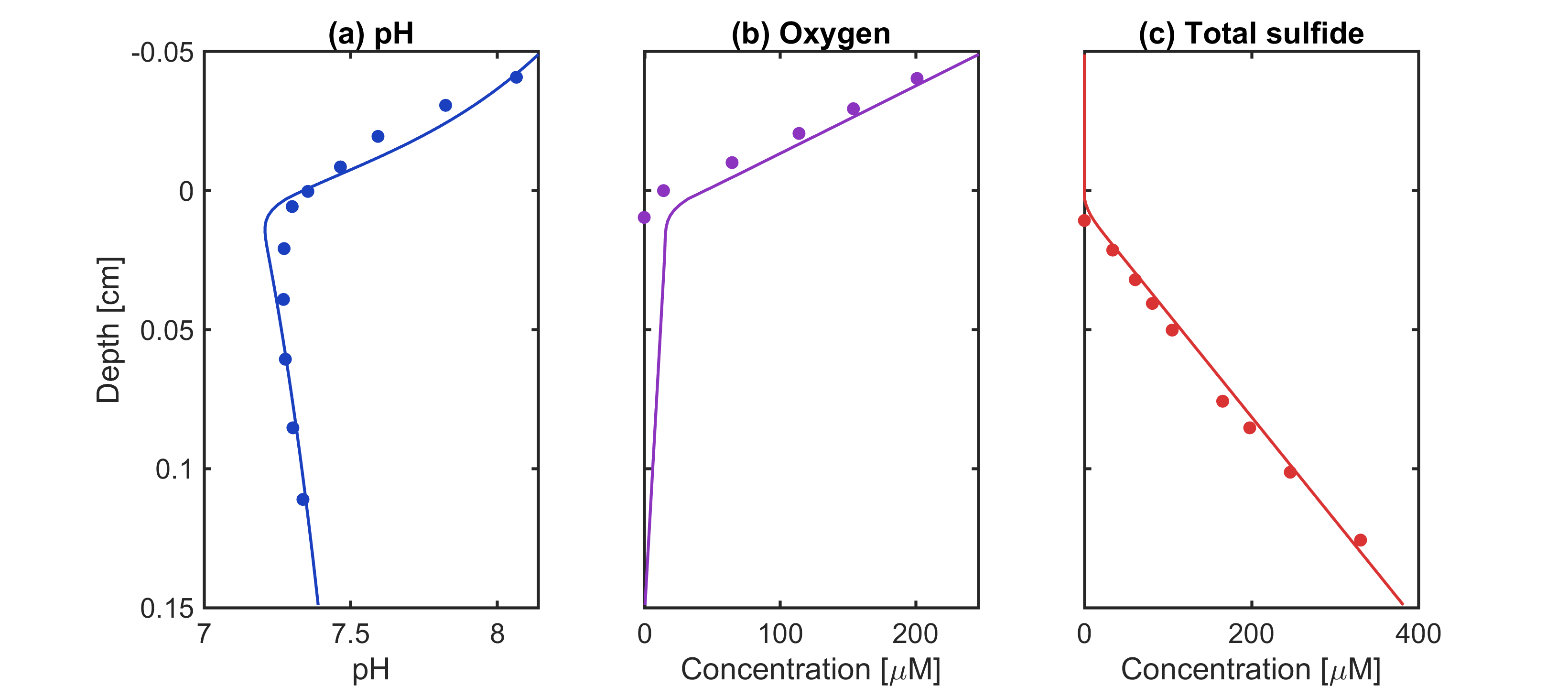}
\caption{Comparison between simulated (lines) and measured (markers) spatial profiles of pH, oxygen and sulfide after 1 hour in a sediment containing sulfide-oxidizing bacteria. In this example, marineGeochemFoam uses the PhreeqcRM coupling package.}\label{fig:6}
\end{figure}

Benchmark 4C further evaluates the PHREEQC coupling package in a problem that also involves redox reaction. The example considers a 1-D domain in which sulfide oxidation takes place in a sediment containing a mat of sulfide-oxidizing bacteria. Oxygen is uniformly distributed in the seawater above the mat and diffuses into the sediment, while hydrogen sulfide is initially present in the sediment and diffuses toward the incoming oxygen front (Figure~\ref{fig:6}). The reaction takes place at the overlapping zone between the oxygen and sulfide fronts and produces acidity, leading to a pH drop at the reactive zone. The simulated profiles overall agree well with the measurements, which confirms the capability of the PHREEQC coupling package for simulating transport, kinetically-controlled reactions, speciation reactions and pH calculation in sediments.

\subsection{Application examples: Simulation of pore water and solid profiles in marine sediments}\label{sec:4.4}

\subsubsection{Hanko submarine groundwater discharge site, Finland}\label{sec:4.4.1}

\begin{figure}[htbp]
\centering
\includegraphics[width=\linewidth,height=.75\textheight,keepaspectratio]{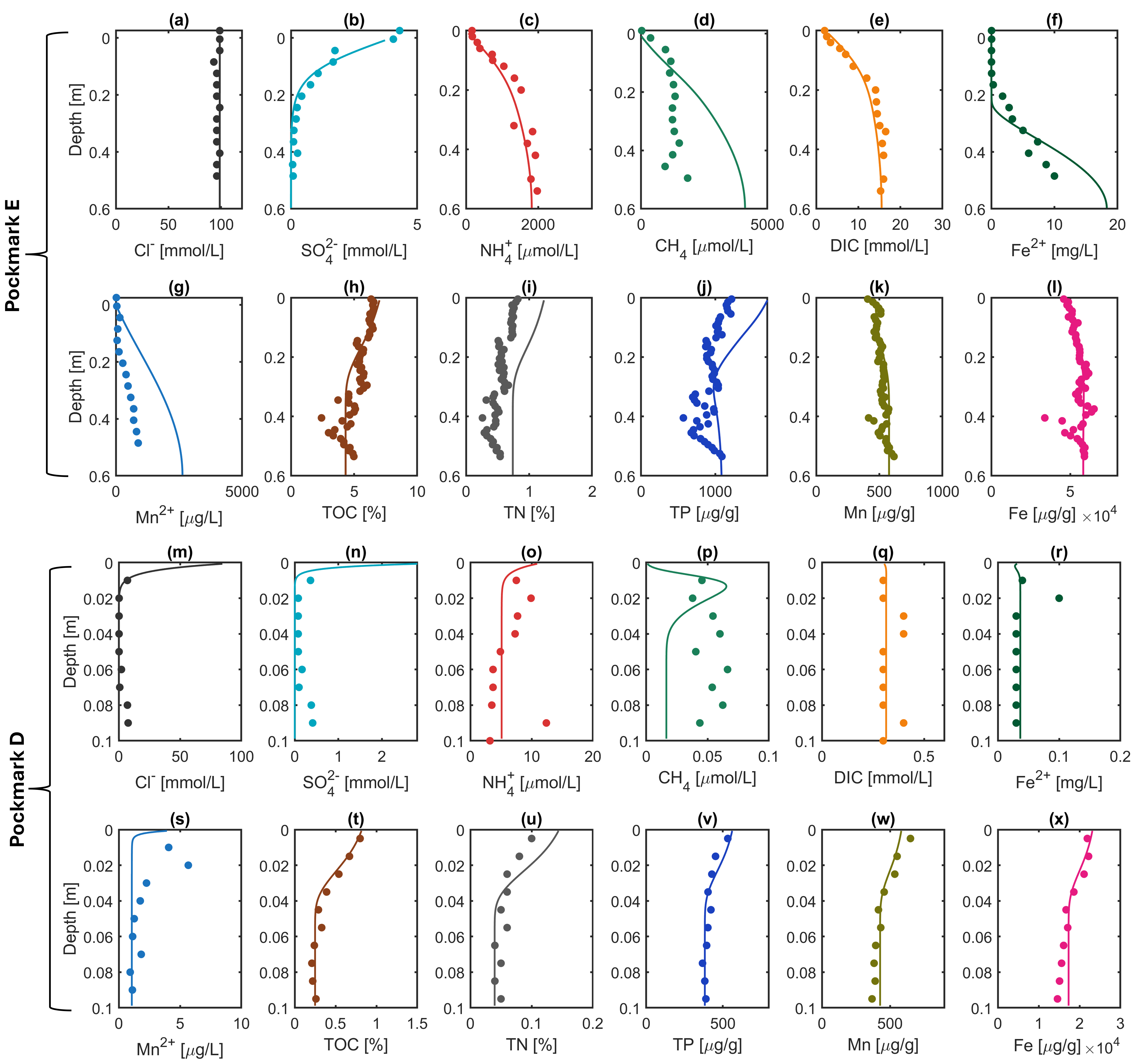}
\caption{Comparison between measured (symbols) and simulated (lines) spatial profiles of solute and solid species at t=50 years in pockmarks E (organic-rich mud) and D (fine sand) in the Hanko submarine groundwater discharge site, Finland, as presented in \citet{Purkamo2022}. The simulations were performed by the multiKineticsLinearized package.}\label{fig:7}
\end{figure}

Once different features of the code were benchmarked, it was applied to reproduce the observed behavior in selected pockmarks in the Hanko SGD site, Finland. For this purpose, we focused on pockmark E (without submarine groundwater discharge) and D (with strong submarine groundwater discharge) as presented in \citet{Purkamo2022}. We included the full reaction network of early diagenetic processes, as presented in their study (Table~\ref{tab:3}). The reactive transport simulation included 14 solute and 15 solid species, with multiple reactivity pools for organic matter and metal oxides. Figure~\ref{fig:7} presents the comparison between the measured and simulated solute and solid species within pockmark E and D after 50 years, whereas the simulated spatial profiles of all solute and solid species for the two pockmarks are given in the Supplementary Material (Figure S5 and S6).

Porewater Cl\textsuperscript{−} concentration in organic-rich mud from pockmark E shows a uniform profile along the core depth with values around 100 mmol/L (Figure~\ref{fig:7}a), which is within the range typically observed in the Baltic Sea water. As Cl\textsuperscript{−} is biogeochemically inert in this system, this pattern suggests that fresh groundwater discharge rate is negligible on pockmark E, and thus solute transport across the organic-rich mud is mainly controlled by diffusion. In contrast, sulfate, iron and methane concentrations show a strong gradient within the sediment, with sulfate concentrations being the highest near the sediment-water interface (i.e., oxic zone) and gradually decrease with depth as it is reduced during organic matter degradation (Figure~\ref{fig:7}b). Moreover, methane concentrations increase at greater depth where oxygen is not available and anaerobic oxidation of organic matter occurs, resulting in the production of methane and its diffusion toward the sulfate-bearing zone (Figure~\ref{fig:7}d). At the sulfate-methane transition zone, where upward-diffusing methane encounters downward-diffusing sulfate, sulfate-mediated anaerobic oxidation of methane occurs. Consequently, in addition to organic matter degradation, methane oxidation contributes to sulfate consumption. The spatial profiles also show the gradual accumulation of reaction products (Fe, NH\textsubscript{4} and DIC) along the depth with their concentrations reaching values as high as approximately 2000, 1500 and 15000 µmol/L, respectively at the deeper locations toward the anoxic zone (Figure~\ref{fig:7}c, e and f). Overall, the concentration profiles reflect strong competition for electron acceptors within the oxic-anoxic transition zone formed along the depth and the progression of early diagenetic processes that shape the biogeochemical structure of the sediment column in pockmark E. The model reproduces the observed spatial profiles reasonably well, with a mismatch for CH\textsubscript{4}, which could be associated with uncertainty in the measurements. In fact, outgassing of CH\textsubscript{4} from the sediment core prior to porewater sampling could have led to the underestimation of the real values below 5 cm as reported in \citet{Purkamo2022} and observed also in other studies \citep[e.g.,][]{Egger2017,Egger2016}\emph{.}

The sediment TOC, TN and TP profiles in pockmark E display a downward-propagating concentration front that penetrates to a depth of approximately 0.3 m, reflecting the combined effect of the sedimentation, burial and degradation processes (Figure~\ref{fig:7}h-i). In contrast, the profiles of Mn and Fe solid-phase species show a mild depletion front in the upper sediment layer due to their reduction during early diagenesis (Figure~\ref{fig:7}k-l). In fact, these species are consumed near the sediment-water interface where oxidizing conditions prevail. Overall, the model successfully reproduces the observed concentration fronts and general trends of solid-phase species in pockmark E. However, some discrepancies between the model and observations remain for TN and TP. These differences may be related to uncertainties in the assumed stoichiometry of organic matter degradation, which directly influences the release and accumulation of nitrogen and phosphorus during early diagenesis.

The comparison between the simulated and measured profiles in pockmark D demonstrates that the model could also reproduce the essential features of the solute and solid-phase profiles in this active pockmark that is strongly influenced by SGD. This is evident in chloride profile (Figure~\ref{fig:7}m), which shows a steep gradient and considerably low concentrations toward the bottom boundary, resulting from the significant influence of groundwater flow that drives the upward advective transport of low-salinity water through the fine sand. A similar trend can also be observed for the electron acceptors like sulfate that penetrate to a much lower depth (\textasciitilde1 cm) in pockmark D compared to pockmark E (\textasciitilde25 cm) due to the upward advective transport (Figure~\ref{fig:7}n). Such contrasting transport regimes have significant impact on the reaction fronts because sulfate-methane transition zone is primarily shifted close to the sediment-water interface in pockmark D. Moreover, the groundwater flow limits the accumulation of reaction products in the sediment and shifts them toward the oxidation zone in the uppermost part of the sediment where they can be further consumed by secondary reactions. This can explain the lower concentration of CH\textsubscript{4}, NH\textsubscript{4} and DIC in pockmark D, reaching as low as 0.05, 5 and 300 µmol/L at the bottom of the sediment core, respectively (Figure~\ref{fig:7}o, p and q). The spatial profiles of the solid-phase species, particularly TOC and TN, also show lower concentrations in pockmark D (Figure~\ref{fig:7}t and u). The solids are mildly depleted with depth and move downwards, but their penetration depth is much lower (\textasciitilde3 cm), which indicates the lower sedimentation rate in this pockmark.

These pockmark examples showcase the capability of \emph{marineGeochemFoam} to simulate transport of multiple solid and solute species via different mechanisms and to handle a comprehensive network of aqueous and solid-phase reactions. In particular, the model shows flexibility to simulate the contrasting biogeochemical behavior of pockmarks D and E by accounting for different transport regimes and degrees of groundwater influence. In addition to such versatile features, the code showed computational efficiency, which is particularly relevant for the reactive transport problems that involve a large reaction network and/or a relatively long simulation time frame. We tested different native packages for these examples and the \emph{multiKineticsLinearized} library with adaptive time stepping provided the most computationally efficient results. For example, the complete simulation could be performed in approximately 1 min without using MPI parallelization on a Dell desktop computer (Precision 3690) equipped with an Intel\textsuperscript{©} Core™ i9-14900K (3.2 GHz) processor and 128 GB of RAM, running Linux Mint 22.2 Cinnamon.

\subsubsection{Biogeochemical and elemental cycling in Santa Barbara basin sediments, California}\label{sec:4.4.2}

\begin{figure}[htbp]
\centering
\includegraphics[width=\linewidth,height=.70\textheight,keepaspectratio]{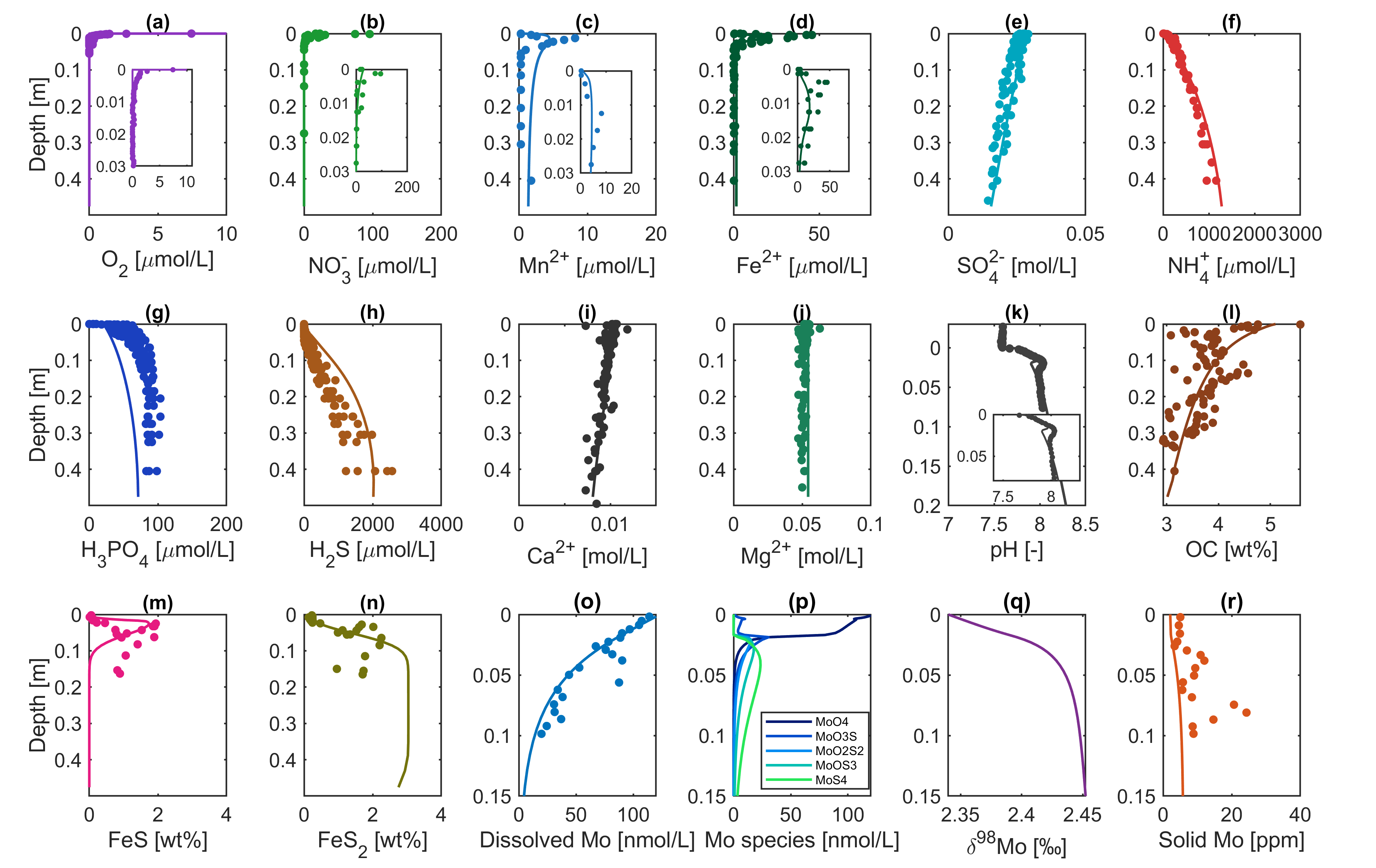}
\caption{Comparison between measured (symbols) and simulated (lines) spatial profiles of solute and solid species under pseudo steady-state condition in organic-rich mud in the Santa Barbara basin, California. Note that the PhreeqcRM coupling package was used for this simulation. The molybdenum isotope composition (δ\textsuperscript{98}Mo), expressed in ‰ relative to the NIST SRM 3134 reference standard, is calculated as δ\textsuperscript{98}Mo =(((\textsuperscript{98}Mo/\textsuperscript{95}Mo\textsubscript{sample})/ (\textsuperscript{98}Mo/\textsuperscript{95}Mo\textsubscript{standard}))-1) × 10\textsuperscript{3} + 0.25. The insets show enlarged views of the spatial profiles near the sediment--water interface.}\label{fig:8}
\end{figure}

This example focused on simulating biogeochemical and elemental (re-)cycling processes in the Santa Barbara basin (SBB) using the PHREEQC coupling package (\emph{phreeqcRM}). The simulation was based on a published comprehensive SBB dataset by \citet{Reimers1996}, which was ideal to test the flexibility of the code for simulating detailed biogeochemistry including speciation reactions, redox couples, and pH that are essential to describe major and trace element cycling in sediments. Figure~\ref{fig:8} shows the comparison between the simulated and measured vertical profiles of solute and solid species.

Overall, the model successfully captures the essential features of the measured profiles and explains the key transport and biogeochemical processes in the sediment. The results show that oxygen penetration is strongly limited and thus anoxic conditions are prevalent across the depth (Figure~\ref{fig:8}a). Such a reducing environment favors organic matter degradation through alternative electron acceptors, which leads to reduction of nitrate, iron, sulfate and manganese and corresponding production/accumulation of reaction products like ammonium, hydrogen sulfide and phosphate (Figure~\ref{fig:8}c-h). The release of byproducts from organic matter decomposition also promotes mineral reactions such as carbonate precipitation that can explain the consumption of calcium along the depth (Figure~\ref{fig:8}i). Similarly, the results show sulfide mineral formations (FeS and FeS\textsubscript{2}) as a result of iron and sulfide liberation during diagenetic processes. Such dynamics in porewater chemistry also affect the pH profile, which shows a gradual increase from the seawater toward the sediment column. This can be related to the reduction of manganese and iron hydroxides, which liberates iron and manganese and consumes protons. Such processes also promote carbonate precipitation, which can lead to pH buffering at deeper locations of the sediment. The local decrease in pH at around 2 cm below the water-sediment interface can be related to FeS precipitation that releases protons (see the inset of Figure~\ref{fig:8}k). The model accurately captures the observed trend of pH and confirms the capability of our code to simulate thermodynamically constrained geochemical reactions. Additionally, the model successfully captures the depletion and burial of organic carbon (Figure~\ref{fig:8}l).

The model also reproduces the measured dissolved Mo profile and calculates its speciation based on the sediment biogeochemistry and redox conditions (Figure~\ref{fig:8}o-r). The results show that under reducing conditions where sulfide concentration increases, the thiomolybdate species become more dominant (Figure~\ref{fig:8}p). The removal of Mo species from the dissolved phase was also included in the model and compared with the measured solid-phase Mo profile. The model predicts an increasing trend in solid Mo with depth but cannot reproduce its accumulation in the upper 10 cm, suggesting that additional Mo inputs, such as Mo supplied by settling particles, may contribute to the observed sedimentary Mo. Additionally, the model simulated two stable Mo isotopes (\textsuperscript{98}Mo and \textsuperscript{95}Mo) and shows a gradual enrichment of the heavier isotope with depth. This trend can be attributed to the preferential removal of isotopically lighter thiomolybdate species from pore water. As thiomolybdate species are depleted in δ\textsuperscript{98}Mo compared to tetra-thiomolybdate, their removal leads to an enrichment of heavier isotope in the residual pore water.

Overall, this example demonstrates the capability of the \emph{phreeqcRM} package to simulate sediment burial, early diagenetic processes and cycling of major and trace elements. It also highlights the flexibility of \emph{marineGeochemFoam} for marine geochemical applications that are beyond simulation of kinetically-controlled redox reactions and require access to aqueous speciation calculations and detailed thermodynamic databases.

\FloatBarrier
\section{Conclusions}\label{sec:5}

In this study, we developed and present \emph{marineGeochemFoam,} an open-source, multi-purpose reactive transport modeling framework, with particular emphasis on sediment burial, early diagenesis, and submarine groundwater discharge processes occurring at surface marine sediments. The model is implemented within the widely used platform OpenFOAM\textsuperscript{®} and therefore benefits from its advanced numerical capabilities. The important features of the code include: i) transport of solute and solid species via advection, dispersion, diffusion and bioturbation, ii) variety of parametrizations for tortuosity, dispersion coefficients, and bioturbation, iii) flexible native reaction package (\emph{multiKinetics}) allowing the definition of arbitrary reaction networks and reaction stoichiometry, with a range of kinetic expression rates, iv) access to geochemical solver PHREEQC, allowing more flexibility for simulation of biogeochemical systems where explicit treatment of speciation and thermodynamic calculations are needed, and v) access to parallel computing, flexible handling of structured and unstructured grids, support for multi-dimensional simulations, variety of numerical schemes.

We first presented a series of benchmark examples with increasing complexity to validate different features of \emph{marineGeochemFoam} against analytical solutions, experimental data and outcomes of other numerical models. Afterwards, we presented three application examples to simulate diagenetic processes in marine sediments from two study sites, the Hanko site (Baltic Sea) in Finland and the Santa Barbara Basin in California. For the Hanko site, we demonstrated the capability of \emph{marineGeochemFoam} to simulate a complex network of 34 reactions, including primary and secondary redox reactions, in pockmarks with and without the influence of submarine groundwater discharge. In the Santa Barbara Basin example, we demonstrated the flexibility of the PHREEQC coupling of the code to extend the biogeochemical network from primary and secondary redox reactions of early diagenesis to trace element (Mo) cycling and isotope fractionation by accounting for speciation reactions and pH dynamics. In both examples, the proposed code could accurately describe the complex system dynamics and reproduced field observations.

\emph{marineGeochemFoam} is primarily designed for simulation of reactive transport processes in marine sediments, particularly under the influence of submarine groundwater discharge. However, the flexibility of the proposed modeling framework also makes it applicable to a broader range of subsurface reactive transport problems. For instance, the code can be applied to simulate: (i) nutrient and carbon cycling in soils, sediments, lakes and wetlands (\citealp{Danielsen2026}; e.g., \citealp{Sulman2024}; \citealp{Wang2024}), (ii) coastal and estuarine biogeochemistry \citep[e.g.,][]{OMeara2024,Seibert2025}, (iii) biogeochemical hotspots at the interface of groundwater-surface water \citep[e.g.,][]{Kaufman2017,Li2020}, (iv) fate and transport of contaminants in the subsurface \citep[e.g.,][]{Battistel2019,Prommer2009,Schafer2020}, and v) model-based interpretation of controlled laboratory and pilot experiments \citep{Cogorno2022,Muniruzzaman2021,Pieretti2022}. Additionally, we envisage several opportunities for further development of \emph{marineGeochemFoam.} In particular, its formulation based on coupled solute and solid continua provides a convenient framework that can be readily extended for other multi-continuum systems. One relevant example includes the extension of its process description toward multiphase systems, in which transport processes can occur within and across different fluid phases \citep[e.g.,][]{Ahmadi2025,Ahmadi2022b,Ahmadi2022a,MouraRibeiro2026,Horgue2015}. Such development would enable the simulation of reactive transport processes under unsaturated conditions for different applications including acid mine drainage from mine wastes \citep[e.g.,][]{Mayer2015,Muniruzzaman2021,Muniruzzaman2020,MuniruzzamanPedretti2021}, greenhouse gas emissions from natural and engineering settings \citep{Ahmadi2021,Ahmadi2020}, and transport of emerging contaminants \citep[e.g.,][]{Cogorno2025}. Furthermore, the existing formulation can also be extended to charged porous media, in which solute transport occurs in charge-balanced free water and charged electrical diffuse layer continua \citep{AppeloWersin2007,LopezVizcaino2022,MuniruzzamanRolle2023,MuniruzzamanRolle2019,Tournassat2019}. The current version of \emph{marineGeochemFoam} supports only prescribed velocity fields, however, it can be readily extended to incorporate Darcy-flow solvers within single or multiple continua, thereby enabling aquifer-scale simulations with relevant field-scale features \citep{Atteia2024,Soulaine2021} as well as explicit treatment of sediment-bottom water interface.

\section*{Acknowledgements}

The authors would like to thank the European Commission and the Research Council of Finland for funding in the frame of the collaborative international consortium SecuCoast financed under the 2023 Joint call of the European Partnership 101060874 Water4All. Additional funding was received through the Research Council of Finland projects NOMAD (project 362378) and HiFi-MuTaNTS (project 364773). Navid Ahmadi would like to acknowledge the financial support from the Global Wetland Center funded by the Novo Nordisk Foundation (research grant NNF23OC0081089) and DHI's research contract with the Danish ministry of Higher Education and Science, Denmark (project 5255-00005B).
\FloatBarrier
\nocite{*}
\bibliographystyle{elsarticle-harv}
\bibliography{references}
\end{document}


\linespread{1.08}\selectfont
\begin{center}%
{\fontsize{20}{24}\selectfont\bfseries Supplementary Material}\par%
\vspace{0.6em}for\par%
\vspace{0.6em}{\fontsize{16}{20}\selectfont\bfseries \textbf{\emph{marineGeochemFoam}: A multipurpose framework for simulating reactive transport and biogeochemical cycling in marine sediments and at seawater-groundwater interfaces}}\par%
\vspace{1em}%
{\normalsize\itshape Navid Ahmadi\textsuperscript{1,2†}, %
    Evgenii Kortunov\textsuperscript{3†}, %
    Karim Norouzi Moghanjoghi\textsuperscript{4}, %
    Joonas J. Virtasalo\textsuperscript{5}, %
    Muhammad Muniruzzaman\textsuperscript{6,7*}}\par%
\vspace{0.7em}%
{\normalsize \textsuperscript{1}Global Wetland Center, Department of Geosciences and Natural Resource Management, University of Copenhagen, Copenhagen, Denmark\par%
\vspace{0.2em}\textsuperscript{2}DHI A/S, Hørsholm, Denmark\par%
\vspace{0.2em}\textsuperscript{3}Energy and Construction Solutions, Geological Survey of Finland (GTK), Espoo, Finland\par%
\vspace{0.2em}\textsuperscript{4}Geosciences Department/MARUM-Center for Marine Environmental Sciences, University of Bremen, Bremen, Germany\par%
\vspace{0.2em}\textsuperscript{5}Marine Geology, Geological Survey of Finland (GTK), Espoo, Finland\par%
\vspace{0.2em}\textsuperscript{6}Institute of Geosciences, University of Bonn, Kirschallee 1-3, 53115 Bonn, Germany\par%
\vspace{0.2em}\textsuperscript{7}Water and Mining Solutions, Geological Survey of Finland (GTK), Espoo, Finland}%
\end{center}
{\normalsize *Corresponding author's emails: \href{mailto:m.muniruzzaman@uni-bonn.de}{m.muniruzzaman@uni-bonn.de}\par \textsuperscript{†} These authors contributed equally to this work.}

\vspace{3.5em}
\noindent\textbf{The file includes:} Tables S1--S15; Figures S1--S6.

\medskip
\vspace{2.5em}
\noindent\textbf{Content}
\begin{itemize}
\item \hyperref[sec:input-parameters]{S1 Input parameters}
\item \hyperref[sec:additional-results]{S2 Additional simulation results}
\end{itemize}

\clearpage
\section{Input parameters}\label{sec:input-parameters}

\subsection{Benchmark examples}

\begingroup%
\linespread{1}\selectfont
\fontsize{10}{12}\selectfont%
\renewcommand{\arraystretch}{1.08}%
\setlength{\tabcolsep}{4pt}%
\setlength{\aboverulesep}{0pt}%
\setlength{\belowrulesep}{0pt}%
\setlength{\parskip}{0pt}%
\setlength{\LTpre}{0pt}
\begin{longtable}{@{}>{\raggedright\arraybackslash}p{\dimexpr0.53848713\linewidth-1\tabcolsep\relax}>{\centering\arraybackslash}p{\dimexpr0.24955634\linewidth-2\tabcolsep\relax}>{\centering\arraybackslash}p{\dimexpr0.21195652\linewidth-1\tabcolsep\relax}@{}}
\caption{Overview of the model input parameters used in the simulation of the conservative solute and solid transport benchmark examples (Benchmark 1).}\label{tab:benchmark-1}\\
\endfirsthead

\multicolumn{3}{l}{\tablename~\thetable{} (continued)}\\ \toprule
\multicolumn{1}{@{}>{\raggedright\arraybackslash}p{\dimexpr0.53848713\linewidth-1\tabcolsep\relax}}{\textbf{Parameters}} & %
    \multicolumn{1}{>{\centering\arraybackslash}p{\dimexpr0.24955634\linewidth-2\tabcolsep\relax}}{\textbf{2-D setup}} & %
    \multicolumn{1}{>{\centering\arraybackslash}p{\dimexpr0.21195652\linewidth-1\tabcolsep\relax}@{}}{\textbf{3-D setup}}\\ \midrule[0.6pt]
\endhead

\midrule\multicolumn{3}{r}{Continued on next page}\\
\endfoot

\endlastfoot

\toprule
\multicolumn{3}{@{}>{\raggedright\arraybackslash}p{\dimexpr1.00000000\linewidth-0\tabcolsep\relax}@{}}{\textbf{Conservative solute transport benchmark example (1A)}}\\*[1.5pt]
\midrule[0.6pt]
\multicolumn{1}{@{}>{\centering\arraybackslash}p{\dimexpr0.53848713\linewidth-1\tabcolsep\relax}}{\textbf{}} & %
    \multicolumn{1}{>{\centering\arraybackslash}p{\dimexpr0.24955634\linewidth-2\tabcolsep\relax}}{\textbf{2-D setup}} & %
    \multicolumn{1}{>{\centering\arraybackslash}p{\dimexpr0.21195652\linewidth-1\tabcolsep\relax}@{}}{\textbf{3-D setup}}\\*[1.5pt]
\cmidrule{2-3}
\multicolumn{1}{@{}>{\centering\arraybackslash}p{\dimexpr0.53848713\linewidth-1\tabcolsep\relax}}{\textbf{Parameters}} & %
    \multicolumn{1}{>{\centering\arraybackslash}p{\dimexpr0.24955634\linewidth-2\tabcolsep\relax}}{\textbf{Value}} & %
    \multicolumn{1}{>{\centering\arraybackslash}p{\dimexpr0.21195652\linewidth-1\tabcolsep\relax}@{}}{\textbf{Value}}\\*[1.5pt]
\midrule[0.6pt]
\multicolumn{1}{@{}>{\raggedright\arraybackslash}p{\dimexpr0.53848713\linewidth-1\tabcolsep\relax}}{Porosity, \(\varepsilon \) [-]} & %
    \multicolumn{1}{>{\centering\arraybackslash}p{\dimexpr0.24955634\linewidth-2\tabcolsep\relax}}{0.4} & %
    \multicolumn{1}{>{\centering\arraybackslash}p{\dimexpr0.21195652\linewidth-1\tabcolsep\relax}@{}}{0.4}\\[1.5pt]
\multicolumn{1}{@{}>{\raggedright\arraybackslash}p{\dimexpr0.53848713\linewidth-1\tabcolsep\relax}}{2-D domain size, length [m]\ensuremath{\times}height [m]} & %
    \multicolumn{1}{>{\centering\arraybackslash}p{\dimexpr0.24955634\linewidth-2\tabcolsep\relax}}{0.28 \ensuremath{\times} 0.1127} & %
    \multicolumn{1}{>{\centering\arraybackslash}p{\dimexpr0.21195652\linewidth-1\tabcolsep\relax}@{}}{-}\\
\end{longtable}
\endgroup

\clearpage
\begingroup%
\linespread{1}\selectfont
\fontsize{10}{12}\selectfont%
\renewcommand{\arraystretch}{1.08}%
\setlength{\tabcolsep}{4pt}%
\setlength{\aboverulesep}{0pt}%
\setlength{\belowrulesep}{0pt}%
\setlength{\parskip}{0pt}%
\setlength{\LTpre}{0pt}
\begin{longtable}{@{}>{\raggedright\arraybackslash}p{\dimexpr0.66881100\linewidth-1\tabcolsep\relax}>{\centering\arraybackslash}p{\dimexpr0.09982254\linewidth-2\tabcolsep\relax}>{\centering\arraybackslash}p{\dimexpr0.23136646\linewidth-1\tabcolsep\relax}@{}}
\caption{Overview of the model input parameters used in examples for benchmarking the native reaction package (Benchmark 2).}\label{tab:benchmark-2}\\
\endfirsthead

\multicolumn{3}{l}{\tablename~\thetable{} (continued)}\\ \toprule
\multicolumn{2}{@{}>{\raggedright\arraybackslash}p{\dimexpr0.76863354\linewidth-1\tabcolsep\relax}}{\textbf{Parameters}} & %
    \multicolumn{1}{>{\centering\arraybackslash}p{\dimexpr0.23136646\linewidth-1\tabcolsep\relax}@{}}{\textbf{Value}}\\ \midrule[0.6pt]
\endhead

\midrule\multicolumn{3}{r}{Continued on next page}\\
\endfoot

\bottomrule
\endlastfoot

\toprule
\multicolumn{1}{@{}>{\raggedright\arraybackslash}p{\dimexpr0.66881100\linewidth-1\tabcolsep\relax}}{\textbf{Parameters}} & %
    \multicolumn{1}{>{\centering\arraybackslash}p{\dimexpr0.09982254\linewidth-2\tabcolsep\relax}}{\textbf{}} & %
    \multicolumn{1}{>{\centering\arraybackslash}p{\dimexpr0.23136646\linewidth-1\tabcolsep\relax}@{}}{\textbf{Value}}\\*[1.5pt]
\midrule[0.6pt]
\multicolumn{1}{@{}>{\raggedright\arraybackslash}p{\dimexpr0.66881100\linewidth-1\tabcolsep\relax}}{\textbf{First-order reaction (A \ensuremath{\rightarrow} 2B), \({R=kc}_{A}\)}} & %
    \multicolumn{1}{>{\centering\arraybackslash}p{\dimexpr0.09982254\linewidth-2\tabcolsep\relax}}{\textbf{}} & %
    \multicolumn{1}{>{\centering\arraybackslash}p{\dimexpr0.23136646\linewidth-1\tabcolsep\relax}@{}}{\textbf{}}\\*[1.5pt]
\midrule[0.6pt]
\multicolumn{1}{@{}>{\raggedright\arraybackslash}p{\dimexpr0.66881100\linewidth-1\tabcolsep\relax}}{Rate coefficient, \(k\) [1/s]} & %
    \multicolumn{1}{>{\centering\arraybackslash}p{\dimexpr0.09982254\linewidth-2\tabcolsep\relax}}{} & %
    \multicolumn{1}{>{\centering\arraybackslash}p{\dimexpr0.23136646\linewidth-1\tabcolsep\relax}@{}}{1 \ensuremath{\times} 10\textsuperscript{-3}}\\[1.5pt]
\multicolumn{1}{@{}>{\raggedright\arraybackslash}p{\dimexpr0.66881100\linewidth-1\tabcolsep\relax}}{Initial A concentration, \({c}_{A0}\)[mol/m\textsuperscript{3}]} & %
    \multicolumn{1}{>{\centering\arraybackslash}p{\dimexpr0.09982254\linewidth-2\tabcolsep\relax}}{} & %
    \multicolumn{1}{>{\centering\arraybackslash}p{\dimexpr0.23136646\linewidth-1\tabcolsep\relax}@{}}{1}\\[1.5pt]
\multicolumn{1}{@{}>{\raggedright\arraybackslash}p{\dimexpr0.66881100\linewidth-1\tabcolsep\relax}}{Initial B concentration, \({c}_{B0}\)[mol/m\textsuperscript{3}]} & %
    \multicolumn{1}{>{\raggedright\arraybackslash}p{\dimexpr0.09982254\linewidth-2\tabcolsep\relax}}{} & %
    \multicolumn{1}{>{\centering\arraybackslash}p{\dimexpr0.23136646\linewidth-1\tabcolsep\relax}@{}}{0}\\
\end{longtable}
\endgroup

\vspace{0.25em} 
\begingroup%
\linespread{1}\selectfont
\fontsize{10}{12}\selectfont%
\renewcommand{\arraystretch}{1.08}%
\setlength{\tabcolsep}{4pt}%
\setlength{\aboverulesep}{0pt}%
\setlength{\belowrulesep}{0pt}%
\setlength{\parskip}{0pt}%
\setlength{\LTpre}{0pt}
\begin{longtable}{@{}>{\raggedright\arraybackslash}p{\dimexpr0.16969831\linewidth-1\tabcolsep\relax}>{\centering\arraybackslash}p{\dimexpr0.19964508\linewidth-2\tabcolsep\relax}>{\centering\arraybackslash}p{\dimexpr0.30944987\linewidth-2\tabcolsep\relax}>{\centering\arraybackslash}p{\dimexpr0.32120674\linewidth-1\tabcolsep\relax}@{}}
\caption{Overview of the model input parameters and initial/boundary conditions used in the simulation of the reactive solute and solid transport benchmark examples performed by the native reaction package (Benchmark 3).}\label{tab:benchmark-3}\\
\endfirsthead

\multicolumn{4}{l}{\tablename~\thetable{} (continued)}\\ \toprule
\multicolumn{3}{@{}>{\raggedright\arraybackslash}p{\dimexpr0.67879326\linewidth-1\tabcolsep\relax}}{\textbf{Parameters}} & %
    \multicolumn{1}{>{\centering\arraybackslash}p{\dimexpr0.32120674\linewidth-1\tabcolsep\relax}@{}}{\textbf{Value}}\\ \midrule[0.6pt]
\endhead

\midrule\multicolumn{4}{r}{Continued on next page}\\
\endfoot

\endlastfoot

\toprule
\multicolumn{2}{@{}>{\raggedright\arraybackslash}p{\dimexpr0.36934339\linewidth-1\tabcolsep\relax}}{\textbf{Parameters}} & %
    \multicolumn{1}{>{\centering\arraybackslash}p{\dimexpr0.30944987\linewidth-2\tabcolsep\relax}}{\textbf{}} & %
    \multicolumn{1}{>{\centering\arraybackslash}p{\dimexpr0.32120674\linewidth-1\tabcolsep\relax}@{}}{\textbf{Value}}\\*[1.5pt]
\midrule[0.6pt]
\multicolumn{2}{@{}>{\raggedright\arraybackslash}p{\dimexpr0.36934339\linewidth-1\tabcolsep\relax}}{Domain size [m]} & %
    \multicolumn{1}{>{\centering\arraybackslash}p{\dimexpr0.30944987\linewidth-2\tabcolsep\relax}}{} & %
    \multicolumn{1}{>{\centering\arraybackslash}p{\dimexpr0.32120674\linewidth-1\tabcolsep\relax}@{}}{1}\\[1.5pt]
\multicolumn{2}{@{}>{\raggedright\arraybackslash}p{\dimexpr0.36934339\linewidth-1\tabcolsep\relax}}{Porosity, [-]} & %
    \multicolumn{1}{>{\centering\arraybackslash}p{\dimexpr0.30944987\linewidth-2\tabcolsep\relax}}{} & %
    \multicolumn{1}{>{\centering\arraybackslash}p{\dimexpr0.32120674\linewidth-1\tabcolsep\relax}@{}}{0.3}\\[1.5pt]
\multicolumn{2}{@{}>{\raggedright\arraybackslash}p{\dimexpr0.36934339\linewidth-1\tabcolsep\relax}}{Flux of solids, \(v\) [m/day]} & %
    \multicolumn{1}{>{\centering\arraybackslash}p{\dimexpr0.30944987\linewidth-2\tabcolsep\relax}}{} & %
    \multicolumn{1}{>{\centering\arraybackslash}p{\dimexpr0.32120674\linewidth-1\tabcolsep\relax}@{}}{-1 \ensuremath{\times} 10\textsuperscript{-5}}\\[1.5pt]
\multicolumn{2}{@{}>{\raggedright\arraybackslash}p{\dimexpr0.36934339\linewidth-1\tabcolsep\relax}}{Flux of solutes, \(u\) [m/day]} & %
    \multicolumn{1}{>{\centering\arraybackslash}p{\dimexpr0.30944987\linewidth-2\tabcolsep\relax}}{} & %
    \multicolumn{1}{>{\centering\arraybackslash}p{\dimexpr0.32120674\linewidth-1\tabcolsep\relax}@{}}{2 \ensuremath{\times} 10\textsuperscript{-6}}\\
\end{longtable}
\endgroup

\clearpage
\begingroup%
\linespread{1}\selectfont
\fontsize{10}{12}\selectfont%
\renewcommand{\arraystretch}{1.08}%
\setlength{\tabcolsep}{4pt}%
\setlength{\aboverulesep}{0pt}%
\setlength{\belowrulesep}{0pt}%
\setlength{\parskip}{0pt}%
\setlength{\LTpre}{0pt}
\begin{longtable}{@{}>{\raggedright\arraybackslash}p{\dimexpr0.15971606\linewidth-1\tabcolsep\relax}>{\centering\arraybackslash}p{\dimexpr0.42923691\linewidth-2\tabcolsep\relax}>{\centering\arraybackslash}p{\dimexpr0.08984028\linewidth-2\tabcolsep\relax}>{\centering\arraybackslash}p{\dimexpr0.32120674\linewidth-1\tabcolsep\relax}@{}}
\caption{Overview of the model input parameters and initial/boundary conditions used in the simulation of the reactive transport benchmark examples performed by the PhreeqcRM coupling (Benchmark 4)}\label{tab:benchmark-4}\\
\endfirsthead

\multicolumn{4}{l}{\tablename~\thetable{} (continued)}\\ \toprule
\multicolumn{3}{@{}>{\raggedright\arraybackslash}p{\dimexpr0.67879326\linewidth-1\tabcolsep\relax}}{\textbf{Parameters}} & %
    \multicolumn{1}{>{\centering\arraybackslash}p{\dimexpr0.32120674\linewidth-1\tabcolsep\relax}@{}}{\textbf{Value}}\\ \midrule[0.6pt]
\endhead

\midrule\multicolumn{4}{r}{Continued on next page}\\
\endfoot

\bottomrule
\endlastfoot

\toprule
\multicolumn{4}{@{}>{\raggedright\arraybackslash}p{\dimexpr1.00000000\linewidth-0\tabcolsep\relax}@{}}{\textbf{Benchmark 4A}}\\*[1.5pt]
\midrule[0.6pt]
\multicolumn{3}{@{}>{\raggedright\arraybackslash}p{\dimexpr0.67879326\linewidth-1\tabcolsep\relax}}{\textbf{Parameters}} & %
    \multicolumn{1}{>{\centering\arraybackslash}p{\dimexpr0.32120674\linewidth-1\tabcolsep\relax}@{}}{\textbf{Value}}\\*[1.5pt]
\midrule[0.6pt]
\multicolumn{3}{@{}>{\raggedright\arraybackslash}p{\dimexpr0.67879326\linewidth-1\tabcolsep\relax}}{Domain size [cm]} & %
    \multicolumn{1}{>{\centering\arraybackslash}p{\dimexpr0.32120674\linewidth-1\tabcolsep\relax}@{}}{8}\\[1.5pt]
\multicolumn{3}{@{}>{\raggedright\arraybackslash}p{\dimexpr0.67879326\linewidth-1\tabcolsep\relax}}{Porosity [-]} & %
    \multicolumn{1}{>{\centering\arraybackslash}p{\dimexpr0.32120674\linewidth-1\tabcolsep\relax}@{}}{0.6}\\[1.5pt]
\multicolumn{3}{@{}>{\raggedright\arraybackslash}p{\dimexpr0.67879326\linewidth-1\tabcolsep\relax}}{Velocity [m/s]} & %
    \multicolumn{1}{>{\centering\arraybackslash}p{\dimexpr0.32120674\linewidth-1\tabcolsep\relax}@{}}{2.7778 \ensuremath{\times} 10\textsuperscript{-6}}\\[1.5pt]
\multicolumn{3}{@{}>{\raggedright\arraybackslash}p{\dimexpr0.67879326\linewidth-1\tabcolsep\relax}}{Dispersivity, \({\alpha }_{l}\) [m]} & %
    \multicolumn{1}{>{\centering\arraybackslash}p{\dimexpr0.32120674\linewidth-1\tabcolsep\relax}@{}}{0.001}\\[1.5pt]
\multicolumn{3}{@{}>{\raggedright\arraybackslash}p{\dimexpr0.67879326\linewidth-1\tabcolsep\relax}}{Cation exchange capacity, X, [eq/L]} & %
    \multicolumn{1}{>{\centering\arraybackslash}p{\dimexpr0.32120674\linewidth-1\tabcolsep\relax}@{}}{0.0011}\\
\end{longtable}
\endgroup

\clearpage

\subsection{Input parameters for simulation of Hanko groundwater discharge site, pockmark E and D}

\begingroup%
\linespread{1}\selectfont
\fontsize{10}{12}\selectfont%
\renewcommand{\arraystretch}{1.1}%
\setlength{\tabcolsep}{3pt}%
\setlength{\parskip}{0pt}%
\setlength{\LTpre}{0pt}
\begin{longtable}{@{}>{\raggedright\arraybackslash}m{\dimexpr0.4\linewidth-8\tabcolsep\relax}>{\raggedleft\arraybackslash}m{0.14\linewidth}>{\raggedleft\arraybackslash}m{0.15\linewidth}>{\raggedleft\arraybackslash}m{0.15\linewidth}>{\raggedleft\arraybackslash}m{0.16\linewidth}@{}}
\caption{Overview of the physical model input parameters used in the simulation of pockmark E and D in the Hanko site}\label{tab:pockmark-flow}\\
\toprule \textbf{Parameter} & %
    \textbf{Symbol} & %
    \textbf{D} & %
    \textbf{E} & %
    \textbf{Units}\\ \midrule[0.6pt]
\endfirsthead

\multicolumn{5}{l}{\tablename~\thetable{} (continued)}\\
\toprule \textbf{Parameter} & %
    \textbf{Symbol} & %
    \textbf{D} & %
    \textbf{E} & %
    \textbf{Units}\\ \midrule[0.6pt]
\endhead

\midrule\multicolumn{5}{r}{Continued on next page}\\
\endfoot

\bottomrule\endlastfoot

Porosity at the SWI & %
    \(\epsilon_0\) & %
    \(0.35\) & %
    \(0.88\) & %
    \(\text{--}\) \\[1.5pt]
Porosity at depth & %
    \(\epsilon_\infty\) & %
    \(0.26\) & %
    \(0.71\) & %
    \(\text{--}\) \\[1.5pt]
Porosity e-folding distance & %
    \(\zeta\) & %
    \(1.75\) & %
    \(14.1\) & %
    \(\mathrm{cm}\) \\[1.5pt]
Sediment density & %
    \(\rho_s\) & %
    \(2.31\) & %
    \(2.31\) & %
    \(\mathrm{g\,cm^{-3}}\) \\[1.5pt]
Temperature & %
    \(T\) & %
    \(8\) & %
    \(8\) & %
    \({}^\circ\mathrm{C}\) \\[1.5pt]
Organic-matter C:N:P & %
    \(\text{--}\) & %
    \(106:16:1\) & %
    \(106:16:1\) & %
    \(\text{--}\) \\[1.5pt]
P:Fe ratio, $\alpha$ & %
    \(\chi^\alpha\) & %
    \(0.1\) & %
    \(0.1\) & %
    \(\text{--}\) \\[1.5pt]
P:Fe ratio, $\beta$ & %
    \(\chi^\beta\) & %
    \(0.055\) & %
    \(0.055\) & %
    \(\text{--}\) \\[1.5pt]
P:Fe ratio, $\gamma$ & %
    \(\chi^\gamma\) & %
    \(0.03\) & %
    \(0.03\) & %
    \(\text{--}\) \\[1.5pt]
Surface bioturbation coefficient & %
    \(D_{b0}\) & %
    \(1.35\times10^{-1}\) & %
    \(2.7\) & %
    \(\mathrm{cm^2\,yr^{-1}}\) \\[1.5pt]
Mixed-layer depth & %
    \(\xi\) & %
    \(1.6\) & %
    \(13\) & %
    \(\mathrm{cm}\) \\[1.5pt]
Solute flux & %
    \(u\) & %
    \(0.31\) & %
    \(0\) & %
    \(\mathrm{cm\,d^{-1}}\) \\[1.5pt]
Solid flux & %
    \(v\) & %
    \(3.6\times10^{-8}\) & %
    \(0.052\) & %
    \(\mathrm{cm\,yr^{-1}}\) \\[1.5pt]
\end{longtable}
\endgroup
\begingroup%
\linespread{1}\selectfont%
\fontsize{9}{11}\selectfont%
\noindent\emph{Notes.} The P:Fe ratios refer to the corresponding \(\mathrm{Fe(OH)_{3}}\) pools.\par\endgroup
\vspace{3em}

\begingroup%
\linespread{1}\selectfont
\fontsize{10}{12}\selectfont%
\renewcommand{\arraystretch}{1.1}%
\setlength{\tabcolsep}{3pt}%
\setlength{\parskip}{0pt}
\begin{longtable}{@{}>{\raggedright\arraybackslash}m{\dimexpr0.38\linewidth-6\tabcolsep\relax}>{\raggedright\arraybackslash}m{0.3\linewidth}>{\raggedright\arraybackslash}m{0.12\linewidth}>{\raggedleft\arraybackslash}m{0.2\linewidth}@{}}
\caption{Aqueous diffusion coefficients used in the simulation of pockmark E and D in Hanko site}\label{tab:pockmark-diffusion}\\
\toprule \textbf{Species} & %
    \textbf{Notation} & %
    \textbf{Phase} & %
    \textbf{$D_{\mathrm{aq}}$ (m$^2$ s$^{-1}$)}\\ \midrule[0.6pt]
\endfirsthead

\multicolumn{4}{l}{\tablename~\thetable{} (continued)}\\
\toprule \textbf{Species} & %
    \textbf{Notation} & %
    \textbf{Phase} & %
    \textbf{$D_{\mathrm{aq}}$ (m$^2$ s$^{-1}$)}\\ \midrule[0.6pt]
\endhead

\midrule\multicolumn{4}{r}{Continued on next page}\\
\endfoot

\bottomrule\endlastfoot

Chloride & %
    \(\mathrm{Cl^-}\) & %
    Solute & %
    \(1.23\times10^{-9}\) \\[1.5pt]
Oxygen & %
    \(\mathrm{O_2}\) & %
    Solute & %
    \(1.42\times10^{-9}\) \\[1.5pt]
Sulfate & %
    \(\mathrm{SO_4^{2-}}\) & %
    Solute & %
    \(6.49\times10^{-10}\) \\[1.5pt]
Manganese & %
    \(\mathrm{Mn^{2+}}\) & %
    Solute & %
    \(4.17\times10^{-10}\) \\[1.5pt]
Iron & %
    \(\mathrm{Fe^{2+}}\) & %
    Solute & %
    \(4.36\times10^{-10}\) \\[1.5pt]
Total sulfide & %
    \(\Sigma\mathrm{H_2S}\) & %
    Solute & %
    \(1.33\times10^{-9}\) \\[1.5pt]
Methane & %
    \(\mathrm{CH_4}\) & %
    Solute & %
    \(1.12\times10^{-9}\) \\[1.5pt]
Total ammonium & %
    \(\Sigma\mathrm{NH_4^+}\) & %
    Solute & %
    \(1.20\times10^{-9}\) \\[1.5pt]
Nitrate & %
    \(\mathrm{NO_3^-}\) & %
    Solute & %
    \(1.15\times10^{-9}\) \\[1.5pt]
Total phosphate & %
    \(\Sigma\mathrm{HPO_4^{2-}}\) & %
    Solute & %
    \(4.45\times10^{-10}\) \\[1.5pt]
Inorganic carbon & %
    \(\mathrm{DIC}\) & %
    Solute & %
    \(7.15\times10^{-10}\) \\[1.5pt]
\end{longtable}
\endgroup
\begingroup%
\linespread{1}\selectfont%
\fontsize{9}{11}\selectfont%
\noindent\emph{Notes.} $\Sigma$ includes the relevant aqueous species and complexes. Aqueous diffusion values are adapted from \citet{Purkamo2022}, including their correction to 8\,$^\circ$C.\par\endgroup
\clearpage%

\begingroup
\linespread{1}\selectfont
\fontsize{10}{12}\selectfont
\renewcommand{\arraystretch}{1.1}
\setlength{\tabcolsep}{3pt}
\setlength{\parskip}{0pt}

\begin{longtable}{@{}
		>{\raggedright\arraybackslash}m{\dimexpr0.24\linewidth-10\tabcolsep\relax}
		>{\centering\arraybackslash}m{0.13\linewidth}
		>{\centering\arraybackslash}m{0.13\linewidth}
		>{\centering\arraybackslash}m{0.15\linewidth}
		>{\centering\arraybackslash}m{0.15\linewidth}
		>{\centering\arraybackslash}m{0.20\linewidth}
		@{}}
	
	\caption{Initial and boundary conditions used in the simulation of pockmarks D and E in the Hanko site.}
	\label{tab:pockmark-boundary}\\
	
	\toprule
	& \multicolumn{2}{c}{\textbf{Initial condition}}
	& \multicolumn{2}{c}{\textbf{Boundary condition}}
	& \multicolumn{1}{c}{}\\*
	\cmidrule(lr){2-3}\cmidrule(lr){4-5}
	\textbf{Species}
	& \textbf{D}
	& \textbf{E}
	& \textbf{D}
	& \textbf{E}
	& \textbf{Units}\\
	\midrule
	\endfirsthead
	
	\multicolumn{6}{l}{\tablename~\thetable{} (continued)}\\
	\toprule
	& \multicolumn{2}{c}{\textbf{Initial condition}}
	& \multicolumn{2}{c}{\textbf{Boundary condition}}
	& \multicolumn{1}{c}{}\\*
	\cmidrule(lr){2-3}\cmidrule(lr){4-5}
	\textbf{Species}
	& \textbf{D}
	& \textbf{E}
	& \textbf{D}
	& \textbf{E}
	& \textbf{Units}\\
	\midrule
	\endhead
	
	\midrule
	\multicolumn{6}{r}{Continued on next page}\\
	\endfoot
	
	\bottomrule
	\endlastfoot
	
	\hline
	\multicolumn{6}{@{}l}{\textbf{Solid species}}\\*[2pt]
	
	\(\mathrm{OM}^\alpha\) & \(0.043313\) & \(637.56\) & \(8.6452\) & \(8316\) & \(\mathrm{mM}\) \\[1.5pt]
	\(\mathrm{OM}^\beta\) & \(0.17325\) & \(2106.7\) & \(34.581\) & \(5197.5\) & \(\mathrm{mM}\) \\[1.5pt]
	\(\mathrm{OM}^\gamma\) & \(433.12\) & \(5544\) & \(1540\) & \(115.5\) & \(\mathrm{mM}\) \\[1.5pt]
	\(\mathrm{Fe(OH)_3}^\alpha\) & \(71.867\) & \(291.06\) & \(0\) & \(0\) & \(\mathrm{mM}\) \\[1.5pt]
	\(\mathrm{Fe(OH)_3}^\beta\) & \(61.6\) & \(249.48\) & \(0\) & \(0\) & \(\mathrm{mM}\) \\[1.5pt]
	\(\mathrm{Fe(OH)_3}^\gamma\) & \(609.84\) & \(2079\) & \(1062.6\) & \(2337.7\) & \(\mathrm{mM}\) \\[1.5pt]
	\(\mathrm{MnO_2}^\alpha\) & \(1.36\times10^{-5}\) & \(1.155\) & \(0.0090667\) & \(0.0029\) & \(\mathrm{mM}\) \\[1.5pt]
	\(\mathrm{MnO_2}^\beta\) & \(1.36\times10^{-5}\) & \(1.155\) & \(0.0090667\) & \(0.0029\) & \(\mathrm{mM}\) \\[1.5pt]
	\(\mathrm{MnO_2}^\gamma\) & \(17.967\) & \(24.255\) & \(24.544\) & \(19.635\) & \(\mathrm{mM}\) \\[1.5pt]
	\(\mathrm{FeS}\) & \(0\) & \(0\) & \(0\) & \(0\) & \(\mathrm{mM}\) \\[1.5pt]
	\(\mathrm{FeS_2}\) & \(2.5987\) & \(0\) & \(10.395\) & \(0\) & \(\mathrm{mM}\) \\[1.5pt]
	\(\mathrm{FeCO_3}\) & \(46.2\) & \(231\) & \(0\) & \(0\) & \(\mathrm{mM}\) \\[1.5pt]
	\(\mathrm{S^0}\) & \(0\) & \(2.31\) & \(0\) & \(0\) & \(\mathrm{mM}\) \\[1.5pt]
	\(\mathrm{Fe}_{\mathrm{ox}}\mathrm{P}\) & \(9.24\) & \(0.0323\) & \(9\) & \(0.0323\) & \(\mathrm{mM}\) \\[1.5pt]
	\(\mathrm{Fe_3(PO_4)_2}\) & \(9.24\) & \(2.31\) & \(9\) & \(0\) & \(\mathrm{mM}\) \\[1.5pt]
	\(\mathrm{P_{org}}\) & \(4.62\) & \(4.62\) & \(0\) & \(0\) & \(\mathrm{mM}\) \\[1.5pt]
	\(\mathrm{P}_{\mathrm{authi,Ca-P}}\) & \(11.55\) & \(11.55\) & \(0\) & \(0\) & \(\mathrm{mM}\) \\[1.5pt]
	\(\mathrm{P_{Detr}}\) & \(2.31\) & \(2.31\) & \(0.057\) & \(0.057\) & \(\mathrm{mM}\) \\[1.5pt]
	
	\hline
	\multicolumn{6}{@{}l}{\textbf{Dissolved species}}\\*[2pt]
	
	\(\mathrm{Cl^-}\) & \(0.01\) & \(0.1\) & \(101.55\) & \(98.73\) & \(\mathrm{mM}\) \\[1.5pt]
	\(\mathrm{O_2}\) & \(0\) & \(0\) & \(0.0018\) & \(0.18\) & \(\mathrm{mM}\) \\[1.5pt]
	\(\mathrm{SO_4^{2-}}\) & \(1\times10^{-10}\) & \(1\times10^{-10}\) & \(4\) & \(4\) & \(\mathrm{mM}\) \\[1.5pt]
	\(\mathrm{Mn^{2+}}\) & \(5\times10^{-6}\) & \(1\times10^{-7}\) & \(10^{-10}\) & \(10^{-10}\) & \(\mathrm{mM}\) \\[1.5pt]
	\(\mathrm{Fe^{2+}}\) & \(1\times10^{-10}\) & \(1\times10^{-10}\) & \(7.16\times10^{-4}\) & \(7.16\times10^{-4}\) & \(\mathrm{mM}\) \\[1.5pt]
	\(\Sigma\mathrm{H_2S}\) & \(1\times10^{-10}\) & \(1\times10^{-10}\) & \(10^{-10}\) & \(10^{-10}\) & \(\mathrm{mM}\) \\[1.5pt]
	\(\mathrm{CH_4}\) & \(1\times10^{-9}\) & \(1\times10^{-9}\) & \(3.64\times10^{-7}\) & \(0\) & \(\mathrm{mM}\) \\[1.5pt]
	\(\Sigma\mathrm{NH_4^+}\) & \(1\times10^{-9}\) & \(1\times10^{-9}\) & \(0.012\) & \(1\times10^{-9}\) & \(\mathrm{mM}\) \\[1.5pt]
	\(\mathrm{NO_3^-}\) & \(0\) & \(1\times10^{-9}\) & \(0.012\) & \(0.01\) & \(\mathrm{mM}\) \\[1.5pt]
	\(\Sigma\mathrm{HPO_4^{2-}}\) & \(1\times10^{-9}\) & \(1\times10^{-9}\) & \(1\times10^{-9}\)& \(1\times10^{-9}\) & \(\mathrm{mM}\) \\[1.5pt]
	\(\mathrm{DIC}\) & \(0.32\) & \(5\) & \(0.3\) & \(2\) & \(\mathrm{mM}\) \\[1.5pt]
	
\end{longtable}
\endgroup

\begingroup
\linespread{1}\selectfont
\fontsize{9}{11}\selectfont
\par
\endgroup
\clearpage


\begingroup%
\linespread{1}\selectfont
\fontsize{10}{12}\selectfont%
\renewcommand{\arraystretch}{1.1}%
\setlength{\tabcolsep}{3pt}%
\setlength{\parskip}{0pt}
\begin{longtable}{@{}>{\raggedright\arraybackslash}m{0.075\linewidth}>{\raggedright\arraybackslash}p{\dimexpr0.925\linewidth-2\tabcolsep\relax}@{}}
\caption{Rate expressions used to describe biogeochemical reactions in pockmark E and D in the Hanko site}\label{tab:pockmark-rates}\\
\toprule \textbf{ID} & %
    \textbf{Rate expression}\\ \midrule[0.6pt]
\endfirsthead

\multicolumn{2}{l}{\tablename~\thetable{} (continued)}\\
\toprule \textbf{ID} & %
    \textbf{Rate expression}\\ \midrule[0.6pt]
\endhead

\midrule\multicolumn{2}{r}{Continued on next page}\\
\endfoot

\bottomrule\endlastfoot

\multicolumn{2}{@{}l}{\textbf{Primary redox reactions}}\\*[2pt]
R1 & %
    \singlelinerate{Purkamo-R1}{R_{1}^{\alpha,\beta}=k_{\alpha,\beta}\text{OM}^{\alpha,\beta}%
        \monodterm{[\text{O}_2]}{K_{O}+[\text{O}_2]}} \\[3pt]
R2 & %
    \singlelinerate{Purkamo-R2}{R_{2}^{\alpha,\beta}=k_{\alpha,\beta}\text{OM}^{\alpha,\beta} %
        \monodterm{[\text{NO}_3^-]}{K_{N}+[\text{NO}_3^-]} %
        \monodterm{K_{L,O}}{K_{L,O}+[\text{O}_2]}} \\[3pt]
R3 & %
    \singlelinerate{Purkamo-R3}{R_{3}^{\alpha,\beta}=k_{\alpha,\beta}\text{OM}^{\alpha,\beta}  %
        \monodterm{[\text{MnO}_2^{\alpha}]}{K_{M}+[\text{MnO}_2^{\alpha}]}  %
        \monodterm{K_{L,N}}{K_{L,N}+[\text{NO}_3^-]} %
        \monodterm{K_{L,O}}{K_{L,O}+[\text{O}_2]}} \\[3pt]
R4 & %
    \singlelinerate{Purkamo-R4}{R_{4}^{\alpha,\beta}=k_{\alpha,\beta}\text{OM}^{\alpha,\beta}  %
        \monodterm{[\text{Fe(OH)}_3^{\alpha}]}{K_{F}+[\text{Fe(OH)}_3^{\alpha}]} %
        \monodterm{K_{L,M}}{K_{L,M}+[\text{MnO}_2^{\alpha}]}  %
        \monodterm{K_{L,N}}{K_{L,N}+[\text{NO}_3^-]} %
        \monodterm{K_{L,O}}{K_{L,O}+[\text{O}_2]}} \\[3pt]
R5 & %
    \singlelinerate{Purkamo-R5}{R_{5}^{\alpha,\beta}=\Psi k_{\alpha,\beta}\text{OM}^{\alpha,\beta} %
        \monodterm{[\text{SO}_4^{2-}]}{K_{S}+[\text{SO}_4^{2-}]}  %
        \monodterm{K_{L,F}}{K_{L,F}+[\text{Fe(OH)}_3^{\alpha}]} %
        \monodterm{K_{L,M}}{K_{L,M}+[\text{MnO}_2^{\alpha}]}  %
        \monodterm{K_{L,N}}{K_{L,N}+[\text{NO}_3^-]} %
        \monodterm{K_{L,O}}{K_{L,O}+[\text{O}_2]}} \\[3pt]
R6 & %
    \singlelinerate{Purkamo-R6}{R_{6}^{\alpha,\beta}=\Psi k_{\alpha,\beta}\text{OM}^{\alpha,\beta} %
        \monodterm{K_{L,S}}{K_{L,S}+[\text{SO}_4^{2-}]}  %
        \monodterm{K_{L,F}}{K_{L,F}+[\text{Fe(OH)}_3^{\alpha}]} %
        \monodterm{K_{L,M}}{K_{L,M}+[\text{MnO}_2^{\alpha}]}  %
        \monodterm{K_{L,N}}{K_{L,N}+[\text{NO}_3^-]} %
        \monodterm{K_{L,O}}{K_{L,O}+[\text{O}_2]}} \\[3pt]
R7 & %
    \singlelinerate{Purkamo-R7}{R_{7}=k_{1}\text{DIC} %
        \monodterm{K_{L,S}}{K_{L,S}+[\text{SO}_4^{2-}]}  %
        \monodterm{K_{L,F}}{K_{L,F}+[\text{Fe(OH)}_3^{\alpha}]} %
        \monodterm{K_{L,M}}{K_{L,M}+[\text{MnO}_2^{\alpha}]}  %
        \monodterm{K_{L,N}}{K_{L,N}+[\text{NO}_3^-]} %
        \monodterm{K_{L,O}}{K_{L,O}+[\text{O}_2]}} \\[3pt]
 \hline
\multicolumn{2}{@{}l}{\textbf{Secondary redox and other reactions}}\\*[2pt]
R8 & %
    \singlelinerate{Purkamo-R8}{R_{8}=k_{2}[\text{O}_2][\text{NH}_4^+]} \\[1.5pt]
R9 & %
    \singlelinerate{Purkamo-R9}{R_{9}=k_{3}[\text{O}_2][\text{Mn}^{2+}]} \\[1.5pt]
R10 & %
    \singlelinerate{Purkamo-R10}{R_{10}=k_{4}[\text{O}_2][\text{Fe}^{2+}]} \\[1.5pt]
R11 & %
    \singlelinerate{Purkamo-R11}{R_{11}=k_{5}[\text{O}_2][\text{FeS}]} \\[1.5pt]
R12 & %
    \singlelinerate{Purkamo-R12}{R_{12}=k_{6}[\text{O}_2][\text{FeS}_2]} \\[1.5pt]
R13 & %
    \singlelinerate{Purkamo-R13}{R_{13}=k_{7}[\text{O}_2][\Sigma\mathrm{H_2S}]} \\[1.5pt]
R14 & %
    \singlelinerate{Purkamo-R14}{R_{14}=k_{8}[\text{O}_2][\text{CH}_4]} \\[1.5pt]
R15 & %
    \singlelinerate{Purkamo-R15}{R_{15}=k_{9}[\text{MnO}_2^{\alpha}][\text{Fe}^{2+}]} \\[1.5pt]
R16 & %
    \singlelinerate{Purkamo-R16}{R_{16}=k_{9}[\text{MnO}_2^{\beta}][\text{Fe}^{2+}]} \\[1.5pt]
R17 & %
    \singlelinerate{Purkamo-R17}{R_{17}=k_{10}[\text{MnO}_2^{\alpha}][\Sigma\mathrm{H_2S}]} \\[1.5pt]
R18 & %
    \singlelinerate{Purkamo-R18}{R_{18}=k_{10}[\text{MnO}_2^{\beta}][\Sigma\mathrm{H_2S}]} \\[1.5pt]
R19 & %
    \singlelinerate{Purkamo-R19}{R_{19}=k_{11}[\text{Fe(OH)}_3^{\alpha}][\Sigma\mathrm{H_2S}]} \\[1.5pt]
R20 & %
    \singlelinerate{Purkamo-R20}{R_{20}=k_{12}[\text{Fe(OH)}_3^{\beta}][\Sigma\mathrm{H_2S}]} \\[1.5pt]
R21 & %
    \singlelinerate{Purkamo-R21}{R_{21}=k_{13}[\text{Fe}^{2+}][\Sigma\mathrm{H_2S}]} \\[1.5pt]
R22 & %
    \singlelinerate{Purkamo-R22}{R_{22}=k_{14}[\text{FeS}][\Sigma\mathrm{H_2S}]} \\[1.5pt]
R23 & %
    \singlelinerate{Purkamo-R23}{R_{23}=k_{15}[\text{S}_0]} \\[1.5pt]
R24 & %
    \singlelinerate{Purkamo-R24}{R_{24}=k_{16}[\text{FeS}][\text{S}_0]} \\[1.5pt]
R25 & %
    \singlelinerate{Purkamo-R25}{R_{25}=k_{17}[\text{SO}_4^{2-}][\text{CH}_4]} \\[1.5pt]
R26 & %
    \singlelinerate{Purkamo-R26}{R_{26}=k_{18}[\text{Fe(OH)}_3^{\alpha,\beta}][\text{CH}_4]} \\[1.5pt]
R27 & %
    \singlelinerate{Purkamo-R27}{R_{27}=k_{19}[\text{Fe(OH)}_3^{\alpha}]} \\[1.5pt]
R28 & %
    \singlelinerate{Purkamo-R28}{R_{28}=k_{20}[\text{Fe(OH)}_3^{\beta}]} \\[1.5pt]
R29 & %
    \singlelinerate{Purkamo-R29}{R_{29}=k_{21}[\text{MnO}_2^{\alpha}]	} \\[1.5pt]
R30 & %
    \singlelinerate{Purkamo-R30}{R_{30}=k_{22}[\text{MnO}_2^{\beta}]} \\[1.5pt]
R31 & %
    \singlelinerate{Purkamo-R31}{R_{31}=k_{23}[\text{Fe}^{2+}][\text{HPO}_4^{2-}]} \\[1.5pt]
R32 & %
    \singlelinerate{Purkamo-R32}{R_{32}=k_{24}[\text{Fe}^{2+}][DIC]} \\[1.5pt]
R33 & %
    \singlelinerate{Purkamo-R33}{R_{33}=k_{25}[\text{FeCO}_3][\Sigma\mathrm{H_2S}]} \\[1.5pt]
R34 & %
    \singlelinerate{Purkamo-R34}{R_{34}=k_{26}[\mathrm{Fe_3(PO_4)_2}][\Sigma\mathrm{H_2S}]} \\[1.5pt]
\end{longtable}
\endgroup

\clearpage%
\begingroup%
\linespread{1}\selectfont
\fontsize{10}{12}\selectfont%
\renewcommand{\arraystretch}{1.1}%
\setlength{\tabcolsep}{3pt}%
\setlength{\parskip}{0pt}%
\setlength{\LTpre}{0pt}
\begin{longtable}{@{}>{\raggedright\arraybackslash}m{\dimexpr0.38\linewidth-8\tabcolsep\relax}>{\raggedright\arraybackslash}m{0.16\linewidth}*{2}{>{\raggedleft\arraybackslash}m{0.14\linewidth}}>{\raggedleft\arraybackslash}m{0.18\linewidth}@{}}
\caption{Rate coefficients and biogeochemical parameters used in the simulation of pockmark E and D in the Hanko site}\label{tab:pockmark-parameters}\\
\toprule \textbf{Parameter} & %
    \textbf{Symbol} & %
    \textbf{D} & %
    \textbf{E} & %
    \textbf{Units}\\ \midrule[0.6pt]
\endfirsthead

\multicolumn{5}{l}{\tablename~\thetable{} (continued)}\\
\toprule \textbf{Parameter} & %
    \textbf{Symbol} & %
    \textbf{D} & %
    \textbf{E} & %
    \textbf{Units}\\ \midrule[0.6pt]
\endhead

\midrule\multicolumn{5}{r}{Continued on next page}\\
\endfoot

\bottomrule\endlastfoot

\pockparameter{ka}{OM decay, $\alpha$}{k_{\alpha}}{0.25}{0.035}{\mathrm{yr}^{-1}}
\pockparameter{kb}{OM decay, $\beta$}{k_{\beta}}{0.0086}{0.006}{\mathrm{yr}^{-1}}
\pockparameter{KO}{Oxygen half-saturation/inhibition}{K_{O},K_{L,O}}{0.02}{0.02}{\mathrm{mM}}
\pockparameter{KN}{Nitrate half-saturation/inhibition}{K_{N},K_{L,N}}{0.004}{0.004}{\mathrm{mM}}
\pockparameter{KM}{Manganese oxide half-saturation/inhibition}{K_{M},K_{L,M}}{4}{4}{\mu\mathrm{mol\,g}^{-1}}
\pockparameter{KF}{Iron oxyhydroxide half-saturation/inhibition}{K_{F},K_{L,F}}{65}{65}{\mu\mathrm{mol\,g}^{-1}}
\pockparameter{KS}{Sulfate half-saturation/inhibition}{K_{S},K_{L,S}}{1.6}{1.6}{\mathrm{mM}}
\pockparameter{Psi}{Anaerobic decay factor}{\Psi}{0.0084}{0.1008}{\text{--}}
\pockparameter{k1}{Rate coefficient (R7)}{k_{1}}{0.00011}{0.0011}{\mathrm{yr}^{-1}}
\pockparameter{k2}{Rate coefficient (R8)}{k_{2}}{10^4}{10^4}{\mathrm{mM}^{-1}\,\mathrm{yr}^{-1}}
\pockparameter{k3}{Rate coefficient (R9)}{k_{3}}{10^4}{10^4}{\mathrm{mM}^{-1}\,\mathrm{yr}^{-1}}
\pockparameter{k4}{Rate coefficient (R10)}{k_{4}}{1.4\times10^5}{1.4\times10^5}{\mathrm{mM}^{-1}\,\mathrm{yr}^{-1}}
\pockparameter{k5}{Rate coefficient (R11)}{k_{5}}{300}{300}{\mathrm{mM}^{-1}\,\mathrm{yr}^{-1}}
\pockparameter{k6}{Rate coefficient (R12)}{k_{6}}{1}{1}{\mathrm{mM}^{-1}\,\mathrm{yr}^{-1}}
\pockparameter{k7}{Rate coefficient (R13)}{k_{7}}{160}{160}{\mathrm{mM}^{-1}\,\mathrm{yr}^{-1}}
\pockparameter{k8}{Rate coefficient (R14)}{k_{8}}{1\times10^7}{1\times10^7}{\mathrm{mM}^{-1}\,\mathrm{yr}^{-1}}
\pockparameter{k9}{Rate coefficient (R15, R16)}{k_{9}}{2}{2}{\mathrm{mM}^{-1}\,\mathrm{yr}^{-1}}
\pockparameter{k10}{Rate coefficient (R17, R18)}{k_{10}}{40}{20}{\mathrm{mM}^{-1}\,\mathrm{yr}^{-1}}
\pockparameter{k11}{Rate coefficient (R19)}{k_{11}}{9.5}{9.5}{\mathrm{mM}^{-1}\,\mathrm{yr}^{-1}}
\pockparameter{k12}{Rate coefficient (R20)}{k_{12}}{0.95}{0.95}{\mathrm{mM}^{-1}\,\mathrm{yr}^{-1}}
\pockparameter{k13}{Rate coefficient (R21)}{k_{13}}{1.2\times10^4}{2.25\times10^6}{\mathrm{mM}^{-1}\,\mathrm{yr}^{-1}}
\pockparameter{k14}{Rate coefficient (R22)}{k_{14}}{3\times10^{-4}}{3\times10^{-4}}{\mathrm{mM}^{-1}\,\mathrm{yr}^{-1}}
\pockparameter{k15}{Rate coefficient (R23)}{k_{15}}{3}{3}{\mathrm{yr}^{-1}}
\pockparameter{k16}{Rate coefficient (R24)}{k_{16}}{1}{1}{\mathrm{mM}^{-1}\,\mathrm{yr}^{-1}}
\pockparameter{k17}{Rate coefficient (R25)}{k_{17}}{14}{0.14}{\mathrm{mM}^{-1}\,\mathrm{yr}^{-1}}
\pockparameter{k18}{Rate coefficient (R26)}{k_{18}}{0.024}{8.8\times10^{-6}}{\mathrm{mM}^{-1}\,\mathrm{yr}^{-1}}
\pockparameter{k19}{Rate coefficient (R27)}{k_{19}}{0.6}{0.6}{\mathrm{yr}^{-1}}
\pockparameter{k20}{Rate coefficient (R28)}{k_{20}}{1.3\times10^{-5}}{1.3\times10^{-5}}{\mathrm{yr}^{-1}}
\pockparameter{k21}{Rate coefficient (R29)}{k_{21}}{1.8}{1.8}{\mathrm{yr}^{-1}}
\pockparameter{k22}{Rate coefficient (R30)}{k_{22}}{1.8\times10^{-3}}{1.8\times10^{-3}}{\mathrm{yr}^{-1}}
\pockparameter{k23}{Rate coefficient (R31)}{k_{23}}{0.052}{5.2\times10^{-3}}{\mathrm{mM}^{-1}\,\mathrm{yr}^{-1}}
\pockparameter{k24}{Rate coefficient (R32)}{k_{24}}{1.2150}{2.7\times10^{-4}}{\mathrm{mM}^{-1}\,\mathrm{yr}^{-1}}
\pockparameter{k25}{Rate coefficient (R33)}{k_{25}}{8\times10^{-4}}{8\times10^{-4}}{\mathrm{mM}^{-1}\,\mathrm{yr}^{-1}}
\pockparameter{k26}{Rate coefficient (R34)}{k_{26}}{8\times10^{-4}}{8\times10^{-4}}{\mathrm{mM}^{-1}\,\mathrm{yr}^{-1}}
\end{longtable}
\endgroup
\begingroup%
\linespread{1}\selectfont%
\fontsize{9}{11}\selectfont%
\noindent\emph{Notes.} $K_X$ and $K_{L,X}$ denote saturation and inhibition constants. Subscripts $O,N,M,F,S$ identify \(\mathrm{O_{2}}\), \(\mathrm{NO_{3}}^{-}\), \(\mathrm{MnO_{2}}^{\alpha}\), \(\mathrm{Fe(OH)_{3}}^{\alpha}\) and \(\mathrm{SO_{4}}^{2-}\), respectively. \par\endgroup

\clearpage
\subsection{Input parameters for Santa Barbara Basin simulation}

\medskip
\begingroup\linespread{1}\selectfont
\fontsize{9.5}{10.6}\selectfont
\renewcommand{\arraystretch}{1.0}
\setlength{\tabcolsep}{3pt}\setlength{\parskip}{0pt}
\begin{longtable}{@{}>{\raggedright\arraybackslash}m{\dimexpr.42\linewidth-6\tabcolsep\relax}>{\raggedright\arraybackslash}m{.14\linewidth}>{\raggedleft\arraybackslash}m{.19\linewidth}>{\raggedright\arraybackslash}m{.25\linewidth}@{}}
\caption{Overview of the physical model input parameters used in the simulation of Santa Barbara Basin}\label{tab:sbb-flow}\\
\toprule \textbf{Parameter} & \textbf{Symbol} & \textbf{Value} & \textbf{Units}\\ \midrule[0.6pt]
\endfirsthead
\multicolumn{4}{l}{\tablename~\thetable{} (continued)}\\
\toprule \textbf{Parameter} & \textbf{Symbol} & \textbf{Value} & \textbf{Units}\\ \midrule[0.6pt]
\endhead
\midrule\multicolumn{4}{r}{Continued on next page}\\
\endfoot\bottomrule\endlastfoot
Sediment depth & %
    \(L\) & %
    \(0.500\) & %
    \(\mathrm{m}\) \\[.3pt]
Number of cells & %
    \(N\) & %
    \(100\) & %
    \(\text{--}\) \\[.3pt]
Mesh end/start cell-size ratio & %
    \(g\) & %
    \(e^{10}\) & %
    \(\text{--}\) \\[.3pt]
Porosity at SWI & %
    \(\epsilon_0\) & %
    \(0.9472\) & %
    \(\text{--}\) \\[.3pt]
Asymptotic porosity & %
    \(\epsilon_\infty\) & %
    \(0.7884\) & %
    \(\text{--}\) \\[.3pt]
Porosity e-folding distance & %
    \(\zeta\) & %
    \(1.073\) & %
    \(\mathrm{m}\) \\[.3pt]
Dry-solid density & %
    \(\rho_s\) & %
    \(2600\) & %
    \(\mathrm{kg\,m^{-3}}\) \\[.3pt]
Solute flux & %
    \(u\) & %
    \(3.317\times10^{-11}\) & %
    \(\mathrm{m\,s^{-1}}\) \\[.3pt]
Solid flux & %
    \(v\) & %
    \(8.904\times10^{-12}\) & %
    \(\mathrm{m\,s^{-1}}\) \\[.3pt]
Sediment mass flux & %
    \(F_{sed}\) & %
    \(0.073\) & %
    \(\mathrm{g\,cm^{-2}\,yr^{-1}}\) \\[.3pt]
Surface bioturbation & %
    \(D_{b0}\) & %
    \(6.671\times10^{-16}\) & %
    \(\mathrm{m^2\,s^{-1}}\) \\[.3pt]
Mixed layer depth & %
    \(\xi\) & %
    \(0.030\) & %
    \(\mathrm{m}\) \\[.3pt]
Dispersivity & %
    \(\alpha_L\) & %
    \(0\) & %
    \(\mathrm{m}\) \\[.3pt]
Archie exponent & %
    \(n\) & %
    \(1\) & %
    \(\text{--}\) \\[.3pt]
Temperature & %
    \(T\) & %
    \(5\) & %
    \({}^\circ\mathrm{C}\) \\[.3pt]
Total simulation time & %
    \(t\) & %
    \(0\text{--}200\) & %
    \(\mathrm{yr}\) \\[.3pt]
Transport time step & %
    \(\Delta t\) & %
    \(1\) & %
    \(\mathrm{d}\) \\[.3pt]
\end{longtable}\endgroup
\begingroup\linespread{1}\fontsize{9}{10.5}\selectfont\noindent\emph{Notes.} Solid age is transported as a conservative solid species and increases by one year per year. \par\endgroup
\medskip
\begingroup\linespread{1}\selectfont
\fontsize{9.5}{10.7}\selectfont
\renewcommand{\arraystretch}{1.0}
\setlength{\tabcolsep}{3pt}\setlength{\parskip}{0pt}
\begin{longtable}{@{}>{\raggedright\arraybackslash}m{\dimexpr.31\linewidth-6\tabcolsep\relax}>{\raggedleft\arraybackslash}m{.23\linewidth}>{\raggedleft\arraybackslash}m{.23\linewidth}>{\raggedleft\arraybackslash}m{.23\linewidth}@{}}
\caption{Overview of initial and boundary conditions used in the simulation of Santa Barbara Basin}\label{tab:sbb-boundary}\label{tab:sbb-solid-boundary}\\
\toprule
 & \textbf{Initial condition} & \multicolumn{2}{c}{\textbf{Boundary condition}}\\*
\cmidrule(lr){2-2}\cmidrule(lr){3-4}
\textbf{Species or site} & \textbf{Input (mol L$^{-1}$)} & \textbf{SWI (mol L$^{-1}$)} & \textbf{Bottom (mol L$^{-1}$)}\\ \midrule[0.6pt]
\endfirsthead
\multicolumn{4}{l}{\tablename~\thetable{} (continued)}\\
\toprule
 & \textbf{Initial condition} & \multicolumn{2}{c}{\textbf{Boundary condition}}\\*
\cmidrule(lr){2-2}\cmidrule(lr){3-4}
\textbf{Species or site} & \textbf{Input (mol L$^{-1}$)} & \textbf{SWI (mol L$^{-1}$)} & \textbf{Bottom (mol L$^{-1}$)}\\ \midrule[0.6pt]
\endhead
\midrule\multicolumn{4}{r}{Continued on next page}\\
\endfoot\bottomrule\endlastfoot
\multicolumn{4}{@{}l}{\textbf{Solid species (per litre of solid phase)}}\\*[1pt]
\(\mathrm{OM}\) & %
    \(0.583/(1-\epsilon)\) & %
    \(11.04\) & %
    \(n\cdot j^s=0\) \\[.2pt]
\(\mathrm{MnO_2}\) & %
    \(9.402\times10^{-3}/(1-\epsilon)\) & %
    \(0.178\) & %
    \(n\cdot j^s=0\) \\[.2pt]
\(\mathrm{FeOOH}\) & %
    \(4.55\times10^{-2}/(1-\epsilon)\) & %
    \(0.862\) & %
    \(n\cdot j^s=0\) \\[.2pt]
\(\mathrm{FeS_2}\) & %
    \(3.761\times10^{-3}/(1-\epsilon)\) & %
    \(7.123\times10^{-2}\) & %
    \(n\cdot j^s=0\) \\[.2pt]
\(\mathrm{FeS}\) & %
    \(1\times10^{-8}/(1-\epsilon)\) & %
    \(1.894\times10^{-7}\) & %
    \(n\cdot j^s=0\) \\[.2pt]
\(\mathrm{S^0}\) & %
    \(1\times10^{-10}/(1-\epsilon)\) & %
    \(1.894\times10^{-9}\) & %
    \(n\cdot j^s=0\) \\[.2pt]
\(\mathrm{CaCO_3}\) & %
    \(1.034\times10^{-1}/(1-\epsilon)\) & %
    \(1.959\) & %
    \(n\cdot j^s=0\) \\[.2pt]
\(\mathrm{MnCO_3}\) & %
    \(0\) & %
    \(0\) & %
    \(n\cdot j^s=0\) \\[.2pt]
\(\mathrm{FeCO_3}\) & %
    \(0\) & %
    \(0\) & %
    \(n\cdot j^s=0\) \\[.2pt]
\(\mathrm{BSi}\) & %
    \(0.376/(1-\epsilon)\) & %
    \(7.123\) & %
    \(n\cdot j^s=0\) \\[.2pt]
\(\mathrm{ASi}\) & %
    \(0\) & %
    \(0\) & %
    \(n\cdot j^s=0\) \\[.2pt]
\(\mathrm{CFA}\) & %
    \(0\) & %
    \(0\) & %
    \(n\cdot j^s=0\) \\[.2pt]
\({}^{95}\mathrm{Mo}_{\mathrm{auth}}\) & %
    \(1.406\times10^{-14}\) & %
    \(1.406\times10^{-14}\) & %
    \(n\cdot j^s=0\) \\[.2pt]
\({}^{98}\mathrm{Mo}_{\mathrm{auth}}\) & %
    \(2.156\times10^{-14}\) & %
    \(2.156\times10^{-14}\) & %
    \(n\cdot j^s=0\) \\[.2pt]
\({}^{95}\mathrm{Mo}_{\mathrm{lith}}\) & %
    \(2.252\times10^{-5}\) & %
    \(2.252\times10^{-5}\) & %
    \(n\cdot j^s=0\) \\[.2pt]
\({}^{98}\mathrm{Mo}_{\mathrm{lith}}\) & %
    \(3.446\times10^{-5}\) & %
    \(3.446\times10^{-5}\) & %
    \(n\cdot j^s=0\) \\[.2pt]
\midrule\multicolumn{4}{@{}l}{\textbf{Dissolved species (per litre of porewater)}}\\*[1pt]
\(\mathrm{O_2}\) & %
    \(1\times10^{-12}\) & %
    \(8.5\times10^{-6}\) & %
    \(n\cdot j^w=0\) \\[.2pt]
\(\mathrm{NO_3^-}\) & %
    \(1\times10^{-10}\) & %
    \(2.8\times10^{-5}\) & %
    \(n\cdot j^w=0\) \\[.2pt]
\(\mathrm{NO_2^-}\) & %
    \(1\times10^{-10}\) & %
    \(2\times10^{-7}\) & %
    \(n\cdot j^w=0\) \\[.2pt]
\(\Sigma\mathrm N(0)\) & %
    \(1\times10^{-12}\) & %
    \(1\times10^{-15}\) & %
    \(n\cdot j^w=0\) \\[.2pt]
\(\Sigma\mathrm{SO_4^{2-}}\) & %
    \(0.015\) & %
    \(0.028\) & %
    \(0.015\) \\[.2pt]
\(\Sigma\mathrm{H_2S}\) & %
    \(2\times10^{-3}\) & %
    \(1\times10^{-15}\) & %
    \(2\times10^{-3}\) \\[.2pt]
\(\mathrm{CH_4}\) & %
    \(9\times10^{-5}\) & %
    \(1\times10^{-15}\) & %
    \(9\times10^{-5}\) \\[.2pt]
\(\Sigma\mathrm{Mn}(II)\) & %
    \(1\times10^{-6}\) & %
    \(6\times10^{-7}\) & %
    \(n\cdot j^w=0\) \\[.2pt]
\(\Sigma\mathrm{Fe}(II)\) & %
    \(1\times10^{-10}\) & %
    \(2.95\times10^{-8}\) & %
    \(n\cdot j^w=0\) \\[.2pt]
\(\Sigma\mathrm{NH_4^+}\) & %
    \(1.3\times10^{-3}\) & %
    \(1\times10^{-6}\) & %
    \(1.3\times10^{-3}\) \\[.2pt]
\(\mathrm{DIC}\) & %
    \(0.015\) & %
    \(2.345\times10^{-3}\) & %
    \(0.015\) \\[.2pt]
\(\Sigma\mathrm P\) & %
    \(2\times10^{-5}\) & %
    \(2\times10^{-5}\) & %
    \(n\cdot j^w=0\) \\[.2pt]
\(\Sigma\mathrm B\) & %
    \(8.706\times10^{-5}\) & %
    \(8.706\times10^{-5}\) & %
    \(n\cdot j^w=0\) \\[.2pt]
\(\Sigma\mathrm F\) & %
    \(4.5\times10^{-5}\) & %
    \(8\times10^{-5}\) & %
    \(4.5\times10^{-5}\) \\[.2pt]
\(\Sigma\mathrm{Si}\) & %
    \(5\times10^{-4}\) & %
    \(2.2\times10^{-4}\) & %
    \(5\times10^{-4}\) \\[.2pt]
\(\mathrm{Mg}\) & %
    \(5.28\times10^{-2}\) & %
    \(5.28\times10^{-2}\) & %
    \(n\cdot j^w=0\) \\[.2pt]
\(\mathrm{Ca}\) & %
    \(8\times10^{-3}\) & %
    \(1.033\times10^{-2}\) & %
    \(8\times10^{-3}\) \\[.2pt]
\(\mathrm K\) & %
    \(1.06\times10^{-2}\) & %
    \(1.06\times10^{-2}\) & %
    \(n\cdot j^w=0\) \\[.2pt]
\(\mathrm{Cl}\) & %
    \(0.566\) & %
    \(0.566\) & %
    \(n\cdot j^w=0\) \\[.2pt]
\(\mathrm{Na}\) & %
    Charge-balanced & %
    Charge-balanced & %
    \(n\cdot j^w=0\) \\[.2pt]
\({}^{95}\mathrm{Mo}\) & %
    \(4.737\times10^{-8}\) & %
    \(4.737\times10^{-8}\) & %
    \(n\cdot j^w=0\) \\[.2pt]
\({}^{98}\mathrm{Mo}\) & %
    \(7.263\times10^{-8}\) & %
    \(7.263\times10^{-8}\) & %
    \(n\cdot j^w=0\) \\[.2pt]
pH & %
    \(7.2\) & %
    \(7.85\) & %
    Calculated \\[.2pt]
\midrule\multicolumn{4}{@{}l}{\textbf{Surface sites (per litre of bulk sediment)}}\\*[1pt]
\(X_{N}\mathrm{OH}\) & %
    \(1.000\) & %
    -- & %
    -- \\[.2pt]
\(X_{M}\mathrm{OH}\) & %
    \(1.000\) & %
    -- & %
    -- \\[.2pt]
\(X_{F}\mathrm{OH}\) & %
    \(1.000\) & %
    -- & %
    -- \\[.2pt]
\end{longtable}\endgroup
\begingroup\linespread{1}\fontsize{9}{10.5}\selectfont\noindent\emph{Notes.} $\Sigma$ includes the dissolved complexes; N(0) is counted as N atoms and O$_2$ as molecules. pH is dimensionless and denotes $-\log_{10}a_{H^+}$ in the inputs. Sodium is charge-balanced. $n\cdot j^{w,s}=0$ specifies zero normal non-advective flux. Initial solid fractions use the depth-dependent $\epsilon(z)$ of Table~\ref{tab:sbb-flow}. Surface sites are used as non-electrostatic surface model in PHREEQC.\par\endgroup
\clearpage
\begingroup\linespread{1}\selectfont
\fontsize{9.5}{11.5}\selectfont
\renewcommand{\arraystretch}{1.0}
\setlength{\tabcolsep}{3pt}\setlength{\parskip}{0pt}
\newcommand{\gfourwrappedrate}[3]{%
\leavevmode
\begingroup\linespread{1}\selectfont%
\sbox{\ratebox}{\(#2\)}%
\ifdim\wd\ratebox>1.12\linewidth\relax%
\def\gfourresult##1{\typeout{G4WRAP #1}\(##1\)}%
\else\def\gfourresult##1{\singlelinerate{#1}{#2}}\fi%
\gfourresult{#3}\endgroup}
\begin{longtable}{@{}>{\raggedright\arraybackslash}m{\dimexpr.075\linewidth-2\tabcolsep\relax}>{\raggedright\arraybackslash}m{.925\linewidth}@{}}
\caption{Rate expressions used for the simulation of biogeochemical processes in the Santa Barbara Basin. All rates are in mol L$^{-1}$ of porewater yr$^{-1}$}\label{tab:sbb-rates}\\
\toprule \textbf{ID} & \textbf{Rate expression}\\ \midrule[0.6pt]
\endfirsthead
\multicolumn{2}{l}{\tablename~\thetable{} (continued)}\\
\toprule \textbf{ID} & \textbf{Rate expression}\\ \midrule[0.6pt]
\endhead
\midrule\multicolumn{2}{r}{Continued on next page}\\
\endfoot\bottomrule\endlastfoot
\multicolumn{2}{@{}l}{\textbf{Primary redox rates}}\\*[1pt]
R1 & %
    \singlelinerate{G4-rate-1}{R_{1}=k_{OM}(\tau)[\mathrm{OM}]\monodterm{[\mathrm{O_2}]}{K_{O}+[\mathrm{O_2}]}} \\[1pt]
R2 & %
    \singlelinerate{G4-rate-2}{R_{2}=k_{OM}(\tau)[\mathrm{OM}]\monodterm{[\mathrm{NO_3^-}]}{K_{N}+[\mathrm{NO_3^-}]}\monodterm{K_{L,T}}{K_{L,T}+[\mathrm{NO_2^-}]}\monodterm{K_{L,O}}{K_{L,O}+[\mathrm{O_2}]}} \\[1pt]
R3 & %
    \singlelinerate{G4-rate-3}{R_{3}=k_{OM}(\tau)[\mathrm{OM}]\monodterm{[\mathrm{NO_2^-}]}{K_{T}+[\mathrm{NO_2^-}]}\monodterm{K_{L,O}}{K_{L,O}+[\mathrm{O_2}]}} \\[1pt]
R4 & %
    \singlelinerate{G4-rate-4}{R_{4}=k_{OM}(\tau)[\mathrm{OM}]\monodterm{[\mathrm{MnO_2}]}{K_{M}+[\mathrm{MnO_2}]}\monodterm{K_{L,N}}{K_{L,N}+[\mathrm{NO_3^-}]}\monodterm{K_{L,T}}{K_{L,T}+[\mathrm{NO_2^-}]}\monodterm{K_{L,O}}{K_{L,O}+[\mathrm{O_2}]}} \\[1pt]
R5 & %
    \singlelinerate{G4-rate-5}{R_{5}=\lambda_Fk_{OM}(\tau)[\mathrm{OM}]\monodterm{[\mathrm{FeOOH}]}{K_{F}+[\mathrm{FeOOH}]}\monodterm{K_{L,M}}{K_{L,M}+[\mathrm{MnO_2}]}\monodterm{K_{L,N}}{K_{L,N}+[\mathrm{NO_3^-}]}\monodterm{K_{L,T}}{K_{L,T}+[\mathrm{NO_2^-}]}\monodterm{K_{L,O}}{K_{L,O}+[\mathrm{O_2}]}} \\[1pt]
R6 & %
    \gfourwrappedrate{G4-rate-6}{R_{6}=f_S(z)k_{OM}(\tau)[\mathrm{OM}]\monodterm{[\mathrm{SO_4^{2-}}]}{K_{S}+[\mathrm{SO_4^{2-}}]}\monodterm{K_{L,F}}{K_{L,F}+[\mathrm{FeOOH}]}\monodterm{K_{L,M}}{K_{L,M}+[\mathrm{MnO_2}]}\monodterm{K_{L,N}}{K_{L,N}+[\mathrm{NO_3^-}]}\monodterm{K_{L,T}}{K_{L,T}+[\mathrm{NO_2^-}]}\monodterm{K_{L,O}}{K_{L,O}+[\mathrm{O_2}]}}{\begin{aligned}[t]R_{6}={}&f_S(z)k_{OM}(\tau)[\mathrm{OM}]\monodterm{[\mathrm{SO_4^{2-}}]}{K_{S}+[\mathrm{SO_4^{2-}}]}\monodterm{K_{L,F}}{K_{L,F}+[\mathrm{FeOOH}]}\monodterm{K_{L,M}}{K_{L,M}+[\mathrm{MnO_2}]}\\ &\quad{}\times\monodterm{K_{L,N}}{K_{L,N}+[\mathrm{NO_3^-}]}\monodterm{K_{L,T}}{K_{L,T}+[\mathrm{NO_2^-}]}\monodterm{K_{L,O}}{K_{L,O}+[\mathrm{O_2}]}\end{aligned}} \\[1pt]
R7 & %
    \gfourwrappedrate{G4-rate-7}{R_{7}=k_{OM}(\tau)[\mathrm{OM}]\monodterm{K_{L,S}}{K_{L,S}+[\mathrm{SO_4^{2-}}]}\monodterm{K_{L,F}}{K_{L,F}+[\mathrm{FeOOH}]}\monodterm{K_{L,M}}{K_{L,M}+[\mathrm{MnO_2}]}\monodterm{K_{L,N}}{K_{L,N}+[\mathrm{NO_3^-}]}\monodterm{K_{L,T}}{K_{L,T}+[\mathrm{NO_2^-}]}\monodterm{K_{L,O}}{K_{L,O}+[\mathrm{O_2}]}}{\begin{aligned}[t]R_{7}={}&k_{OM}(\tau)[\mathrm{OM}]\monodterm{K_{L,S}}{K_{L,S}+[\mathrm{SO_4^{2-}}]}\monodterm{K_{L,F}}{K_{L,F}+[\mathrm{FeOOH}]}\monodterm{K_{L,M}}{K_{L,M}+[\mathrm{MnO_2}]}\\ &\quad{}\times\monodterm{K_{L,N}}{K_{L,N}+[\mathrm{NO_3^-}]}\monodterm{K_{L,T}}{K_{L,T}+[\mathrm{NO_2^-}]}\monodterm{K_{L,O}}{K_{L,O}+[\mathrm{O_2}]}\end{aligned}} \\[1pt]
\midrule\multicolumn{2}{@{}l}{\textbf{Secondary reaction rates}}\\*[1pt]
R8 & %
    \singlelinerate{G4-rate-8}{R_{8}=k_{8}[\mathrm{O_2}][\mathrm{NO_2^-}]} \\[0pt]
R9 & %
    \singlelinerate{G4-rate-9}{R_{9}=k_{9}[\mathrm{O_2}][\Sigma\mathrm{NH_4^+}]} \\[0pt]
R10 & %
    \singlelinerate{G4-rate-10}{R_{10}=k_{10}[\mathrm{O_2}][\Sigma\mathrm{Mn}(II)]} \\[0pt]
R11 & %
    \singlelinerate{G4-rate-11}{R_{11}=k_{11}[\mathrm{O_2}][\Sigma\mathrm{Fe}(II)]} \\[0pt]
R12 & %
    \singlelinerate{G4-rate-12}{R_{12}=k_{12}[\mathrm{O_2}][\Sigma\mathrm{H_2S}]} \\[0pt]
R13 & %
    \singlelinerate{G4-rate-13}{R_{13}=k_{13}[\mathrm{O_2}][\mathrm{FeS}]} \\[0pt]
R14 & %
    \singlelinerate{G4-rate-14}{R_{14}=k_{14}[\mathrm{O_2}][\mathrm{CH_4}]} \\[0pt]
R15 & %
    \singlelinerate{G4-rate-15}{R_{15}=k_{15}[\mathrm{NO_2^-}][\Sigma\mathrm{NH_4^+}]} \\[0pt]
R16 & %
    \singlelinerate{G4-rate-16}{R_{16}=k_{16}[\mathrm{NO_3^-}][\Sigma\mathrm{Mn}(II)]} \\[0pt]
R17 & %
    \singlelinerate{G4-rate-17}{R_{17}=k_{17}[\mathrm{NO_3^-}][\Sigma\mathrm{Fe}(II)]} \\[0pt]
R18 & %
    \singlelinerate{G4-rate-18}{R_{18}=k_{18}[\mathrm{CH_4}]\monodterm{[\mathrm{SO_4^{2-}}]}{K_A+[\mathrm{SO_4^{2-}}]}} \\[0pt]
R19 & %
    \singlelinerate{G4-rate-19}{R_{19}=k_{19}[\mathrm{MnO_2}][\Sigma\mathrm{Fe}(II)]} \\[0pt]
R20 & %
    \singlelinerate{G4-rate-20}{R_{20}=k_{20}[\mathrm{MnO_2}][\Sigma\mathrm{H_2S}]} \\[0pt]
R21 & %
    \singlelinerate{G4-rate-21}{R_{21}=k_{21}f_F(z)[\mathrm{FeOOH}][\Sigma\mathrm{H_2S}]} \\[0pt]
R22 & %
    \singlelinerate{G4-rate-22}{R_{22}=k_{22}[\mathrm{FeS}][\Sigma\mathrm{H_2S}]} \\[0pt]
R23 & %
    \singlelinerate{G4-rate-23}{R_{23}=k_{23}[\mathrm{FeS}](1-\Omega_{FeS})_+} \\[0pt]
R24 & %
    \singlelinerate{G4-rate-24}{R_{24}=\chi f_P(z)k_{24}(\Omega_{FeS}-1)_+} \\[0pt]
\midrule\multicolumn{2}{@{}l}{\textbf{Mineral and isotope-transfer rates}}\\*[1pt]
R25 & %
    \singlelinerate{G4-rate-25}{R_{25}=k^d_{cal}[\mathrm{CaCO_3}](1-\Omega_{cal})_+-\chi k^p_{cal}(\Omega_{cal}-1)_+} \\[0pt]
R26 & %
    \singlelinerate{G4-rate-26}{R_{26}=k^d_{rho}[\mathrm{MnCO_3}](1-\Omega_{rho})_+-\chi k^p_{rho}(\Omega_{rho}-1)_+} \\[0pt]
R27 & %
    \singlelinerate{G4-rate-27}{R_{27}=k^d_{sid}[\mathrm{FeCO_3}](1-\Omega_{sid})_+-\chi k^p_{sid}(\Omega_{sid}-1)_+} \\[0pt]
R28 & %
    \singlelinerate{G4-rate-28}{R_{28}=k_B e^{-z/z_B}[\mathrm{BSi}](1-\Omega_B)_+} \\[0pt]
R29 & %
    \singlelinerate{G4-rate-29}{R_{29}=\chi k_A(\Omega_A-1)_+} \\[0pt]
R30 & %
    \singlelinerate{G4-rate-30}{R_{30}=\frac{k_{CFA}}{\phi}(\Omega_{CFA}-1)_+} \\[0pt]
R31 & %
    \singlelinerate{G4-rate-31}{R_{31}=4k_{rm}\sum_{j=1}^{4}[{}^i\mathrm{MoO}_{4-j}\mathrm S_j]\quad(i=95,98)} \\[0pt]
\end{longtable}\endgroup
\begingroup\linespread{1}\fontsize{9}{10.5}\selectfont\noindent\emph{Notes.} Brackets denote concentrations per litre of porewater, including solid species after volume conversion. $\chi=(1-\epsilon)/\epsilon$ and $(x)_+=\max(x,0)$. The OM decay coefficient is $k_{OM}(\tau_{age})=\nu/(a_0+\tau)$, with $\nu=0.12$, $a_0=3.2$ yr and $\tau_{age}$ being the local transported solid age (yr) (age is transported as a conservative solid species). $\Sigma$ denotes dissolved component totals. Coefficients and calibrated depth functions are in Table~\ref{tab:sbb-coefficients}.\par\endgroup
\clearpage
\begingroup\linespread{1}\selectfont
\fontsize{9.5}{11}\selectfont
\renewcommand{\arraystretch}{1.0}
\setlength{\tabcolsep}{3pt}\setlength{\parskip}{0pt}
\begin{longtable}{@{}>{\raggedright\arraybackslash}m{\dimexpr.41\linewidth-6\tabcolsep\relax}>{\raggedright\arraybackslash}m{.18\linewidth}>{\raggedleft\arraybackslash}m{.18\linewidth}>{\raggedright\arraybackslash}m{.23\linewidth}@{}}
\caption{Rate coefficients and biogeochemical parameters used in the Santa Barbara Basin simulation}\label{tab:sbb-coefficients}\\
\toprule \textbf{Parameter} & \textbf{Symbol} & \textbf{Value} & \textbf{Units}\\ \midrule[0.6pt]
\endfirsthead
\multicolumn{4}{l}{\tablename~\thetable{} (continued)}\\
\toprule \textbf{Parameter} & \textbf{Symbol} & \textbf{Value} & \textbf{Units}\\ \midrule[0.6pt]
\endhead
\midrule\multicolumn{4}{r}{Continued on next page}\\
\endfoot\bottomrule\endlastfoot
\multicolumn{4}{@{}l}{\textbf{OM and stoichiometry}}\\*[1pt]
OM carbon coefficient & %
    \(a\) & %
    \(1\) & %
    \(\text{--}\) \\[.5pt]
OM N:C coefficient & %
    \(b\) & %
    \(\dfrac{3}{20}\) & %
    \(\text{--}\) \\[.5pt]
OM P:C coefficient & %
    \(c\) & %
    \(\dfrac{3}{200}\) & %
    \(\text{--}\) \\[.5pt]
Reactive-continuum exponent & %
    \(\nu\) & %
    \(0.120\) & %
    \(\text{--}\) \\[.5pt]
Initial OM reactivity age & %
    \(a_0\) & %
    \(3.2\) & %
    \(\mathrm{yr}\) \\[.5pt]
FeOOH primary-rate multiplier & %
    \(\lambda_F\) & %
    \(0.25\) & %
    \(\text{--}\) \\[.5pt]
\midrule\multicolumn{4}{@{}l}{\textbf{Monod and inhibition constants}}\\*[1pt]
Oxygen saturation/inhibition & %
    \(K_{O},K_{L,O}\) & %
    \(1\times10^{-6}\) & %
    \(\mathrm{mol\,L^{-1}}\) \\[.5pt]
Nitrate saturation/inhibition & %
    \(K_{N},K_{L,N}\) & %
    \(1\times10^{-5}\) & %
    \(\mathrm{mol\,L^{-1}}\) \\[.5pt]
Nitrite saturation/inhibition & %
    \(K_{T},K_{L,T}\) & %
    \(1\times10^{-5}\) & %
    \(\mathrm{mol\,L^{-1}}\) \\[.5pt]
Sulfate saturation/inhibition & %
    \(K_{S},K_{L,S}\) & %
    \(1\times10^{-4}\) & %
    \(\mathrm{mol\,L^{-1}}\) \\[.5pt]
MnO$_2$ saturation/inhibition & %
    \(K_M,K_{L,M}\) & %
    \(4.160\times10^{-2}\chi\) & %
    \(\mathrm{mol\,L^{-1}}\) \\[.5pt]
FeOOH saturation/inhibition & %
    \(K_F,K_{L,F}\) & %
    \(1.300\chi\) & %
    \(\mathrm{mol\,L^{-1}}\) \\[.5pt]
AOM sulfate saturation & %
    \(K_A\) & %
    \(0.001\) & %
    \(\mathrm{mol\,L^{-1}}\) \\[.5pt]
\midrule\multicolumn{4}{@{}l}{\textbf{Secondary kinetic coefficients}}\\*[1pt]
Rate coefficient (R8) & %
    \(k_{8}\) & %
    \(1\times10^{7}\) & %
    \(\mathrm{L\,mol^{-1}\,yr^{-1}}\) \\[.5pt]
Rate coefficient (R9) & %
    \(k_{9}\) & %
    \(1\times10^{7}\) & %
    \(\mathrm{L\,mol^{-1}\,yr^{-1}}\) \\[.5pt]
Rate coefficient (R10) & %
    \(k_{10}\) & %
    \(5\times10^{6}\) & %
    \(\mathrm{L\,mol^{-1}\,yr^{-1}}\) \\[.5pt]
Rate coefficient (R11) & %
    \(k_{11}\) & %
    \(5\times10^{8}\) & %
    \(\mathrm{L\,mol^{-1}\,yr^{-1}}\) \\[.5pt]
Rate coefficient (R12) & %
    \(k_{12}\) & %
    \(1\times10^{5}\) & %
    \(\mathrm{L\,mol^{-1}\,yr^{-1}}\) \\[.5pt]
Rate coefficient (R13) & %
    \(k_{13}\) & %
    \(1\times10^{5}\) & %
    \(\mathrm{L\,mol^{-1}\,yr^{-1}}\) \\[.5pt]
Rate coefficient (R14) & %
    \(k_{14}\) & %
    \(1\times10^{10}\) & %
    \(\mathrm{L\,mol^{-1}\,yr^{-1}}\) \\[.5pt]
Rate coefficient (R15) & %
    \(k_{15}\) & %
    \(1\times10^{8}\) & %
    \(\mathrm{L\,mol^{-1}\,yr^{-1}}\) \\[.5pt]
Rate coefficient (R16) & %
    \(k_{16}\) & %
    \(1\times10^{5}\) & %
    \(\mathrm{L\,mol^{-1}\,yr^{-1}}\) \\[.5pt]
Rate coefficient (R17) & %
    \(k_{17}\) & %
    \(1.5\times10^{5}\) & %
    \(\mathrm{L\,mol^{-1}\,yr^{-1}}\) \\[.5pt]
Rate coefficient (R18) & %
    \(k_{18}\) & %
    \(0.04\) & %
    \(\mathrm{yr^{-1}}\) \\[.5pt]
Rate coefficient (R19) & %
    \(k_{19}\) & %
    \(1\times10^{7}\) & %
    \(\mathrm{L\,mol^{-1}\,yr^{-1}}\) \\[.5pt]
Rate coefficient (R20) & %
    \(k_{20}\) & %
    \(1\times10^{5}\) & %
    \(\mathrm{L\,mol^{-1}\,yr^{-1}}\) \\[.5pt]
Rate coefficient (R21) & %
    \(k_{21}\) & %
    \(4000\) & %
    \(\mathrm{L\,mol^{-1}\,yr^{-1}}\) \\[.5pt]
Rate coefficient (R22) & %
    \(k_{22}\) & %
    \(300\) & %
    \(\mathrm{L\,mol^{-1}\,yr^{-1}}\) \\[.5pt]
Rate coefficient (R23) & %
    \(k_{23}\) & %
    \(1\times10^{-3}\) & %
    \(\mathrm{yr^{-1}}\) \\[.5pt]
Rate coefficient (R24) & %
    \(k_{24}\) & %
    \(0.52\) & %
    \(\mathrm{mol\,L_s^{-1}\,yr^{-1}}\) \\[.5pt]
\midrule\multicolumn{4}{@{}l}{\textbf{Mineral parameters}}\\*[1pt]
Carbonate dissolution (cal) & %
    \(k^d_{cal}\) & %
    \(0.02\) & %
    \(\mathrm{yr^{-1}}\) \\[.5pt]
Carbonate precipitation (cal) & %
    \(k^p_{cal}\) & %
    \(1\times10^{-4}\) & %
    \(\mathrm{mol\,L_s^{-1}\,yr^{-1}}\) \\[.5pt]
Carbonate dissolution (rho) & %
    \(k^d_{rho}\) & %
    \(0.8\) & %
    \(\mathrm{yr^{-1}}\) \\[.5pt]
Carbonate precipitation (rho) & %
    \(k^p_{rho}\) & %
    \(0.26\) & %
    \(\mathrm{mol\,L_s^{-1}\,yr^{-1}}\) \\[.5pt]
Carbonate dissolution (sid) & %
    \(k^d_{sid}\) & %
    \(0.5\) & %
    \(\mathrm{yr^{-1}}\) \\[.5pt]
Carbonate precipitation (sid) & %
    \(k^p_{sid}\) & %
    \(1\times10^{-4}\) & %
    \(\mathrm{mol\,L_s^{-1}\,yr^{-1}}\) \\[.5pt]
Biogenic silica dissolution & %
    \(k_B\) & %
    \(4.38\times10^{-2}\) & %
    \(\mathrm{yr^{-1}}\) \\[.5pt]
BSi dissolution attenuation & %
    \(z_B\) & %
    \(1.700\) & %
    \(\mathrm{cm}\) \\[.5pt]
Authigenic silica precipitation & %
    \(k_A\) & %
    \(8.76\times10^{-6}\) & %
    \(\mathrm{mol\,L_s^{-1}\,yr^{-1}}\) \\[.5pt]
CFA precipitation & %
    \(k_{CFA}\) & %
    \(2.7\times10^{-8}\) & %
    \(\mathrm{mol\,L_b^{-1}\,yr^{-1}}\) \\[.5pt]
CFA carbonate exponent & %
    \(\eta_C\) & %
    \(1.1307\) & %
    \(\text{--}\) \\[.5pt]
\noalign{\penalty-10000}
\midrule\multicolumn{4}{@{}l}{\textbf{Mo isotope parameters}}\\*[1pt]
Case 1 Mo removal coefficient & %
    \(k_{rm}\) & %
    \(2\) & %
    \(\mathrm{yr^{-1}}\) \\[.5pt]
Isotope standard ratio & %
    \(R_{std}\) & %
    \(15303/10000\) & %
    \(\text{--}\) \\[.5pt]
Isotope reporting offset & %
    \(\delta_0\) & %
    \(0.25\) & %
    \(\text{\textperthousand}\) \\[.5pt]
Mo equilibrium isotope enrichment & %
    \(\epsilon_1\) & %
    \(1.4\) & %
    \(\text{\textperthousand}\) \\[.5pt]
Mo equilibrium isotope enrichment & %
    \(\epsilon_2\) & %
    \(2.8\) & %
    \(\text{\textperthousand}\) \\[.5pt]
Mo equilibrium isotope enrichment & %
    \(\epsilon_3\) & %
    \(4.55\) & %
    \(\text{\textperthousand}\) \\[.5pt]
Mo equilibrium isotope enrichment & %
    \(\epsilon_4\) & %
    \(6.3\) & %
    \(\text{\textperthousand}\) \\[.5pt]
SWI total Mo & %
    \([\mathrm{Mo}]_{SWI}\) & %
    \(1.2\times10^{-7}\) & %
    \(\mathrm{mol\,L^{-1}}\) \\[.5pt]
SWI Mo isotope composition & %
    \(\delta^{98}\mathrm{Mo}_{SWI}\) & %
    \(2.34\) & %
    \(\text{\textperthousand}\) \\[.5pt]
Lithogenic Mo isotope composition & %
    \(\delta^{98}\mathrm{Mo}_{lith}\) & %
    \(0.30\) & %
    \(\text{\textperthousand}\) \\[.5pt]
\midrule\multicolumn{4}{@{}l}{\textbf{Depth functions ($z$ in cm)}}\\*[1pt]
FeOOH--sulfide & \(f_F(z)\) & \multicolumn{2}{l}{\(4+20\exp[-((z-2)/0.7)^2]\)}\\[1pt]
FeS precipitation & \(f_P(z)\) & \multicolumn{2}{l}{\(0.25+3.75/(1+\exp{-(z-3.15)/0.3})\)}\\[1pt]
Deep sulfate reduction & \(f_S(z)\) & \multicolumn{2}{l}{\(1.16-0.16/(1+\exp{(z-3.2)/0.6})\)}\\[1pt]
\end{longtable}\endgroup
\begingroup\linespread{1}\fontsize{9}{10.5}\selectfont\noindent\emph{Notes.} $L_s$, $L_b$ and $L$ denote litres of solids, bulk sediment and porewater, respectively. $\chi=(1-\epsilon)/\epsilon$ converts solid-volume Monod constants to the porewater basis. Displayed values are rounded, whereas model inputs include full precision. $K_X$ and $K_{L,X}$ have equal values for each listed acceptor. Isotope enrichments satisfy $K^{95}_j/K^{98}_j=1+\epsilon_j/1000$; $\delta^{98}\mathrm{Mo}=1000[({}^{98}\mathrm{Mo}/{}^{95}\mathrm{Mo})/({}^{98}\mathrm{Mo}/{}^{95}\mathrm{Mo})_{std}-1]+\delta_0$. \par\endgroup
\medskip
\begingroup\linespread{1}\selectfont
\fontsize{9.5}{11}\selectfont
\renewcommand{\arraystretch}{1.05}
\setlength{\tabcolsep}{3pt}\setlength{\parskip}{0pt}
\begin{longtable}{@{}>{\raggedright\arraybackslash}m{\dimexpr.13\linewidth-6\tabcolsep\relax}>{\raggedright\arraybackslash}m{.14\linewidth}>{\raggedleft\arraybackslash}m{.16\linewidth}>{\raggedright\arraybackslash}m{.57\linewidth}@{}}
\caption{Surface and mineral reaction constants used in the simulation of the Santa Barbara simulation }\label{tab:sbb-equilibria}\\
\toprule \textbf{ID} & \textbf{Symbol} & \textbf{$\log_{10}K$} & \textbf{Definition}\\ \midrule[0.6pt]
\endfirsthead
\multicolumn{4}{l}{\tablename~\thetable{} (continued)}\\
\toprule \textbf{ID} & \textbf{Symbol} & \textbf{$\log_{10}K$} & \textbf{Definition}\\ \midrule[0.6pt]
\endhead
\midrule\multicolumn{4}{r}{Continued on next page}\\
\endfoot\bottomrule\endlastfoot
\multicolumn{4}{@{}l}{\textbf{Surface equilibria (5 $^\circ$C)}}\\*[1pt]
R32 & %
    \(K_{32}\) & %
    \(-0.381\) & %
    Surface formation \\[1.5pt]
R33 & %
    \(K_{33}\) & %
    \(0.862\) & %
    Surface formation \\[1.5pt]
R34 & %
    \(K_{34}\) & %
    \(1.843\) & %
    Surface formation \\[1.5pt]
\midrule\multicolumn{4}{@{}l}{\textbf{Apparent saturation constants used by the kinetic laws}}\\*[1pt]
R23, R24 & %
    \(K_{FeS}\) & %
    \(-2.200\) & %
    \singlelinerate{G4-saturation-R23-R24}{\Omega_{FeS}=m_{Fe^{2+}}m_{HS^-}/(m_{H^+}K_{FeS})} \\[3pt]
R25 & %
    \(K_{cal}\) & %
    \(-6.324\) & %
    \singlelinerate{G4-saturation-R25}{\Omega_{cal}=m_{Ca^{2+}}m_{CO_3^{2-}}/K_{cal}} \\[3pt]
R26 & %
    \(K_{rho}\) & %
    \(-9.477\) & %
    \singlelinerate{G4-saturation-R26}{\Omega_{rho}=m_{Mn^{2+}}m_{CO_3^{2-}}/K_{rho}} \\[3pt]
R27 & %
    \(K_{sid}\) & %
    \(-9.230\) & %
    \singlelinerate{G4-saturation-R27}{\Omega_{sid}=m_{Fe^{2+}}m_{CO_3^{2-}}/K_{sid}} \\[3pt]
R29 & %
    \(K_A^{sat}\) & %
    \(-3.699\) & %
    \singlelinerate{G4-saturation-R29}{\Omega_A=m_{H_4SiO_4}/K_A^{sat}} \\[3pt]
R30 & %
    \(K_{CFA}\) & %
    \(-54.938\) & %
    \singlelinerate{G4-saturation-R30}{\Omega_{CFA}=m_{Ca^{2+}}^{9.54}m_{PO_4^{3-}}^{4.8}m_{F^-}^{2.48}/(m_{CO_3^{2-}}^{\eta_C}K_{CFA})} \\[3pt]
R28 & %
    \(K_B^{sat}(z)\) & %
    Depth-dependent & %
    \(\Omega_B=m_{H_4SiO_4}/K_B^{sat}(z)\) \\[2pt]
\end{longtable}\endgroup
\begingroup\linespread{1}\fontsize{9}{10.5}\selectfont\noindent\emph{Notes.} $m_X$ is the free-species molality returned by PHREEQC. These kinetic saturation thresholds are apparent numerical constants, distinct from thermodynamic PHREEQC phase log$K$ values. With $z$ in cm and $T=278.15$ K, $K_B^{sat}(z)=2.754\exp[(-2229.664+0.020z-2.7\times10^{-6}z^2+1.46\times10^{-10}z^3)/T]$. The surface assemblages have capacity 1 mol L$_b^{-1}$ per site. \par\endgroup
\clearpage
\begingroup\linespread{1}\selectfont
\fontsize{9.5}{11}\selectfont
\renewcommand{\arraystretch}{1.0}
\setlength{\tabcolsep}{3pt}\setlength{\parskip}{0pt}
\begin{longtable}{@{}>{\raggedright\arraybackslash}m{\dimexpr.07\linewidth-4\tabcolsep\relax}>{\raggedright\arraybackslash}m{.79\linewidth}>{\raggedleft\arraybackslash}m{.14\linewidth}@{}}
\caption{Aqueous speciation reactions (R35--R111) used in the Santa Barbara Basin, with log$K$ values at 5 $^\circ$C}\label{tab:sbb-speciation}\\
\toprule \textbf{ID} & \textbf{Reaction} & \textbf{$\log_{10}K$}\\ \midrule[0.6pt]
\endfirsthead
\multicolumn{3}{l}{\tablename~\thetable{} (continued)}\\
\toprule \textbf{ID} & \textbf{Reaction} & \textbf{$\log_{10}K$}\\ \midrule[0.6pt]
\endhead
\midrule\multicolumn{3}{r}{Continued on next page}\\
\endfoot\bottomrule\endlastfoot
\multicolumn{3}{@{}l}{\textbf{Water, carbonate, nutrients, sulfur, boron, fluoride and silica}}\\*[1pt]
\hypertarget{sbb-reaction:35}{R35} & %
    \singlelinerate{G4-speciation-35}{\mathrm{H_2O}+0.01e^-\rightleftharpoons\mathrm{H_2O}^{-0.01}} & %
    \(-9.000\) \\[.8pt]
\hypertarget{sbb-reaction:36}{R36} & %
    \singlelinerate{G4-speciation-36}{\mathrm{H_{2}O} \rightleftharpoons \mathrm{OH}^{-} + \mathrm{H}^{+}} & %
    \(-14.732\) \\[.8pt]
\hypertarget{sbb-reaction:37}{R37} & %
    \singlelinerate{G4-speciation-37}{2\,\mathrm{H_{2}O} \rightleftharpoons \mathrm{O_{2}} + 4\,\mathrm{H}^{+} + 4\,\mathrm{e}^{-}} & %
    \(-93.184\) \\[.8pt]
\hypertarget{sbb-reaction:38}{R38} & %
    \singlelinerate{G4-speciation-38}{2\,\mathrm{H}^{+} + 2\,\mathrm{e}^{-} \rightleftharpoons \mathrm{H_{2}}} & %
    \(-3.057\) \\[.8pt]
\hypertarget{sbb-reaction:39}{R39} & %
    \singlelinerate{G4-speciation-39}{\mathrm{H}^{+} + \mathrm{CO_{3}}^{2-} \rightleftharpoons \mathrm{HCO_{3}}^{-}} & %
    \(9.258\) \\[.8pt]
\hypertarget{sbb-reaction:40}{R40} & %
    \singlelinerate{G4-speciation-40}{2\,\mathrm{H}^{+} + \mathrm{CO_{3}}^{2-} \rightleftharpoons \mathrm{CO_{2}} + \mathrm{H_{2}O}} & %
    \(15.286\) \\[.8pt]
\hypertarget{sbb-reaction:41}{R41} & %
    \singlelinerate{G4-speciation-41}{\mathrm{H}^{+} + \mathrm{SO_{4}}^{2-} \rightleftharpoons \mathrm{HSO_{4}}^{-}} & %
    \(1.815\) \\[.8pt]
\hypertarget{sbb-reaction:42}{R42} & %
    \singlelinerate{G4-speciation-42}{\mathrm{HS}^{-} \rightleftharpoons \mathrm{S}^{2-} + \mathrm{H}^{+}} & %
    \(-13.556\) \\[.8pt]
\hypertarget{sbb-reaction:43}{R43} & %
    \singlelinerate{G4-speciation-43}{\mathrm{H}^{+} + \mathrm{HS}^{-} \rightleftharpoons \mathrm{H_{2}S}} & %
    \(7.255\) \\[.8pt]
\hypertarget{sbb-reaction:44}{R44} & %
    \singlelinerate{G4-speciation-44}{\mathrm{NH_{4}}^{+} \rightleftharpoons \mathrm{NH_{3}} + \mathrm{H}^{+}} & %
    \(-9.904\) \\[.8pt]
\hypertarget{sbb-reaction:45}{R45} & %
    \singlelinerate{G4-speciation-45}{\mathrm{NH_{4}}^{+} + \mathrm{SO_{4}}^{2-} \rightleftharpoons \mathrm{NH_{4}SO_{4}}^{-}} & %
    \(1.110\) \\[.8pt]
\hypertarget{sbb-reaction:46}{R46} & %
    \singlelinerate{G4-speciation-46}{\mathrm{H_{3}BO_{3}} \rightleftharpoons \mathrm{H_{2}BO_{3}}^{-} + \mathrm{H}^{+}} & %
    \(-9.410\) \\[.8pt]
\hypertarget{sbb-reaction:47}{R47} & %
    \singlelinerate{G4-speciation-47}{\mathrm{F}^{-} + \mathrm{H_{3}BO_{3}} \rightleftharpoons \mathrm{BF(OH)_{3}}^{-}} & %
    \(-0.498\) \\[.8pt]
\hypertarget{sbb-reaction:48}{R48} & %
    \singlelinerate{G4-speciation-48}{2\,\mathrm{F}^{-} + \mathrm{H}^{+} + \mathrm{H_{3}BO_{3}} \rightleftharpoons \mathrm{BF_{2}(OH)_{2}}^{-} + \mathrm{H_{2}O}} & %
    \(7.545\) \\[.8pt]
\hypertarget{sbb-reaction:49}{R49} & %
    \singlelinerate{G4-speciation-49}{3\,\mathrm{F}^{-} + 2\,\mathrm{H}^{+} + \mathrm{H_{3}BO_{3}} \rightleftharpoons \mathrm{BF_{3}OH}^{-} + 2\,\mathrm{H_{2}O}} & %
    \(13.755\) \\[.8pt]
\hypertarget{sbb-reaction:50}{R50} & %
    \singlelinerate{G4-speciation-50}{4\,\mathrm{F}^{-} + 3\,\mathrm{H}^{+} + \mathrm{H_{3}BO_{3}} \rightleftharpoons \mathrm{BF_{4}}^{-} + 3\,\mathrm{H_{2}O}} & %
    \(20.377\) \\[.8pt]
\hypertarget{sbb-reaction:51}{R51} & %
    \singlelinerate{G4-speciation-51}{\mathrm{H}^{+} + \mathrm{PO_{4}}^{3-} \rightleftharpoons \mathrm{HPO_{4}}^{2-}} & %
    \(12.532\) \\[.8pt]
\hypertarget{sbb-reaction:52}{R52} & %
    \singlelinerate{G4-speciation-52}{2\,\mathrm{H}^{+} + \mathrm{PO_{4}}^{3-} \rightleftharpoons \mathrm{H_{2}PO_{4}}^{-}} & %
    \(19.791\) \\[.8pt]
\hypertarget{sbb-reaction:53}{R53} & %
    \singlelinerate{G4-speciation-53}{\mathrm{F}^{-} + \mathrm{H}^{+} \rightleftharpoons \mathrm{HF}} & %
    \(3.027\) \\[.8pt]
\hypertarget{sbb-reaction:54}{R54} & %
    \singlelinerate{G4-speciation-54}{2\,\mathrm{F}^{-} + \mathrm{H}^{+} \rightleftharpoons \mathrm{HF_{2}}^{-}} & %
    \(3.520\) \\[.8pt]
\hypertarget{sbb-reaction:55}{R55} & %
    \singlelinerate{G4-speciation-55}{\mathrm{H_{4}SiO_{4}} \rightleftharpoons \mathrm{H_{3}SiO_{4}}^{-} + \mathrm{H}^{+}} & %
    \(-10.177\) \\[.8pt]
\hypertarget{sbb-reaction:56}{R56} & %
    \singlelinerate{G4-speciation-56}{\mathrm{H_{4}SiO_{4}} \rightleftharpoons \mathrm{H_{2}SiO_{4}}^{2-} + 2\,\mathrm{H}^{+}} & %
    \(-23.982\) \\[.8pt]
\hypertarget{sbb-reaction:57}{R57} & %
    \singlelinerate{G4-speciation-57}{6\,\mathrm{F}^{-} + 4\,\mathrm{H}^{+} + \mathrm{H_{4}SiO_{4}} \rightleftharpoons \mathrm{SiF_{6}}^{2-} + 4\,\mathrm{H_{2}O}} & %
    \(31.037\) \\[.8pt]
\midrule\multicolumn{3}{@{}l}{\textbf{Major-cation complexes}}\\*[1pt]
\hypertarget{sbb-reaction:58}{R58} & %
    \singlelinerate{G4-speciation-58}{\mathrm{Ca}^{2+} + \mathrm{H_{2}O} \rightleftharpoons \mathrm{CaOH}^{+} + \mathrm{H}^{+}} & %
    \(-12.780\) \\[.8pt]
\hypertarget{sbb-reaction:59}{R59} & %
    \singlelinerate{G4-speciation-59}{\mathrm{Ca}^{2+} + \mathrm{CO_{3}}^{2-} \rightleftharpoons \mathrm{CaCO_{3}}} & %
    \(3.128\) \\[.8pt]
\hypertarget{sbb-reaction:60}{R60} & %
    \singlelinerate{G4-speciation-60}{\mathrm{Ca}^{2+} + \mathrm{H}^{+} + \mathrm{CO_{3}}^{2-} \rightleftharpoons \mathrm{CaHCO_{3}}^{+}} & %
    \(11.455\) \\[.8pt]
\hypertarget{sbb-reaction:61}{R61} & %
    \singlelinerate{G4-speciation-61}{\mathrm{Ca}^{2+} + \mathrm{SO_{4}}^{2-} \rightleftharpoons \mathrm{CaSO_{4}}} & %
    \(2.213\) \\[.8pt]
\hypertarget{sbb-reaction:62}{R62} & %
    \singlelinerate{G4-speciation-62}{\mathrm{Ca}^{2+} + \mathrm{HSO_{4}}^{-} \rightleftharpoons \mathrm{CaHSO_{4}}^{+}} & %
    \(1.080\) \\[.8pt]
\hypertarget{sbb-reaction:63}{R63} & %
    \singlelinerate{G4-speciation-63}{\mathrm{Ca}^{2+} + \mathrm{PO_{4}}^{3-} \rightleftharpoons \mathrm{CaPO_{4}}^{-}} & %
    \(6.296\) \\[.8pt]
\hypertarget{sbb-reaction:64}{R64} & %
    \singlelinerate{G4-speciation-64}{\mathrm{Ca}^{2+} + \mathrm{HPO_{4}}^{2-} \rightleftharpoons \mathrm{CaHPO_{4}}} & %
    \(2.565\) \\[.8pt]
\hypertarget{sbb-reaction:65}{R65} & %
    \singlelinerate{G4-speciation-65}{\mathrm{Ca}^{2+} + \mathrm{H_{2}PO_{4}}^{-} \rightleftharpoons \mathrm{CaH_{2}PO_{4}}^{+}} & %
    \(1.229\) \\[.8pt]
\hypertarget{sbb-reaction:66}{R66} & %
    \singlelinerate{G4-speciation-66}{\mathrm{Ca}^{2+} + \mathrm{F}^{-} \rightleftharpoons \mathrm{CaF}^{+}} & %
    \(0.723\) \\[.8pt]
\hypertarget{sbb-reaction:67}{R67} & %
    \singlelinerate{G4-speciation-67}{\mathrm{H_{2}O} + \mathrm{Mg}^{2+} \rightleftharpoons \mathrm{MgOH}^{+} + \mathrm{H}^{+}} & %
    \(-12.281\) \\[.8pt]
\hypertarget{sbb-reaction:68}{R68} & %
    \singlelinerate{G4-speciation-68}{\mathrm{Mg}^{2+} + \mathrm{CO_{3}}^{2-} \rightleftharpoons \mathrm{MgCO_{3}}} & %
    \(2.846\) \\[.8pt]
\hypertarget{sbb-reaction:69}{R69} & %
    \singlelinerate{G4-speciation-69}{\mathrm{H}^{+} + \mathrm{Mg}^{2+} + \mathrm{CO_{3}}^{2-} \rightleftharpoons \mathrm{MgHCO_{3}}^{+}} & %
    \(11.604\) \\[.8pt]
\hypertarget{sbb-reaction:70}{R70} & %
    \singlelinerate{G4-speciation-70}{\mathrm{Mg}^{2+} + \mathrm{SO_{4}}^{2-} \rightleftharpoons \mathrm{MgSO_{4}}} & %
    \(2.130\) \\[.8pt]
\hypertarget{sbb-reaction:71}{R71} & %
    \singlelinerate{G4-speciation-71}{\mathrm{Mg}^{2+} + \mathrm{PO_{4}}^{3-} \rightleftharpoons \mathrm{MgPO_{4}}^{-}} & %
    \(6.426\) \\[.8pt]
\hypertarget{sbb-reaction:72}{R72} & %
    \singlelinerate{G4-speciation-72}{\mathrm{HPO_{4}}^{2-} + \mathrm{Mg}^{2+} \rightleftharpoons \mathrm{MgHPO_{4}}} & %
    \(2.696\) \\[.8pt]
\hypertarget{sbb-reaction:73}{R73} & %
    \singlelinerate{G4-speciation-73}{\mathrm{H_{2}PO_{4}}^{-} + \mathrm{Mg}^{2+} \rightleftharpoons \mathrm{MgH_{2}PO_{4}}^{+}} & %
    \(1.334\) \\[.8pt]
\hypertarget{sbb-reaction:74}{R74} & %
    \singlelinerate{G4-speciation-74}{\mathrm{F}^{-} + \mathrm{Mg}^{2+} \rightleftharpoons \mathrm{MgF}^{+}} & %
    \(1.651\) \\[.8pt]
\hypertarget{sbb-reaction:75}{R75} & %
    \singlelinerate{G4-speciation-75}{\mathrm{H_{2}O} + \mathrm{Na}^{+} \rightleftharpoons \mathrm{NaOH} + \mathrm{H}^{+}} & %
    \(-14.180\) \\[.8pt]
\hypertarget{sbb-reaction:76}{R76} & %
    \singlelinerate{G4-speciation-76}{\mathrm{Na}^{+} + \mathrm{CO_{3}}^{2-} \rightleftharpoons \mathrm{NaCO_{3}}^{-}} & %
    \(0.800\) \\[.8pt]
\hypertarget{sbb-reaction:77}{R77} & %
    \singlelinerate{G4-speciation-77}{\mathrm{HCO_{3}}^{-} + \mathrm{Na}^{+} \rightleftharpoons \mathrm{NaHCO_{3}}} & %
    \(-0.250\) \\[.8pt]
\hypertarget{sbb-reaction:78}{R78} & %
    \singlelinerate{G4-speciation-78}{\mathrm{Na}^{+} + \mathrm{SO_{4}}^{2-} \rightleftharpoons \mathrm{NaSO_{4}}^{-}} & %
    \(0.641\) \\[.8pt]
\hypertarget{sbb-reaction:79}{R79} & %
    \singlelinerate{G4-speciation-79}{\mathrm{HPO_{4}}^{2-} + \mathrm{Na}^{+} \rightleftharpoons \mathrm{NaHPO_{4}}^{-}} & %
    \(0.290\) \\[.8pt]
\hypertarget{sbb-reaction:80}{R80} & %
    \singlelinerate{G4-speciation-80}{\mathrm{F}^{-} + \mathrm{Na}^{+} \rightleftharpoons \mathrm{NaF}} & %
    \(-0.240\) \\[.8pt]
\hypertarget{sbb-reaction:81}{R81} & %
    \singlelinerate{G4-speciation-81}{\mathrm{H_{2}O} + \mathrm{K}^{+} \rightleftharpoons \mathrm{KOH} + \mathrm{H}^{+}} & %
    \(-14.460\) \\[.8pt]
\hypertarget{sbb-reaction:82}{R82} & %
    \singlelinerate{G4-speciation-82}{\mathrm{K}^{+} + \mathrm{SO_{4}}^{2-} \rightleftharpoons \mathrm{KSO_{4}}^{-}} & %
    \(0.684\) \\[.8pt]
\hypertarget{sbb-reaction:83}{R83} & %
    \singlelinerate{G4-speciation-83}{\mathrm{HPO_{4}}^{2-} + \mathrm{K}^{+} \rightleftharpoons \mathrm{KHPO_{4}}^{-}} & %
    \(0.290\) \\[.8pt]
\midrule\multicolumn{3}{@{}l}{\textbf{Iron(II) complexes}}\\*[1pt]
\hypertarget{sbb-reaction:84}{R84} & %
    \singlelinerate{G4-speciation-84}{\mathrm{Fe}^{2+} + \mathrm{H_{2}O} \rightleftharpoons \mathrm{FeOH}^{+} + \mathrm{H}^{+}} & %
    \(-10.196\) \\[.8pt]
\hypertarget{sbb-reaction:85}{R85} & %
    \singlelinerate{G4-speciation-85}{\mathrm{Cl}^{-} + \mathrm{Fe}^{2+} \rightleftharpoons \mathrm{FeCl}^{+}} & %
    \(0.140\) \\[.8pt]
\hypertarget{sbb-reaction:86}{R86} & %
    \singlelinerate{G4-speciation-86}{\mathrm{Fe}^{2+} + \mathrm{CO_{3}}^{2-} \rightleftharpoons \mathrm{FeCO_{3}}} & %
    \(4.380\) \\[.8pt]
\hypertarget{sbb-reaction:87}{R87} & %
    \singlelinerate{G4-speciation-87}{\mathrm{Fe}^{2+} + \mathrm{HCO_{3}}^{-} \rightleftharpoons \mathrm{FeHCO_{3}}^{+}} & %
    \(2.000\) \\[.8pt]
\hypertarget{sbb-reaction:88}{R88} & %
    \singlelinerate{G4-speciation-88}{\mathrm{Fe}^{2+} + \mathrm{SO_{4}}^{2-} \rightleftharpoons \mathrm{FeSO_{4}}} & %
    \(2.080\) \\[.8pt]
\hypertarget{sbb-reaction:89}{R89} & %
    \singlelinerate{G4-speciation-89}{\mathrm{Fe}^{2+} + \mathrm{HSO_{4}}^{-} \rightleftharpoons \mathrm{FeHSO_{4}}^{+}} & %
    \(1.080\) \\[.8pt]
\hypertarget{sbb-reaction:90}{R90} & %
    \singlelinerate{G4-speciation-90}{\mathrm{Fe}^{2+} + 2\,\mathrm{HS}^{-} \rightleftharpoons \mathrm{Fe(HS)_{2}}} & %
    \(8.950\) \\[.8pt]
\hypertarget{sbb-reaction:91}{R91} & %
    \singlelinerate{G4-speciation-91}{\mathrm{Fe}^{2+} + 3\,\mathrm{HS}^{-} \rightleftharpoons \mathrm{Fe(HS)_{3}}^{-}} & %
    \(10.987\) \\[.8pt]
\hypertarget{sbb-reaction:92}{R92} & %
    \singlelinerate{G4-speciation-92}{\mathrm{Fe}^{2+} + \mathrm{HPO_{4}}^{2-} \rightleftharpoons \mathrm{FeHPO_{4}}} & %
    \(3.600\) \\[.8pt]
\hypertarget{sbb-reaction:93}{R93} & %
    \singlelinerate{G4-speciation-93}{\mathrm{Fe}^{2+} + \mathrm{H_{2}PO_{4}}^{-} \rightleftharpoons \mathrm{FeH_{2}PO_{4}}^{+}} & %
    \(2.700\) \\[.8pt]
\hypertarget{sbb-reaction:94}{R94} & %
    \singlelinerate{G4-speciation-94}{\mathrm{F}^{-} + \mathrm{Fe}^{2+} \rightleftharpoons \mathrm{FeF}^{+}} & %
    \(1.000\) \\[.8pt]
\midrule\multicolumn{3}{@{}l}{\textbf{Manganese(II) complexes}}\\*[1pt]
\hypertarget{sbb-reaction:95}{R95} & %
    \singlelinerate{G4-speciation-95}{\mathrm{H_{2}O} + \mathrm{Mn}^{2+} \rightleftharpoons \mathrm{MnOH}^{+} + \mathrm{H}^{+}} & %
    \(-11.349\) \\[.8pt]
\hypertarget{sbb-reaction:96}{R96} & %
    \singlelinerate{G4-speciation-96}{\mathrm{Cl}^{-} + \mathrm{Mn}^{2+} \rightleftharpoons \mathrm{MnCl}^{+}} & %
    \(0.610\) \\[.8pt]
\hypertarget{sbb-reaction:97}{R97} & %
    \singlelinerate{G4-speciation-97}{2\,\mathrm{Cl}^{-} + \mathrm{Mn}^{2+} \rightleftharpoons \mathrm{MnCl_{2}}} & %
    \(0.250\) \\[.8pt]
\hypertarget{sbb-reaction:98}{R98} & %
    \singlelinerate{G4-speciation-98}{3\,\mathrm{Cl}^{-} + \mathrm{Mn}^{2+} \rightleftharpoons \mathrm{MnCl_{3}}^{-}} & %
    \(-0.310\) \\[.8pt]
\hypertarget{sbb-reaction:99}{R99} & %
    \singlelinerate{G4-speciation-99}{\mathrm{Mn}^{2+} + \mathrm{CO_{3}}^{2-} \rightleftharpoons \mathrm{MnCO_{3}}} & %
    \(4.900\) \\[.8pt]
\hypertarget{sbb-reaction:100}{R100} & %
    \singlelinerate{G4-speciation-100}{\mathrm{HCO_{3}}^{-} + \mathrm{Mn}^{2+} \rightleftharpoons \mathrm{MnHCO_{3}}^{+}} & %
    \(1.950\) \\[.8pt]
\hypertarget{sbb-reaction:101}{R101} & %
    \singlelinerate{G4-speciation-101}{\mathrm{Mn}^{2+} + \mathrm{SO_{4}}^{2-} \rightleftharpoons \mathrm{MnSO_{4}}} & %
    \(2.072\) \\[.8pt]
\hypertarget{sbb-reaction:102}{R102} & %
    \singlelinerate{G4-speciation-102}{\mathrm{Mn}^{2+} + 2\,\mathrm{NO_{3}}^{-} \rightleftharpoons \mathrm{Mn(NO_{3})_{2}}} & %
    \(0.621\) \\[.8pt]
\hypertarget{sbb-reaction:103}{R103} & %
    \singlelinerate{G4-speciation-103}{\mathrm{F}^{-} + \mathrm{Mn}^{2+} \rightleftharpoons \mathrm{MnF}^{+}} & %
    \(0.840\) \\[.8pt]
\midrule\multicolumn{3}{@{}l}{\textbf{Molybdenum isotope reactions}}\\*[1pt]
\hypertarget{sbb-reaction:104}{R104} & %
    \singlelinerate{G4-speciation-104}{{}^{95}\mathrm{MoO_{4}}^{2-} + \mathrm{H_{2}S} \rightleftharpoons {}^{95}\mathrm{MoO_{3}S}^{2-} + \mathrm{H_{2}O}} & %
    \(5.191\) \\[.8pt]
\hypertarget{sbb-reaction:105}{R105} & %
    \singlelinerate{G4-speciation-105}{{}^{95}\mathrm{MoO_{4}}^{2-} + 2\,\mathrm{H_{2}S} \rightleftharpoons {}^{95}\mathrm{MoO_{2}S_{2}}^{2-} + 2\,\mathrm{H_{2}O}} & %
    \(9.991\) \\[.8pt]
\hypertarget{sbb-reaction:106}{R106} & %
    \singlelinerate{G4-speciation-106}{{}^{95}\mathrm{MoO_{4}}^{2-} + 3\,\mathrm{H_{2}S} \rightleftharpoons {}^{95}\mathrm{MoOS_{3}}^{2-} + 3\,\mathrm{H_{2}O}} & %
    \(14.992\) \\[.8pt]
\hypertarget{sbb-reaction:107}{R107} & %
    \singlelinerate{G4-speciation-107}{{}^{95}\mathrm{MoO_{4}}^{2-} + 4\,\mathrm{H_{2}S} \rightleftharpoons {}^{95}\mathrm{MoS_{4}}^{2-} + 4\,\mathrm{H_{2}O}} & %
    \(19.873\) \\[.8pt]
\hypertarget{sbb-reaction:108}{R108} & %
    \singlelinerate{G4-speciation-108}{{}^{98}\mathrm{MoO_{4}}^{2-} + \mathrm{H_{2}S} \rightleftharpoons {}^{98}\mathrm{MoO_{3}S}^{2-} + \mathrm{H_{2}O}} & %
    \(5.190\) \\[.8pt]
\hypertarget{sbb-reaction:109}{R109} & %
    \singlelinerate{G4-speciation-109}{{}^{98}\mathrm{MoO_{4}}^{2-} + 2\,\mathrm{H_{2}S} \rightleftharpoons {}^{98}\mathrm{MoO_{2}S_{2}}^{2-} + 2\,\mathrm{H_{2}O}} & %
    \(9.990\) \\[.8pt]
\hypertarget{sbb-reaction:110}{R110} & %
    \singlelinerate{G4-speciation-110}{{}^{98}\mathrm{MoO_{4}}^{2-} + 3\,\mathrm{H_{2}S} \rightleftharpoons {}^{98}\mathrm{MoOS_{3}}^{2-} + 3\,\mathrm{H_{2}O}} & %
    \(14.990\) \\[.8pt]
\hypertarget{sbb-reaction:111}{R111} & %
    \singlelinerate{G4-speciation-111}{{}^{98}\mathrm{MoO_{4}}^{2-} + 4\,\mathrm{H_{2}S} \rightleftharpoons {}^{98}\mathrm{MoS_{4}}^{2-} + 4\,\mathrm{H_{2}O}} & %
    \(19.870\) \\[.8pt]
\end{longtable}\endgroup
\begingroup\linespread{1}\fontsize{9}{10.5}\selectfont\noindent\emph{Notes.} $\log K$ values use database analytical temperature expressions, or enthalpy corrections when provided. These reactions are incorporated in \texttt{phreeqc.dat} to prepare a customized database. 
\par\smallskip\noindent
For the forward thiolation reactions R104--R111, the molybdate-to-thiomolybdate equilibrium isotope fractionation is defined as
\[
\alpha_j=
\frac{[{}^{98}\mathrm{MoO}_4^{2-}]/[{}^{95}\mathrm{MoO}_4^{2-}]}
{[{}^{98}\mathrm{MoO}_{4-j}\mathrm{S}_j^{2-}]/[{}^{95}\mathrm{MoO}_{4-j}\mathrm{S}_j^{2-}]}
=\frac{K_j^{95}}{K_j^{98}},\qquad j=1,\ldots,4.
\]
The isotope-specific constants follow the form used in Du \citep{Du2023}:
\[
\log_{10}K_j^{95}=\log_{10}K_j^{98}+\log_{10}\alpha_j,
\qquad \alpha_j=1+\epsilon_j/1000.
\]
The reference $\log_{10}K_j^{98}$ values are $(5.190,9.990,14.990,19.870)$ for $j=1,\ldots,4$; the enrichments $\epsilon_j$ are listed in Table~\ref{tab:sbb-coefficients}.\par\endgroup
\clearpage
\section{Additional simulation results}\label{sec:additional-results}

\begingroup\setlength{\intextsep}{0pt}
\begin{figure}[H]
\centering
\includegraphics[width=1.0\linewidth,height=0.66\textheight,keepaspectratio]{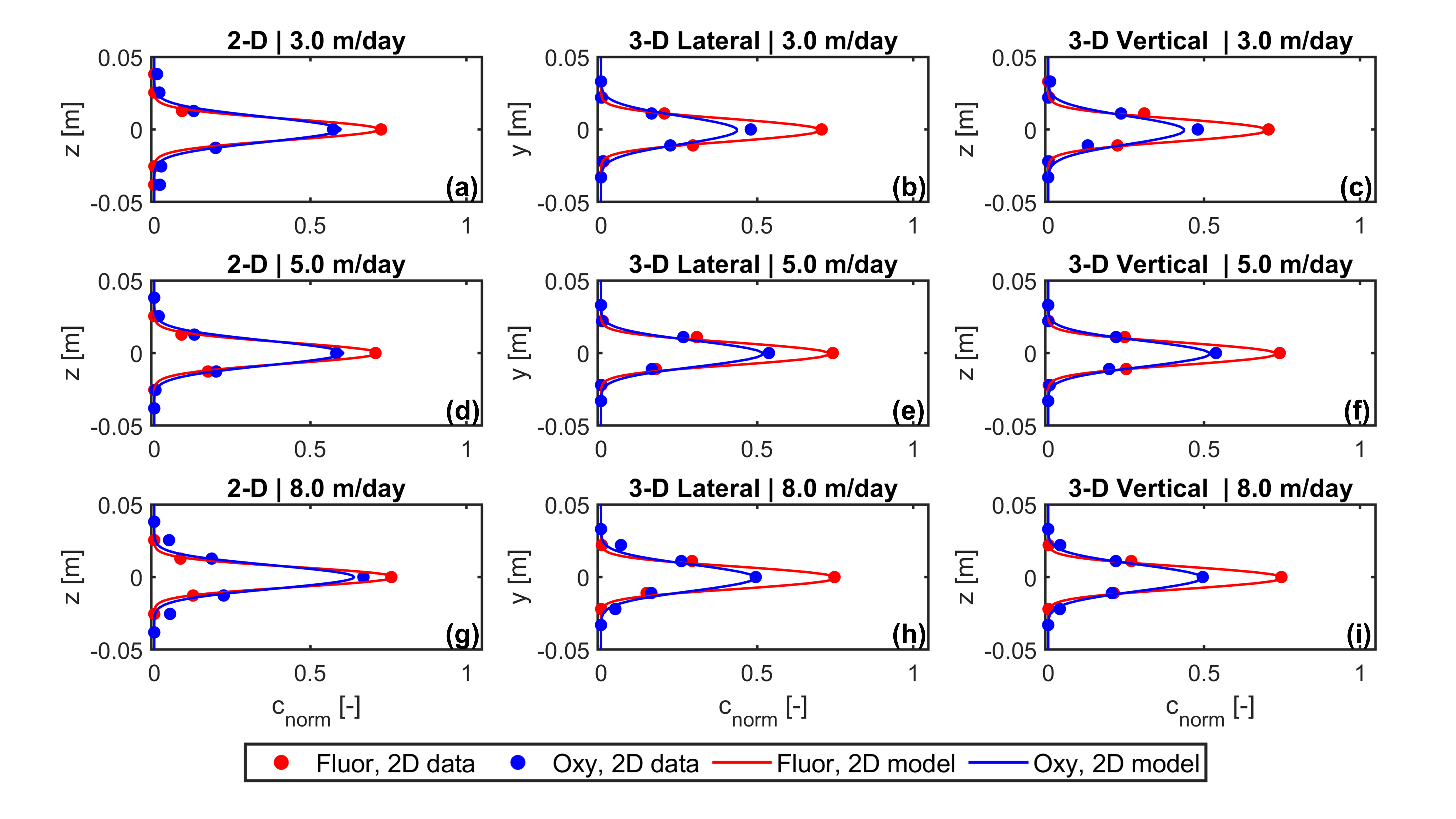}
\normalsize\let\footnotesize\normalsize
\caption{Measured (markers) versus simulated (solid lines) spatial profiles of fluorescein and oxygen at the outlet of the quasi 2-D setup and the 3-D system along the horizontal and vertical directions for the seepage velocity of 3 m/d (a-c), 5 m/d (d-f) and 8 m/d (g-i).}\label{fig:si-1}
\end{figure}\endgroup

\clearpage
\begingroup\setlength{\intextsep}{0pt}
\begin{figure}[H]
\centering
\includegraphics[width=1.0\linewidth,height=0.66\textheight,keepaspectratio]{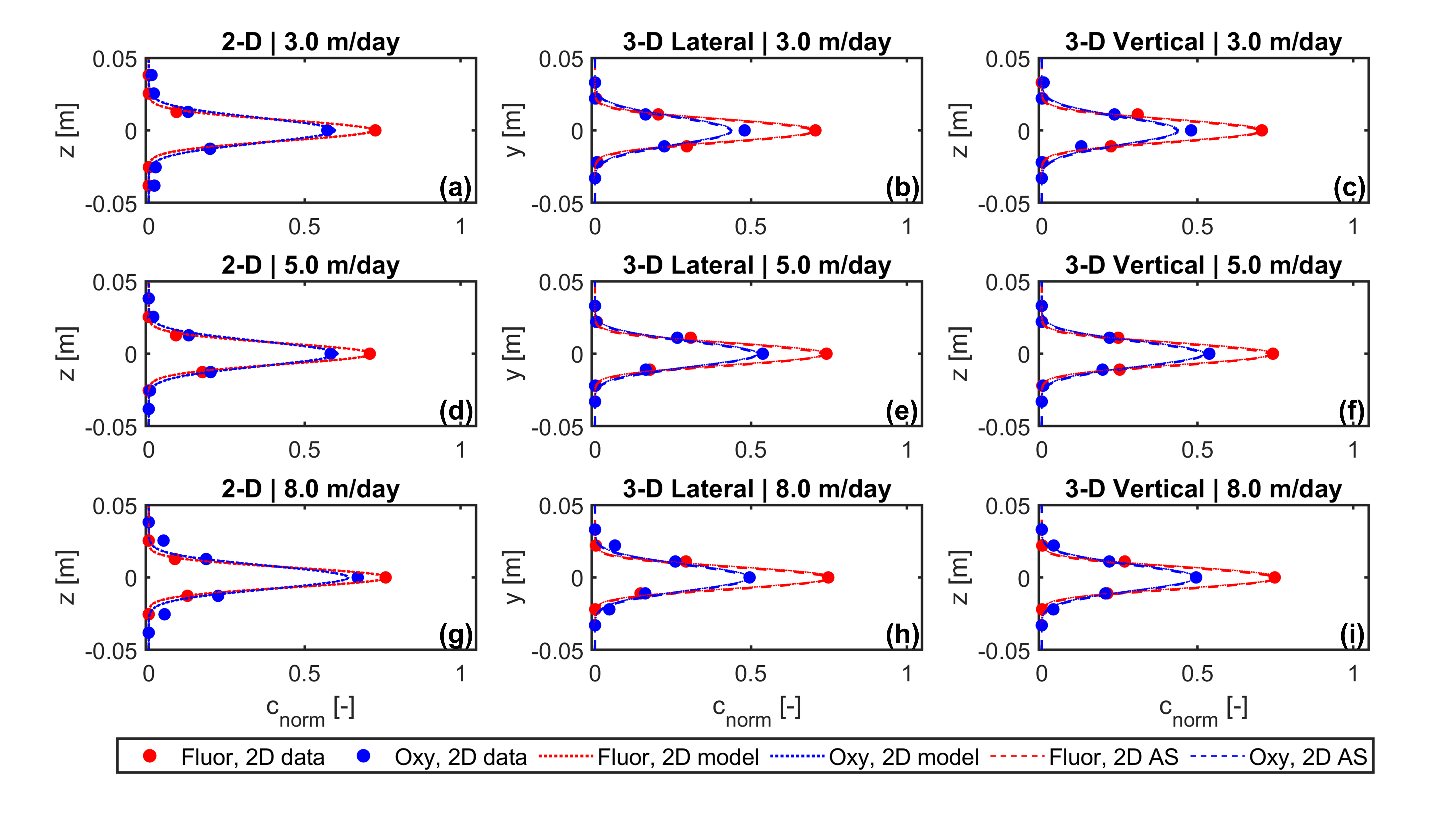}
\normalsize\let\footnotesize\normalsize
\caption{Comparison between the simulation results of marineGeochemFoam (solid lines), analytical solutions (dashed lines) and experimental data (markers) for the quasi 2-D setup and the 3-D system along the horizontal and vertical directions for the seepage velocity of 3 m/d (a-c), 5 m/d (d-f) and 8 m/d (g-i).}\label{fig:si-2}
\end{figure}\endgroup

\clearpage
\begingroup\setlength{\intextsep}{0pt}
\begin{figure}[H]
\centering
\includegraphics[width=1.0\linewidth,height=0.66\textheight,keepaspectratio]{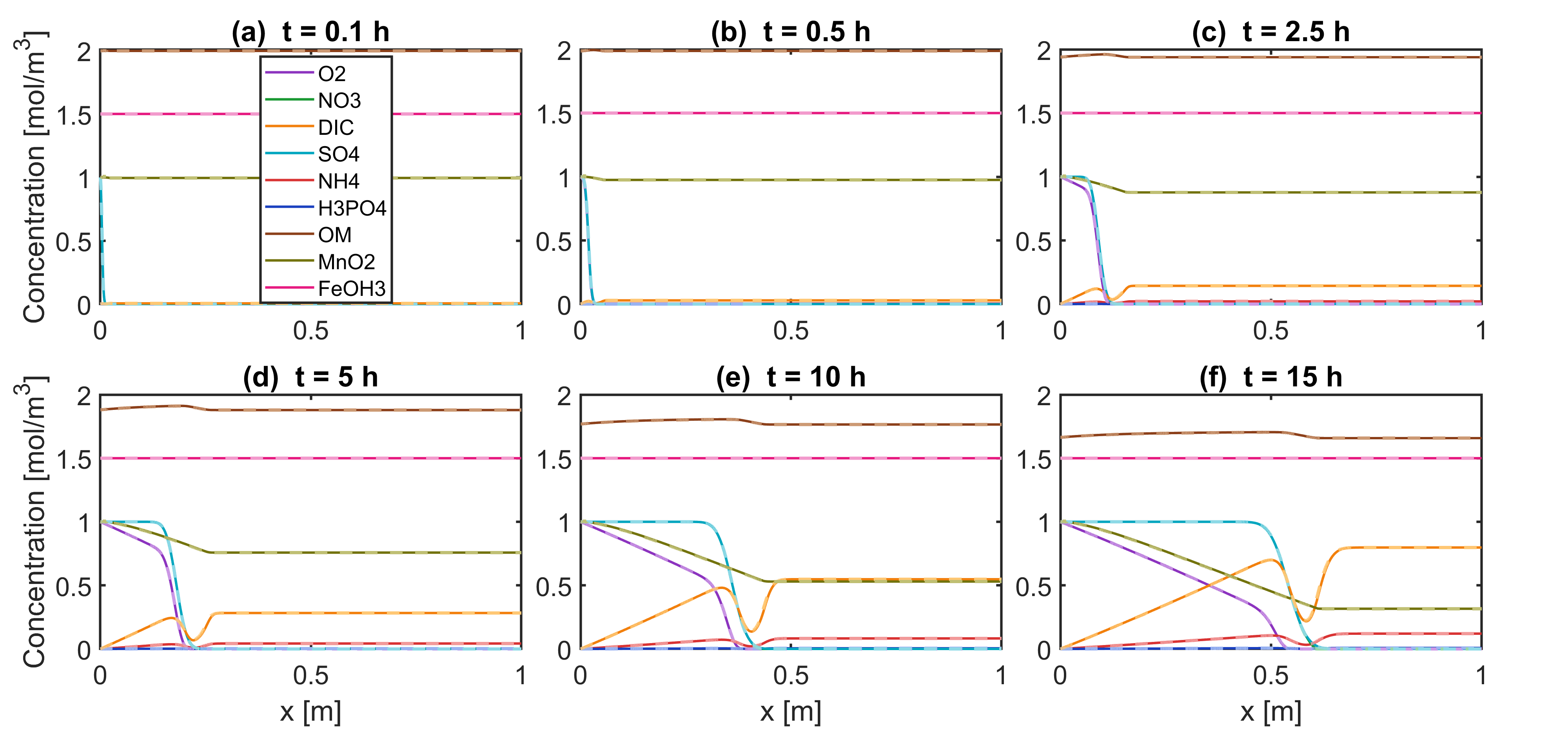}
\normalsize\let\footnotesize\normalsize
\caption{Simulated solute and solid species profiles at t=0.1, 0.5, 2.5, 5, 10 and 15 hours (a-f) using marineGeochemFoam (solid lines) and COMSOL Multiphysics (dashed lines). Solutes enter from the left boundary, whereas solids are immobile (Benchmark 3A). Note that the NO\textsubscript{3} and SO\textsubscript{4} profiles overlap in this example due to their similar reactive transport behavior.}\label{fig:si-3}
\end{figure}\endgroup

\vspace{3em}
Figure \ref{fig:si-3} shows the results from Benchmark 3A. The evolution of solute species shows further advancement of the sulfate and nitrate front compared to oxygen. In fact, as the electron acceptors propagate, their different energetic favorability represented by the inhibition terms in the reactive transport equations leads to their sequential consumption through oxidation of organic matter. This results in the spatial and temporal evolution of redox zonation and propagation of distinct reaction fronts for oxygen, nitrate and sulfate. Although organic matter oxidation is primarily controlled by the front of incoming solutes, the presence of MnO\textsubscript{2} as the solid electron acceptor generates a different zone for OM oxidation. This leads to the development of gradients at the moving interface between these two oxidation zones, which also shifts DIC production pattern across the two oxidation zones. Unlike MnO\textsubscript{2}, reduction of Fe(OH)\textsubscript{3} remains limited in this example due to the continuous supply of more energetically favorable electron acceptors. The excellent agreement between the two simulators for all species demonstrates that \emph{marineGeochemFoam} is capable of simulating reactive transport processes involving both solid and solute species and competitive redox reactions.

\clearpage
Figure \ref{fig:si-4} shows the results from Benchmark 4A, in which we simulated a 1-D problem involving ion exchange and advective/\allowbreak dispersive transport in a column. The exchanger with a cation exchange capacity (CEC) of 0.0011 eq/L is initially in equilibrium with a solution of 1 mM NaNO\textsubscript{3} and 0.2 mM KNO\textsubscript{3}. The column is continuously flushed with a 0.6 mM CaCl\textsubscript{2} solution, which promotes the competitive exchange of Ca\textsuperscript{+2}, Na\textsuperscript{+} and K\textsuperscript{+} as they migrate along the column. Figure~\ref{fig:si-4} shows the temporal evolution of the ions at the column outlet.

\begingroup\setlength{\intextsep}{0pt}
\begin{figure}[H]
\centering
\includegraphics[width=0.82\linewidth,height=0.66\textheight,keepaspectratio]{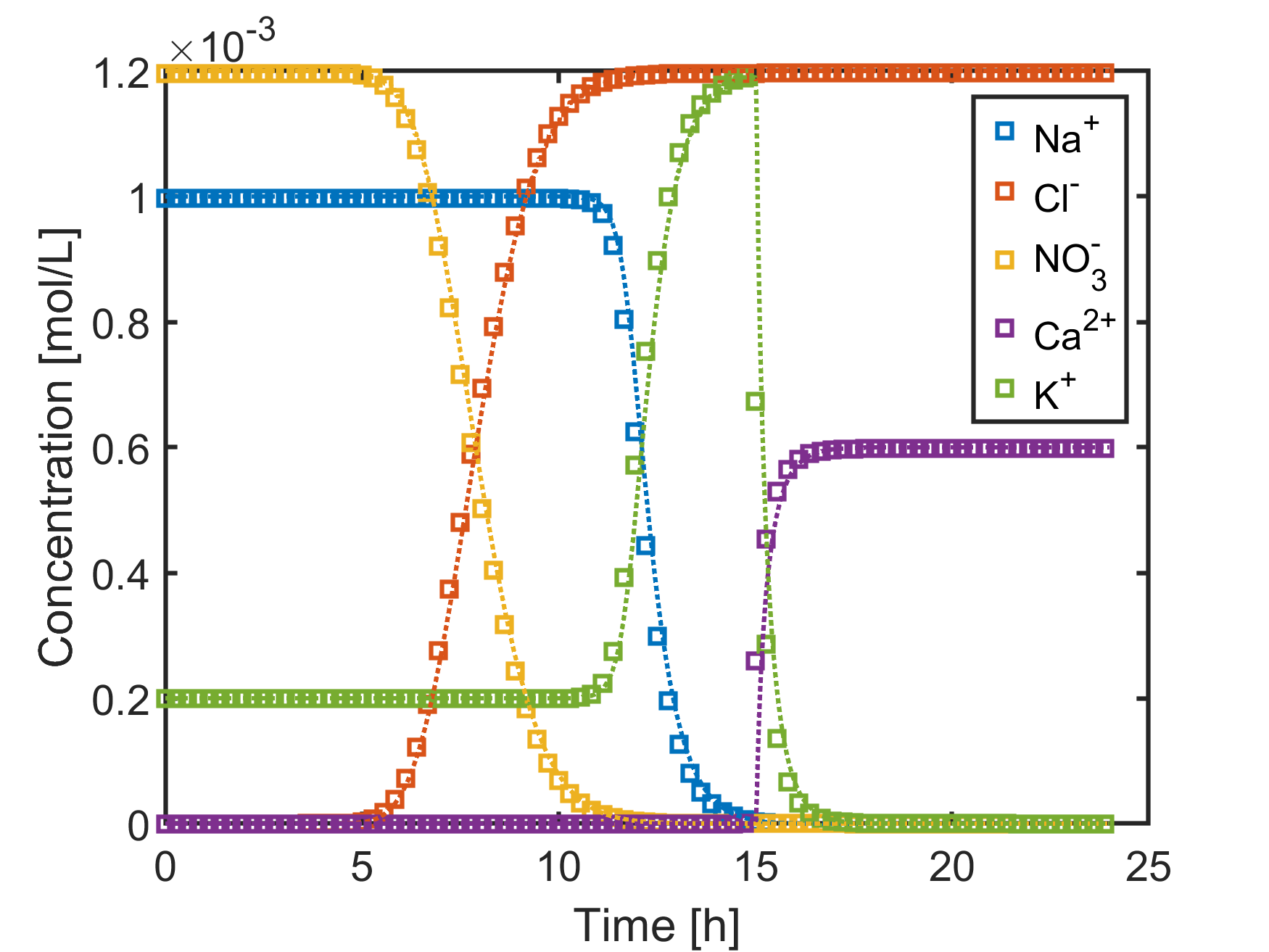}
\normalsize\let\footnotesize\normalsize
\caption{Simulated temporal evolution of aqueous species at the column outlet using marineGeochemFoam (dashed lines) and PHREEQC (dotted lines). In this example (Benchmark 4A), marineGeochemFoam uses the PhreeqcRM coupling package.}\label{fig:si-4}
\end{figure}\endgroup

\vspace{3em}
The temporal evolution of cation concentrations at the column outlet is controlled by the equilibrium exchange reactions between the cations and the exchanger. As the injected solution migrates through the column, Ca\textsuperscript{2+} progressively displaces Na\textsuperscript{+} and K\textsuperscript{+} from the exchange sites. Due to the weak affinity of sodium for the exchanger, it is eluted first, followed by potassium. Calcium, however, is preferentially retained on the exchanger and therefore reaches the outlet at the later times. The results of the two simulators show a very good agreement, which validates the implementation of the OpenFOAM-PhreeqcRM coupling framework as an additional feature of \emph{marineGeochemFoam} for simulation of reactive transport problems involving equilibrium reactions.

\clearpage
\begingroup\setlength{\intextsep}{0pt}
\begin{figure}[H]
\centering
\includegraphics[width=1.0\linewidth,height=0.66\textheight,keepaspectratio]{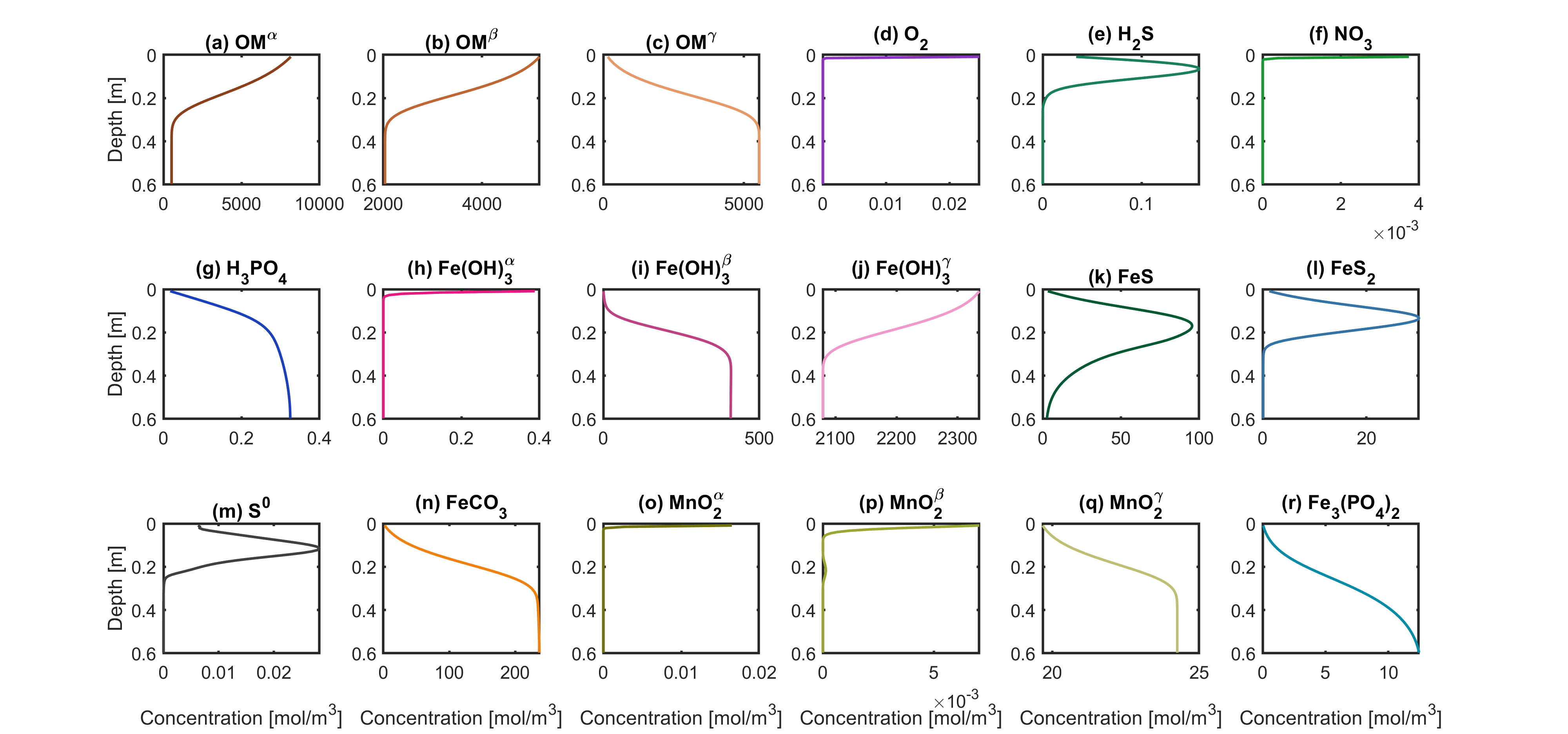}
\normalsize\let\footnotesize\normalsize
\caption{Simulated spatial profiles of solute and solid species at t=50 years in pockmarks E in Hanko site, Finland. The simulations were performed by the multiKineticsLinearized package.}\label{fig:si-5}
\end{figure}\endgroup

\begingroup\setlength{\intextsep}{0pt}
\begin{figure}[H]
\centering
\includegraphics[width=1.0\linewidth,height=0.66\textheight,keepaspectratio]{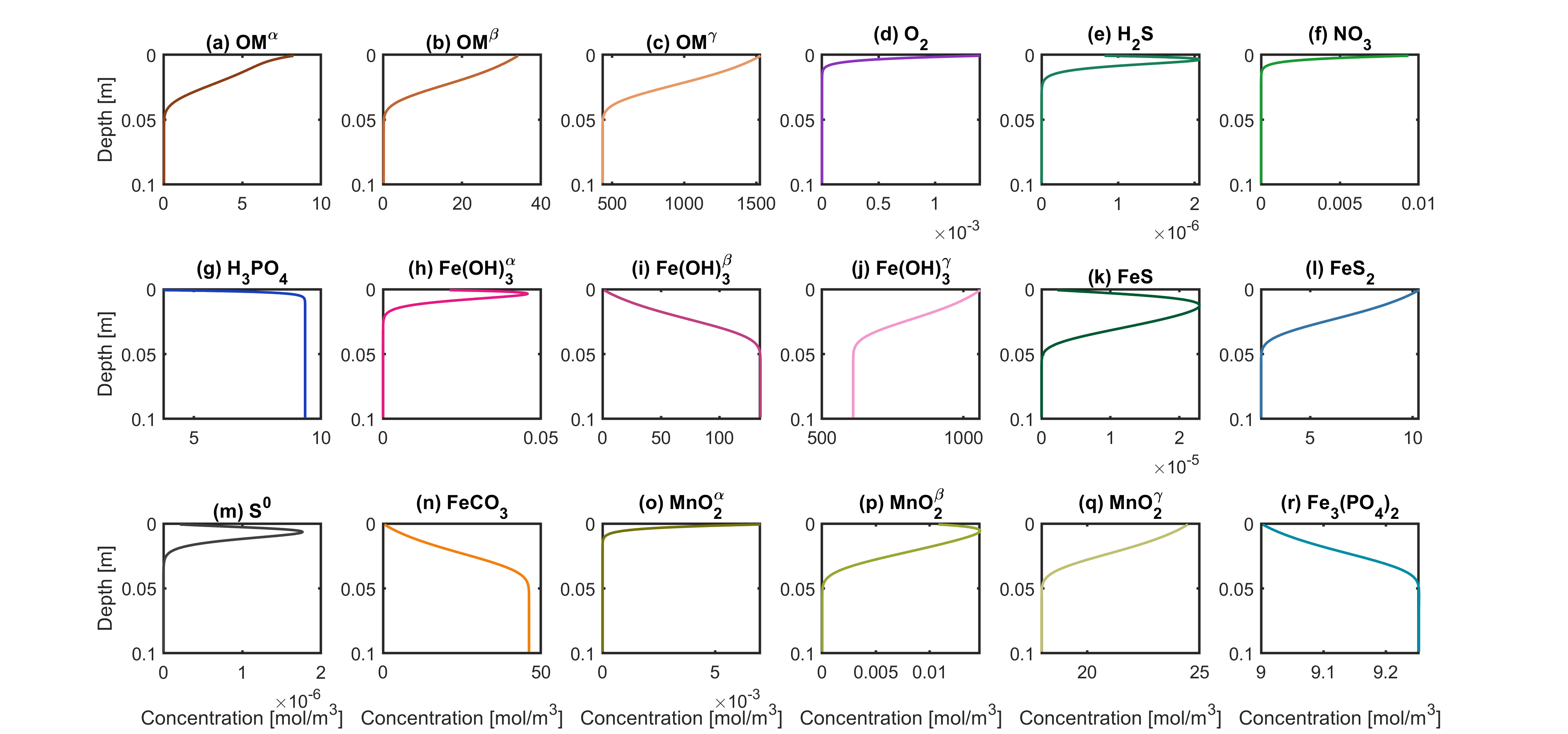}
\normalsize\let\footnotesize\normalsize
\caption{Simulated spatial profiles of solute and solid species at t=50 years in pockmarks D in Hanko site, Finland. The simulations were performed by the multiKineticsLinearized package.}\label{fig:si-6}
\end{figure}\endgroup

\clearpage

\nocite{Meysman2003}
\bibliographystyle{elsarticle-num-names}
\bibliography{si_references}